\pdfoutput=1
\documentclass{jfm}
\usepackage[utf8x]{inputenc}
\usepackage{graphicx}
\usepackage{float}
\usepackage{amsmath}
\usepackage{amssymb}
\usepackage{natbib}
\usepackage{bm}
\usepackage{color}
\usepackage[breaklinks=true]{hyperref}
\usepackage{breakcites}
\hypersetup{
    colorlinks=true,
    linkcolor=black,
    citecolor=black,
    filecolor=black,
    urlcolor=black,
}

\ifCUPmtlplainloaded \else
  \checkfont{eurm10}
  \iffontfound
    \IfFileExists{upmath.sty}
      {\typeout{^^JFound AMS Euler Roman fonts on the system,
                   using the 'upmath' package.^^J}%
       \usepackage{upmath}}
      {\typeout{^^JFound AMS Euler Roman fonts on the system, but you
                   dont seem to have the}%
       \typeout{'upmath' package installed. JFM.cls can take advantage
                 of these fonts,^^Jif you use 'upmath' package.^^J}%
      }
  \else
  \fi
\fi

\ifCUPmtlplainloaded \else
  \checkfont{msam10}
  \iffontfound
    \IfFileExists{amssymb.sty}
      {\typeout{^^JFound AMS Symbol fonts on the system, using the
                'amssymb' package.^^J}%
       \usepackage{amssymb}%
         \let\leq=\leqslant
         \let\geq=\geqslant
      }{}
  \fi
\fi

\ifCUPmtlplainloaded \else
  \IfFileExists{amsbsy.sty}
    {\typeout{^^JFound the 'amsbsy' package on the system, using it.^^J}%
     \usepackage{amsbsy}}
    {\providecommand\boldsymbol[1]{\mbox{\boldmath $##1$}}}
\fi

\newcommand{\figs}{./figs}

\title[Reynolds-number effects on inertial particle dynamics, Part I.]
      {The effect of Reynolds number on inertial particle dynamics in isotropic turbulence.
       Part I: Simulations without gravitational effects.}

\author[P. J. Ireland et al.]%
{Peter J. Ireland,\ns Andrew D. Bragg\thanks{Present address:  Now with the Applied Mathematics \& Plasma Physics Group, Los Alamos National Laboratory, Los Alamos, NM  87545, USA.},\ns and Lance R. Collins\thanks{Email address for correspondence: lc246@cornell.edu}}

\affiliation{Sibley School of Mechanical and Aerospace Engineering, Cornell University, Ithaca, NY  14853, USA\\[\affilskip]
International Collaboration for Turbulence Research}

\pubyear{2014}
\volume{}
\pagerange{}
\date{?; revised ?; accepted ?. - To be entered by editorial office}

\begin{document}

\maketitle

\begin{abstract}
In this study, we analyze the statistics of both individual inertial particles and inertial particle pairs
in direct numerical simulations of homogeneous isotropic turbulence in the absence of gravity.
The effect of the Taylor microscale Reynolds number $R_\lambda$ on the particle
statistics is examined over the largest range to date (from $R_\lambda = 88-597$),
at small, intermediate, and large Kolmogorov-scale Stokes numbers $St$.
We first explore the effect of preferential sampling on the single-particle statistics
and find that low-$St$ inertial particles are ejected from both vortex tubes
and vortex sheets (the latter becoming increasingly prevalent 
at higher Reynolds numbers) and preferentially accumulate in regions of irrotational dissipation.
We use this understanding of preferential sampling to 
provide a physical explanation for many of the trends in the 
particle velocity gradients, kinetic energies, and accelerations at low $St$,
which are well-represented by the model of \cite{chun05}.
As $St$ increases, inertial filtering effects become more important, causing the particle kinetic
energies and accelerations to decrease. The effect of inertial 
filtering on the particle kinetic energies and accelerations
diminishes with increasing Reynolds number and 
is well-captured by the models of \cite{abrahamson75} and \cite{zaichik08}, respectively.

We then consider particle-pair statistics, and focus our attention
on the relative velocities and radial distribution functions (RDFs) of the particles,
with the aim of understanding the underlying physical mechanisms contributing to particle collisions.
The relative velocity statistics indicate that preferential-sampling effects are important
for $St \lesssim 0.1$ and that path-history/non-local effects become increasingly
important for $St \gtrsim 0.2$.
While higher-order relative velocity statistics are influenced
by the increased intermittency of the turbulence at high Reynolds numbers,
the lower-order relative velocity statistics are only weakly
sensitive to changes in Reynolds number at low $St$. 
The Reynolds-number trends in these quantities at intermediate and large $St$ are explained
based on the influence of the available flow scales on the path-history and inertial filtering effects.
We find that the RDFs peak near $St$ of order unity, that they exhibit
power-law scaling for low and intermediate $St$, 
and that they are largely independent of Reynolds number for low and intermediate $St$.
We use the model of \cite{zaichik09} to explain the physical mechanisms
responsible for these trends, and find that this model is able to
capture the quantitative behavior of the RDFs extremely well
when DNS data for the structure functions are specified, in agreement with \cite{bragg14}.
We also observe that at large $St$, changes in the RDF are related to
changes the scaling exponents of the relative velocity variances.
The particle collision kernel closely matches that computed by \cite{rosa13} 
and is found to be largely insensitive to the flow Reynolds number.
This suggests that relatively low-Reynolds-number simulations
may be able to capture much of the relevant physics of
droplet collisions and growth in the adiabatic cores of atmospheric clouds.
\end{abstract}

\section{Introduction}
\label{sec:introduction}
Since the pioneering study of \cite{orszag72} over forty years ago,
direct numerical simulation (DNS) has been widely used to study turbulent flows.
Previous DNS studies have provided a wealth of information about
the underlying turbulent flow field, much of which is very difficult to
obtain experimentally, including Lagrangian statistics \citep{yeung89}, 
pressure fluctuations \citep{spalart88}, and velocity gradient tensors \citep{ashurst87}.

Only within the last ten years, however, with the advent of tera- and petascale
computing, have DNS at Reynolds numbers comparable to those in
the largest laboratory experiments become possible. The highest-Reynolds-number
simulations to date (with Taylor microscale Reynolds numbers $R_\lambda \sim 1000$)
have been of isotropic turbulence in tri-periodic domains
and have considered both the Eulerian
dynamics of the turbulent flow field and the Lagrangian dynamics of inertialess
tracer (i.e., fluid) particles advected by the flow \citep{kaneda03,ishihara07,ishihara09,yeung12}.

Many industrial and environmental
turbulent flows, however, are laden with dense, inertial particles,
which can display profoundly different dynamics
than inertialess fluid particles.  The degree to which the dynamics of inertial particles differ from those
of fluid particles depends on their Stokes number $St$, a non-dimensional
measure of particle inertia, which we define based on Kolmogorov-scale turbulence.
We summarize the relevant physical mechanisms at small, intermediate, and large values of $St$ below.

It is well-known from both computational and experimental studies that 
inertial particles preferentially
sample certain regions of the flow \citep[e.g., see][]{balachandar10}. 
This preferential sampling is often attributed
to the fact that heavy particles are centrifuged out of vortex cores
and accumulate in low-vorticity and high-strain regions \citep{maxey87b,squires91a,eaton94}, 
leading to higher collision rates \citep{sundaram4}. However, this centrifuge mechanism
is mainly important for small-$St$ particles which are strongly coupled to the underlying flow.
As $St$ is increased, the particle dynamics become less coupled to the local fluid velocity field
and the influence of their path-history interactions with the turbulence becomes increasingly
important \citep[e.g., see][]{bragg14b}.
Particles with sufficiently large $St$ can therefore come together from different regions of the flow
with large relative velocities, increasing their collision rate \citep{wilkinson06,falkovich07}.
Such a process is referred to as `caustics' \citep{wilkinson06} and
the `sling effect' \citep{falkovich07}.
At high values of $St$, several studies \citep[e.g.,][]{bec06a,sathya08a} 
have shown that particles have a modulated
response to the underlying turbulence as they filter out high-frequency
flow features (i.e., features with timescales significantly below the particle response time),
and they therefore have lower kinetic energies and lower accelerations.

Despite recent advances in simulating high-Reynolds-number
turbulent flows, current studies of inertial
particles in turbulence are primarily at low and moderate Reynolds numbers
($R_\lambda \lesssim 500$),
and only recently have DNSs been conducted of inertial particles
in turbulence with a well-defined inertial range \citep{bec10a,bec10b,pan11,ray11,rosa13,pan13}.
It is vital to understand the effect of Reynolds number on the mechanisms above (preferential sampling,
path-history interactions, and inertial filtering), particularly at higher Reynolds numbers which are
more representative of those in nature.  We give two examples to emphasize the importance of 
developing such an understanding.

The first example, cloud formation, is the primary motivation for this work.
For reviews on this subject, see \cite{shaw03,devenish12,grabowski13}; here we provide a brief overview.
It is well-known that standard microphysical cloud models
over-predict the time required for the onset of precipitation in warm cumulus
clouds \citep[e.g., see][]{shaw03}. At early stages of cloud formation,
particles experience condensational growth. This process slows down quickly
with increasing droplet diameter, making condensational growth effective only for droplets with 
diameters less than about $30 \mu m$ \citep{grabowski13}.
Moreover, gravity is only able to significantly enhance collisional growth
for particles with diameters above $80 \mu m$ \citep{prupp97,grabowski13},
leaving a `size gap' where neither condensational growth nor gravitational
coalescence is very effective.
For particles between these two limits, it has been proposed that turbulence-induced
collisions are primarily responsible for droplet growth.

It is unclear, however, the extent to which particle collision rates
are affected by changes in Reynolds number at conditions
representative of those in cumulus clouds \citep[which have $R_\lambda \sim 10,000$, see][]{siebert06}.
\cite{sundaram4} showed that particle collision rates depend on 
both the degree of clustering and on the relative velocities between particles,
and thus many subsequent analyses have considered the Reynolds-number dependence of both of these statistics.
While the early study of \cite{wwz00} suggested that clustering increases
with $R_\lambda$, later investigations \citep{keswani04,bec10a,ray11,rosa13} indicate
that clustering saturates at higher Reynolds numbers. 
Other researchers have suggested that caustics
become more prevalent at high Reynolds numbers, leading to larger relative velocities and thus
more frequent particle collisions \citep{falkovich02,wilkinson06}.
The findings of \cite{bec10a} and \cite{rosa13}, however, do not seem to support
that trend.  In all cases, 
the Reynolds-number range ($R_\lambda \lesssim 500$) leaves open the question
of whether the results apply to atmospheric conditions at much higher Reynolds numbers.

The second example relates to planetesimal formation. Planetesimals begin to form 
when small dust grains collide and coalesce in turbulent protoplanetary nebulae \citep{pan10}.
\cite{cuzzi01} estimated that the turbulence in such nebulae is characterized by 
$R_\lambda \sim 10^4-10^6$. It is unclear to what
extent the rate of coalescence depends on the Reynolds number, and studies
at progressively higher Reynolds numbers are necessary to develop scaling relations
for particle collision rates at conditions representative of nebula turbulence.
\cite{pan10} noted that the range of relevant particle
sizes in the planetesimal formation process spans about nine orders of magnitude,
and therefore we expect that the collision rates will be affected by 
preferential sampling (for small, medium, and large particles),
path-history interactions (for medium and large particles),
and inertial filtering (for the largest particles).

In this study, we use high-performance computing resources provided by the
U. S. National Center for Atmospheric Research \citep{yellowstone} to simulate 
inertial particles in isotropic turbulence over the range $88 \leq R_\lambda \leq 597$.
To our knowledge, the top value represents the highest Reynolds-number flow with
particles simulated to date.
The overall goal is to improve predictions for the collision kernel at Reynolds numbers
more representative of those in atmospheric clouds. Gravitational forces
are neglected in this study, but will be considered in detail in Part II of this study
\citep{ireland15b}.

The paper is organized as follows: \textsection \ref{sec:dns} provides a summary of the numerical
methods used and the relevant fluid and particle parameters.
In \textsection \ref{sec:single_particle},
we study single-particle statistics (small-scale velocity gradients, large-scale velocity fluctuations,
and accelerations). Many of the results from this section help explain the particle-pair
statistics presented in \textsection \ref{sec:two_particle_stats}. These statistics include
the particle relative velocities, radial distribution functions, and collision kernels.
Finally, in \textsection \ref{sec:conclusions}, we summarize our results
and suggest practical implications for the turbulence and cloud physics communities.
\section{Overview of simulations}
\label{sec:dns}
A brief summary of the simulation parameters and numerical methods is provided below.
Refer to \cite{ireland13} for a more detailed description of the code, including integration techniques, 
parallelization strategies, and interpolation methods.

\subsection{Fluid phase}
\label{sec:fluid_phase}

We perform DNS of isotropic turbulence on a cubic, 
tri-periodic domain of length $\mathcal{L}=2\pi$ with $N^3$ grid points.
A pseudospectral method \citep{orszag} is used to evaluate 
the continuity and momentum equations for an incompressible flow,
\begin{equation}
 \nabla \cdot \bm{u} = 0 \mathrm{,}
 \label{eq:continuity}
\end{equation}
\begin{equation}
 \frac{\partial \bm{u}}{\partial t} + \boldsymbol{\omega} \times \bm{u} 
   + \nabla \left( \frac{p}{\rho_f} + \frac{u^2}{2} \right) = \nu \nabla^2 \bm{u} + \bm{f} \mathrm{.}
 \label{eq:momentum}
\end{equation}
Here, $\bm{u}$ is the fluid velocity, $\boldsymbol{\omega} \equiv \nabla \times \bm{u}$ is the vorticity,
$p$ is the pressure, $\rho_f$ is the fluid density, $\nu$ is the kinematic viscosity, 
and $\bm{f}$ is a large-scale forcing term that is added to make the 
flow field statistically stationary. For our simulations, we added forcing to wavenumbers with magnitude $\kappa = \sqrt{2}$
in Fourier space in a deterministic fashion to compensate precisely for the energy lost to viscous
dissipation \citep{witkowska97}.

\begin{figure}
 \centering
 \includegraphics[width=2.6in]{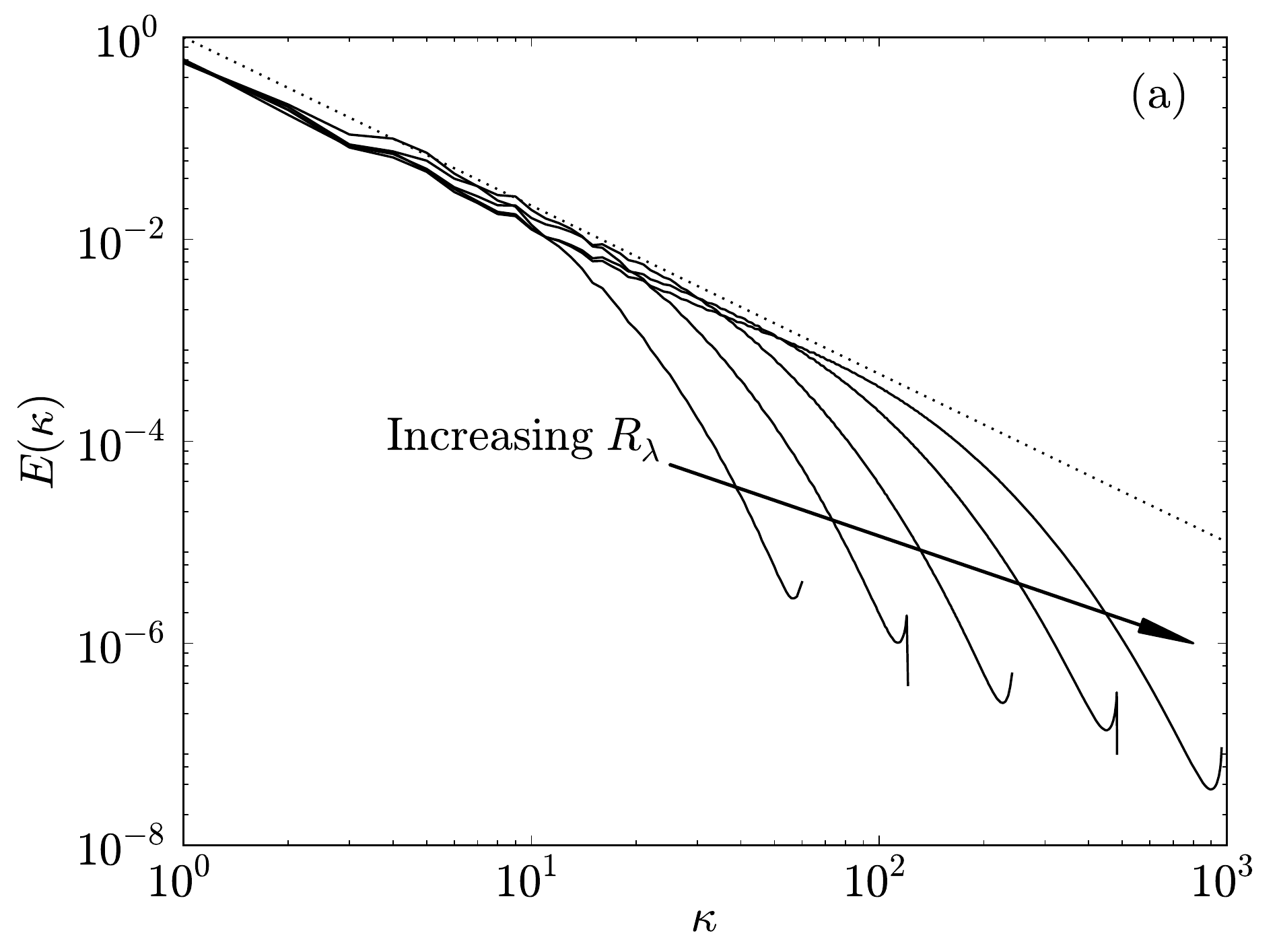}
 \includegraphics[width=2.6in]{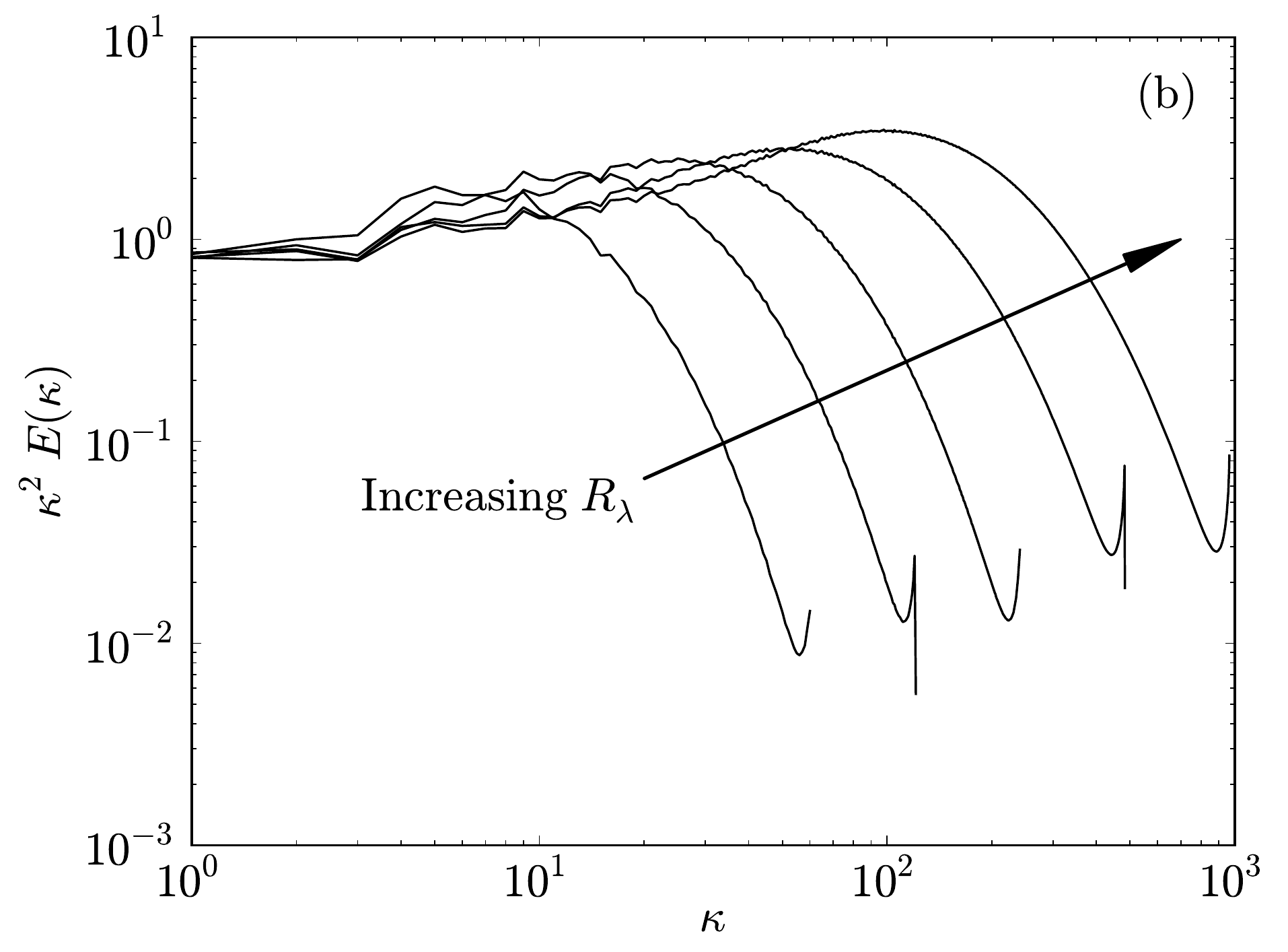}
 \caption{(a) Energy (a) and (b) dissipation spectra for the different simulations described in table~\ref{tab:parameters}.
 The diagonal dotted line in (a) has a slope of $-5/3$, the expected spectral scaling in the inertial subrange.
 All values are in arbitrary units.}
 \label{fig:energy_and_dissipation}
\end{figure}

We perform a series of five different simulations, with Taylor microscale Reynolds numbers 
$R_\lambda \equiv 2 k \sqrt{5/\left(3 \nu \epsilon \right)}$ ranging from $88$ to $597$,
where $k$ denotes the turbulent kinetic energy and $\epsilon$ the turbulent energy dissipation rate.
Details of the simulations are given in table~\ref{tab:parameters}.
The simulations are parameterized to have similar large scales, but different dissipation (small) scales.
The small-scale resolution for the simulations was held constant, with 
$\kappa_\mathrm{max} \eta \approx 1.6-1.7$,
where $\kappa_\mathrm{max} \equiv \sqrt{2}N/3$ is the maximum resolved wavenumber and 
$\eta \equiv \left(\nu^3/\epsilon \right)^{1/4}$ is the Kolmogorov lengthscale.
Time-averaged energy and dissipation spectra for all five simulations
are shown in figure~\ref{fig:energy_and_dissipation}.
A clear $-5/3$ spectral slope is evident for the three highest Reynolds-number cases ($R_\lambda \geq 224$),
indicating the presence of a well-defined inertial subrange.
The simulations are performed in parallel on $N_\mathrm{proc}$ processors, and the P3DFFT library
\citep{p3dfft} is used for efficient parallel computation of three-dimensional fast Fourier transforms.

\begin{table}
 \centering
 \caption{Flow parameters for the DNS study. All dimensional parameters are in arbitrary units, and
 all statistics are averaged over time $T$. All quantities are defined in the text in \textsection 
 \ref{sec:fluid_phase} and \textsection \ref{sec:particle_phase}.}
 \label{tab:parameters}
 \begin{tabular}{ l  l  l  l  l  l }
  Simulation& I & II & III & IV & V \\ 
 $R_\lambda$ & 88 & 140 & 224 & 398 & 597 \\ 
 $\nu$ & 0.005 & 0.002 & 0.0008289 & 0.0003 & 0.00013 \\ 
 $\epsilon$ & 0.270 & 0.267 & 0.253 & 0.223 & 0.228 \\ 
 $\ell$ & 1.46 & 1.41 & 1.40 & 1.45 & 1.43 \\ 
 $\ell/\eta$ & 55.8 & 107 & 204 & 436 & 812 \\
 $u'$ & 0.914 & 0.914 & 0.915 & 0.915 & 0.915 \\
 $u'/u_\eta$ & 4.77 & 6.01 & 7.60 & 10.1 & 12.4 \\
 $T_L$ & 1.60 & 1.54 & 1.53 & 1.58 & 1.57 \\
 $T_L/\tau_\eta$ & 11.7 & 17.7 & 26.8 & 43.0 & 65.4 \\
 $T/T_L$ & 15.0 & 10.4 & 11.4 & 11.1 & 5.75 \\
 $k_\mathrm{max} \eta$ & 1.59 & 1.59 & 1.66 & 1.60 & 1.70 \\
 $N$ & 128 & 256 & 512 & 1024 & 2048 \\
 $N_p$ & 262,144 & 262,144 & 2,097,152 & 16,777,216 & 134,217,728 \\
 $N_\mathrm{tracked}$ & 32,768 & 32,768 & 262,144 & 2,097,152 & 16,777,216 \\
 $N_\mathrm{proc}$ & 16 & 16 & 64 & 1024 & 16,384 \\
 \end{tabular}
\end{table}

\subsection{Particle phase}
\label{sec:particle_phase}

We simulate the motion of small ($d/\eta \ll 1$, where $d$ is the particle diameter),
heavy ($\rho_p/\rho_f \gg 1$, where $\rho_p$ is the particle density), spherical particles.
18 different particle classes are simulated with Stokes numbers $St$ ranging from $0$ to $30$.
$St \equiv \tau_p / \tau_\eta$ is a non-dimensional measure of a particle's inertia,
comparing the response time of the particle $\tau_p \equiv \rho_p d^2 / \left( 18 \rho_f \nu \right)$
to the Kolmogorov timescale $\tau_\eta \equiv (\nu/\epsilon)^{1/2}$.

We assume that the particles are subjected to only
linear drag forces, which is a reasonable approximation when the particle
Reynolds number $Re_p \equiv \vert \bm{u}(\bm{x}^p(t),t)-\bm{v}^p(t) \vert / \nu < 0.5$
\citep{elghobashi92}. Here, $\bm{u}(\bm{x}^p(t),t)$ denotes the undisturbed fluid velocity at the particle position
$\bm{x}^p(t)$, and $\bm{v}^p(t)$ denotes the velocity of the particle.
(Throughout this study, we use the superscript $p$ on $\bm{x}$, $\bm{u}$, and $\bm{v}$ to
denote time-dependent, Lagrangian variables defined along particle trajectories.
Phase-space positions and velocities are denoted without the superscript $p$.)
Though particles with large $St$ experience non-negligible nonlinear drag forces \citep[e.g.,][]{wang93},
the use of a linear drag model for large-$St$ particles provides a useful first approximation
and facilitates comparison between several theoretical models 
that make the same assumption \citep[e.g.,][]{chun05,zaichik09,gustavsson11}.
The present study also neglects the influence of gravity. Part II of this study \citep{ireland15b}
will address the combined effects of gravity and turbulence on particle motion.
Finally, since a primary motivation is to understand droplet dynamics in atmospheric clouds,
where the particle mass and volume loadings are low \citep{shaw03}, we assume that the
particle loadings are sufficiently dilute such that inter-particle interactions and two-way coupling 
between the phases are negligible \citep{elghobashi93,sundaram6}.

Under these assumptions, each inertial particle obeys a simplified Maxey-Riley equation \citep{maxey83},
\begin{equation}
\label{eq:maxey_riley}
 \frac{d^2 \bm{x}^p}{dt^2} = \frac{d \bm{v}^p}{dt} = 
 \frac{\bm{u}\left(\bm{x}^p(t),t\right)-\bm{v}^p(t)}{\tau_p} \mathrm{,}
\end{equation}
and each fluid (i.e., inertialess) particle is tracked by solving
\begin{equation}
 \label{eq:fluid_velocities}
 \frac{d\bm{x}^p}{dt} = \bm{u}(\bm{x}^p(t),t) \mathrm{.}
\end{equation}
To compute $\bm{u}^p(t)=\bm{u}(\bm{x}^p(t),t)$, we need to interpolate from the Eulerian grid to the particle location.
While other studies \citep[e.g., see][]{bec10a,durham13} have done so using tri-linear interpolation,
\cite{ireland13} showed that such an approach can lead to errors in the interpolated velocity
which are orders of magnitude above the local time-stepping error.
In addition, \cite{vanhinsberg13} demonstrated that tri-linear interpolation,
which possesses only $C^0$ continuity, leads to artificial high frequency oscillations
in the computed particle accelerations. 
\cite{ray13} noted that the relative motion of particles at small separations
will depend strongly on the interpolation scheme. Since a main focus of this
paper is particle motion near-contact and its influence on particle collisions, it is crucial to calculate
$\bm{u}(\bm{x}^p(t),t)$ as accurately as possible. To that end, we use an eight-point B-spline interpolation
scheme (with $C^6$ continuity) based on the algorithm in \cite{vanhinsberg12}.

The particles were initially placed in the flow with a uniform distribution and velocities $\bm{v}^p$ equal to
the underlying fluid velocity $\bm{u}^p$. We began computing particle statistics
once the particle distributions and velocities became statistically stationary, usually about
$5$ large-eddy turnover times $T_L \equiv \ell / u'$ (where $\ell$ is the integral lengthscale and 
$u' \equiv \sqrt{2k/3}$) after the particles were introduced into the flow. 
Particle statistics were calculated at a frequency of 2-3 times per $T_L$ and were
time-averaged over the duration of the run $T$.

For a subset $N_\mathrm{tracked}$ of the total number of particles in each class $N_{p}$, 
we stored particle positions, velocities, and velocity gradients every $0.1 \tau_\eta$ for
a duration of about $100 \tau_\eta$.
These data are used to compute Lagrangian correlations, accelerations, and timescales of the particles.
\section{Single-particle statistics}
\label{sec:single_particle}
We first consider single-particle statistics from our simulations.  These statistics will provide
a basis for our understanding of the two-particle statistics presented in \textsection \ref{sec:two_particle_stats}.
We explore velocity gradient (i.e., small-scale velocity) statistics in \textsection \ref{sec:topology},
kinetic energy (i.e., large-scale velocity) statistics in \textsection \ref{sec:particle_kinetic_energy},
and acceleration statistics in \textsection \ref{sec:particle_acceleration}.
In each case, we study the
effect of the underlying flow topology on these statistics.

\subsection{Velocity gradient statistics}
\label{sec:topology}
We consider the gradients of the underlying fluid velocity at the particle locations,
$\bm{A}(\bm{x}^p(t),t) \equiv \nabla\bm{u}(\bm{x}^p(t),t)$.
These statistics provide us with information about the small-scale velocity
field experienced by the particles. 
(Refer to \cite{meneveau11} for a recent review on this subject.)
In particular, to understand the interaction of particles with specific topological features of
the turbulence, we decompose
$\bm{A}(\bm{x}^p(t),t)$ into a symmetric strain rate tensor 
$\bm{\mathcal{S}}(\bm{x}^p(t),t) \equiv [\bm{A}(\bm{x}^p(t),t) + \bm{A}^\intercal(\bm{x}^p(t),t)]/2$
and an antisymmetric rotation rate tensor 
$\bm{\mathcal{R}}(\bm{x}^p(t),t) \equiv [\bm{A}(\bm{x}^p(t),t) - \bm{A}^\intercal(\bm{x}^p(t),t)]/2$.

Due to their inertia, heavy particles are ejected out of regions of high rotation rate and accumulate in regions of high strain rate
\citep[e.g.,][]{maxey87b,squires91a,eaton94}, and this is associated with a `preferential sampling' of $\bm{A}(\bm{x},t)$.
For particles with low inertia ($St \ll 1$), preferential sampling is the dominant mechanism
affecting the particle motion \citep[e.g., see][]{chun05}.
As the particle inertia increases, the particle motion becomes increasingly
decoupled from the local fluid turbulence, and the effect of the preferential sampling on the particle dynamics decreases.
At the other limit ($St \gg 1$), preferential sampling vanishes
and the particles have a damped response to the underlying flow which
leads them to sample the turbulence more uniformly \citep[e.g., see][]{bec06a}.

We first consider the average of the second invariants of the strain rate and rotation rate tensors evaluated at the inertial particle positions
\begin{equation}
 \langle \mathcal{S}^2 \rangle^p \equiv 
 \langle \boldsymbol{\mathcal{S}}(\bm{x}^p(t),t) \colon \boldsymbol{\mathcal{S}}(\bm{x}^p(t),t) \rangle \mathrm{,}
\end{equation}
and
\begin{equation}
 \langle \mathcal{R}^2 \rangle^p \equiv \langle \boldsymbol{\mathcal{R}}(\bm{x}^p(t),t) \colon \boldsymbol{\mathcal{R}}(\bm{x}^p(t),t) \rangle \mathrm{.}
\end{equation}
By definition, for fully mixed fluid particles ($St = 0$) in homogeneous turbulence,
$\tau_\eta^2 \langle \mathcal{S}^2 \rangle^p = \tau_\eta^2 \langle \mathcal{R}^2 \rangle^p = 0.5$.

\begin{figure}
 \centering
 \includegraphics[width=2.6in]{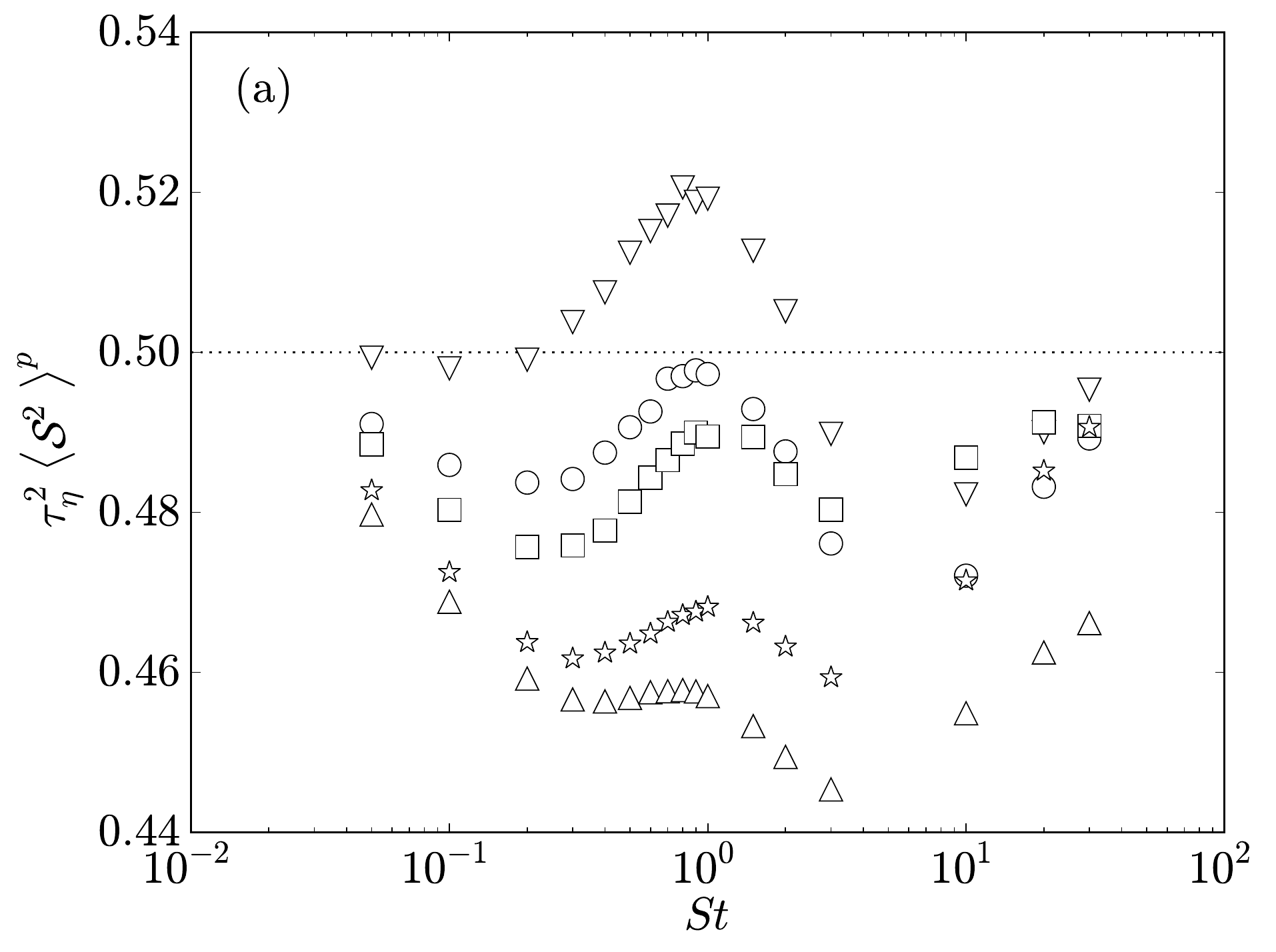}
 \includegraphics[width=2.6in]{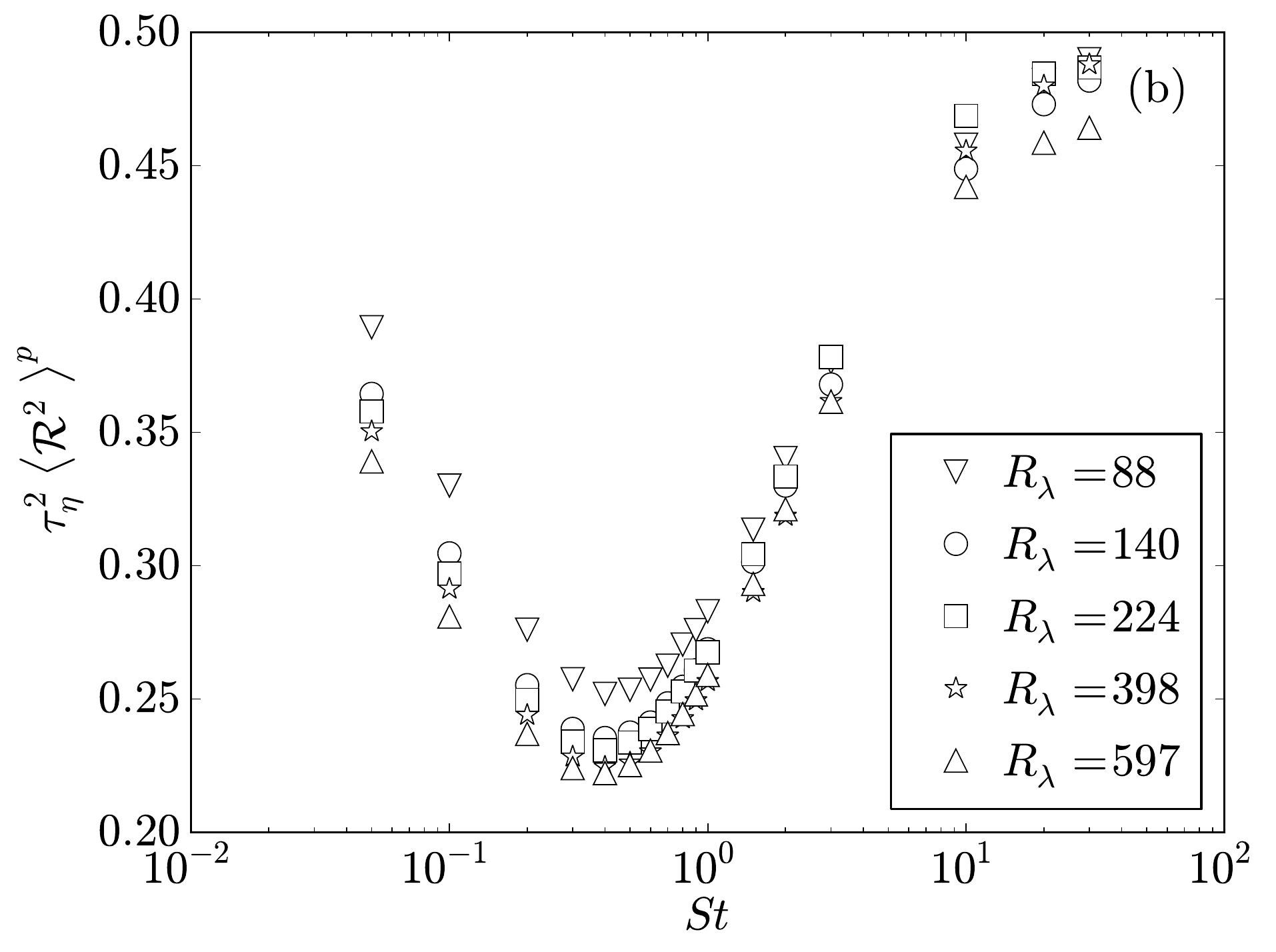}
 \includegraphics[width=2.6in]{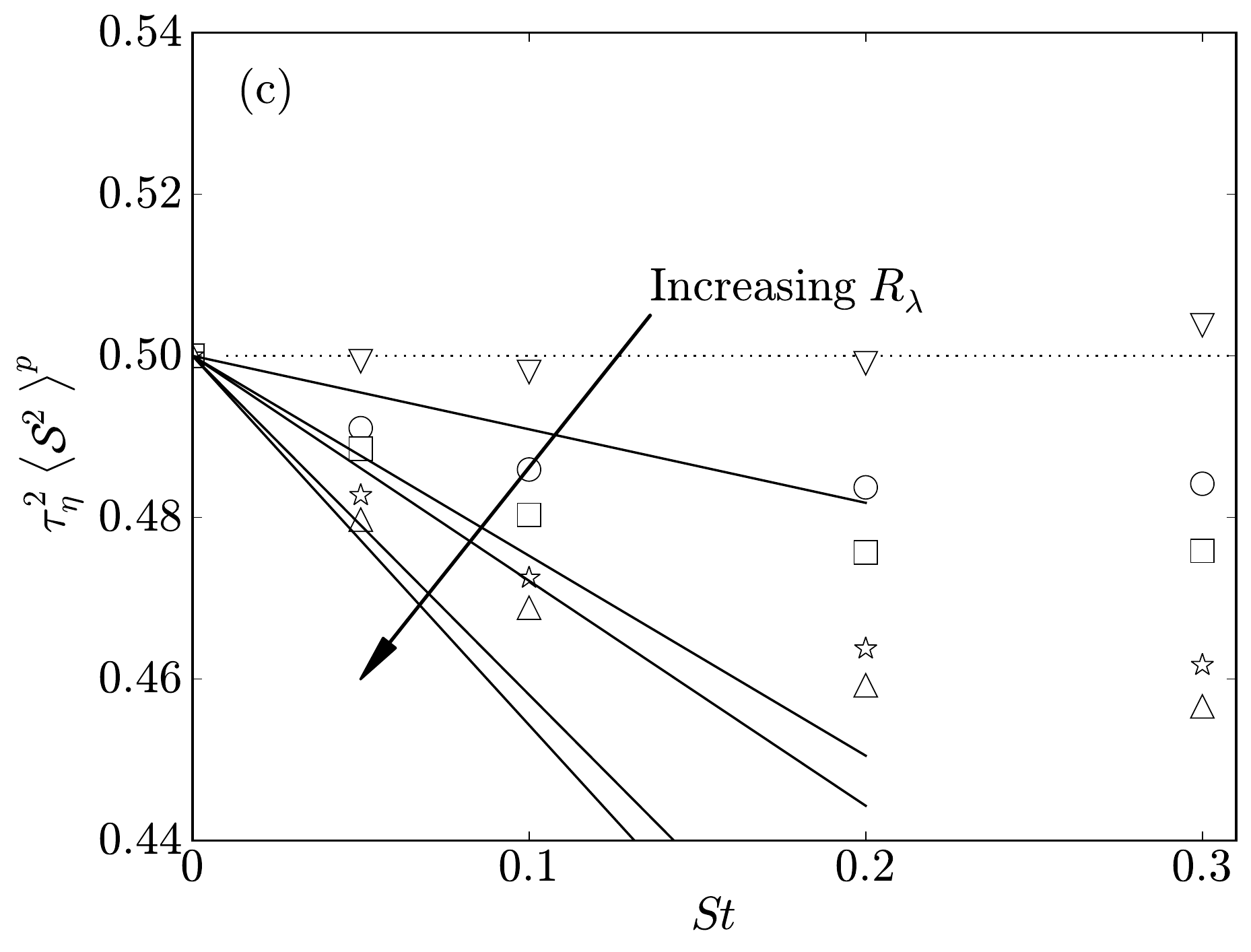}
 \includegraphics[width=2.6in]{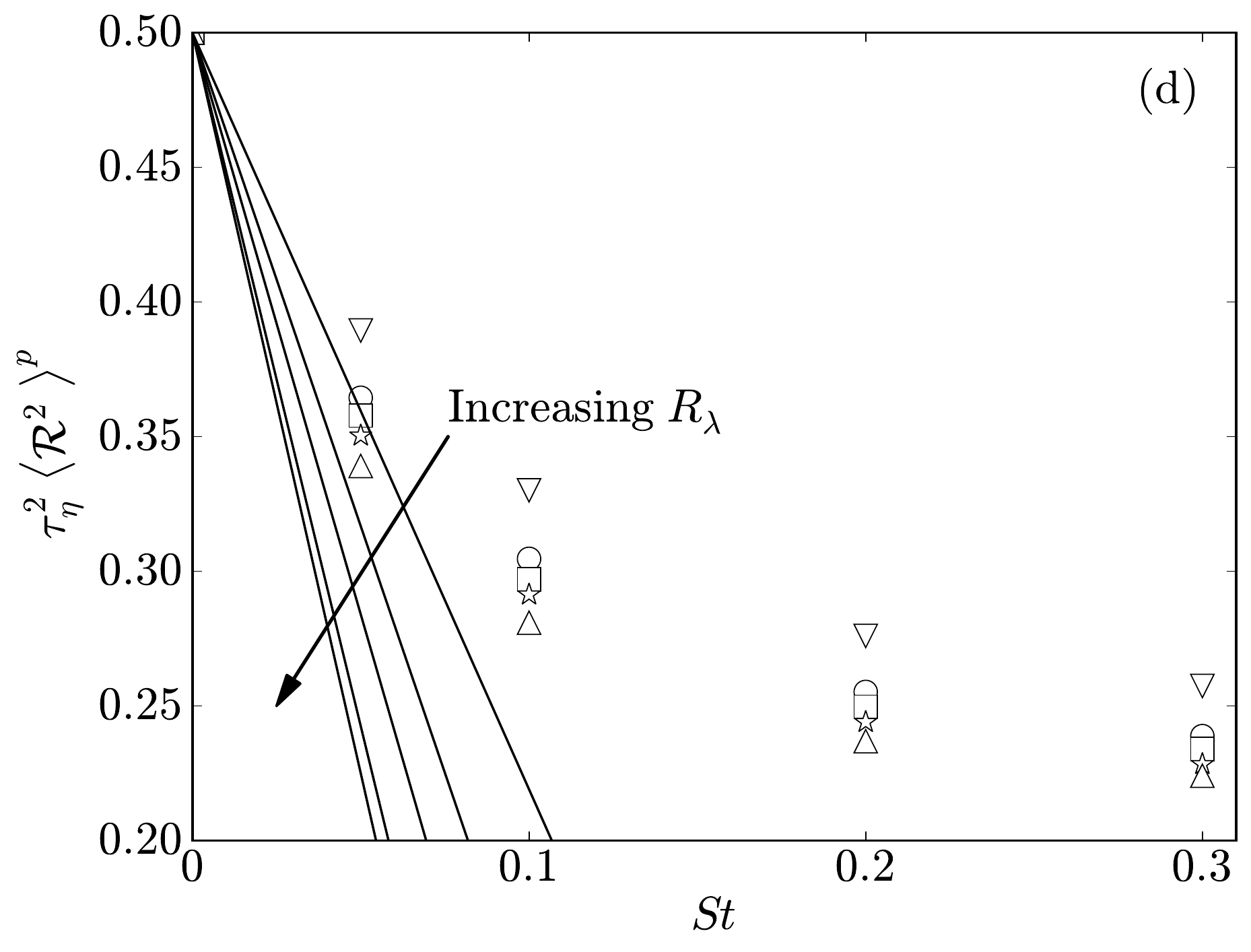}
 \caption{Data for $\langle\mathcal{S}^2\rangle^p$ (a,c) and $\langle\mathcal{R}^2\rangle^p$ (b,d) 
 sampled at inertial particle positions as function of $St$ for different values of $R_\lambda$.
 The data are shown at low $St$ in (c,d) to highlight the effect of preferential sampling in this regime.
 The solid lines in (c) and (d) are the predictions from (\ref{eq:chun_theory})
 for $St \ll 1$. DNS data are shown with symbols.}
 \label{fig:S2_R2}
\end{figure}

Since small-$St$ particles are centrifuged out of regions of high rotation, we expect that
$\tau_\eta^2 \langle \mathcal{R}^2 \rangle^p$ will decrease with increasing $St$;
their accumulation in high strain regions would also lead to the expectation that $\tau_\eta^2 \langle \mathcal{S}^2 \rangle^p$ 
will increase with increasing $St$. 
In figure~\ref{fig:S2_R2} we see that 
while $\tau_\eta^2 \langle \mathcal{R}^2 \rangle^p$ is more strongly affected by changes in $R_\lambda$
than is $\tau_\eta^2 \langle \mathcal{S}^2 \rangle^p$, both quantities decrease with increasing $St$ (for $St \ll 1$).
This surprising result is consistent with other DNS \citep{keswani04,chun05,salazar12b}.
Our data also show that both $\tau_\eta^2 \langle \mathcal{S}^2 \rangle^p$ and $\tau_\eta^2 \langle \mathcal{R}^2 \rangle^p$
decrease with increasing $R_\lambda$ for $St \ll 1$, in agreement with \cite{keswani04}.

We use the formulation given in \cite{chun05} \citep[and re-derived in][]{salazar12b}
to model the effect of preferential sampling on $\tau_\eta^2 \langle \mathcal{S}^2 \rangle^p$ and
$\tau_\eta^2 \langle \mathcal{R}^2 \rangle^p$ in limit of $St \ll 1$.
\cite{chun05,salazar12b} showed that for an arbitrary quantity $\phi$,
the average value of $\phi$ sampled along a particle trajectory 
$\langle \phi \rangle^p$ can be reconstructed entirely from fluid particle statistics using the relation,
\begin{equation}
\label{eq:chun_theory}
 \langle \phi(St) \rangle^p = \langle \phi(St=0) \rangle^p + \tau_\eta \sigma^p_\phi St
 \left( \rho^p_{\mathcal{S}^2 \phi} \sigma^p_{\mathcal{S}^2} T^p_{\mathcal{S}^2 \phi} - \rho^p_{\mathcal{R}^2 \phi} \sigma^p_{\mathcal{R}^2} T^p_{\mathcal{R}^2 \phi} \right) \mathrm{.}
\end{equation}
Here, $\sigma^p_Y$ denotes the standard deviation of a variable $Y$ along a fluid particle trajectory,
$\rho^p_{YZ}$ is the correlation coefficient between $Y$ and $Z$,
\begin{equation}
 \rho^p_{YZ} \equiv \frac{\Big\langle \left[ Y(\bm{x}^p(t),t) - \langle Y(\bm{x}^p(t),t) \rangle \right]
 \left[ Z(\bm{x}^p(t),t) - \langle Z(\bm{x}^p(t),t) \rangle \right] \Big \rangle}{\sigma^p_Y \sigma^p_Z} \mathrm{,}
\end{equation}
and $T^p_{YZ}$ is the Lagrangian correlation time,
\begin{equation}
\label{eq:timescales}
 T^p_{YZ} \equiv \frac{ \displaystyle{\int_0^\infty \Big\langle \big[ Y(\bm{x}^p(0),0) - \langle Y(\bm{x}^p(t),t) \rangle \big]
 \big[ Z(\bm{x}^p(t'),t') - \langle Z(\bm{x}^p(t),t) \rangle \big] \Big\rangle\ dt'}}
 {\Big\langle \big[ Y(\bm{x}^p(t),t) - \langle Y(\bm{x}^p(t),t) \rangle \big] 
 \big[ Z(\bm{x}^p(t),t) - \langle Z(\bm{x}^p(t),t) \rangle \big] \Big\rangle} \mathrm{.}
\end{equation}

The predictions from (\ref{eq:chun_theory})
for small $St$ are shown by the solid lines in figure~\ref{fig:S2_R2}(c) and figure~\ref{fig:S2_R2}(d).
In the limit of small $St$, this model is able to
capture the decrease in both $\tau_\eta^2 \langle \mathcal{S}^2 \rangle^p$ 
and $\tau_\eta^2 \langle \mathcal{R}^2 \rangle^p$ with increasing $St$,
and also the decrease in these quantities with increasing $R_\lambda$.
It is uncertain whether the quantitative differences between the DNS data and the model
are due to shortcomings of the model or the fact that the smallest inertial particles ($St = 0.05$)
are too large for the model (which assumes $St \ll 1$) to hold.

\begin{figure}
 \centering
 \includegraphics[width=2.6in]{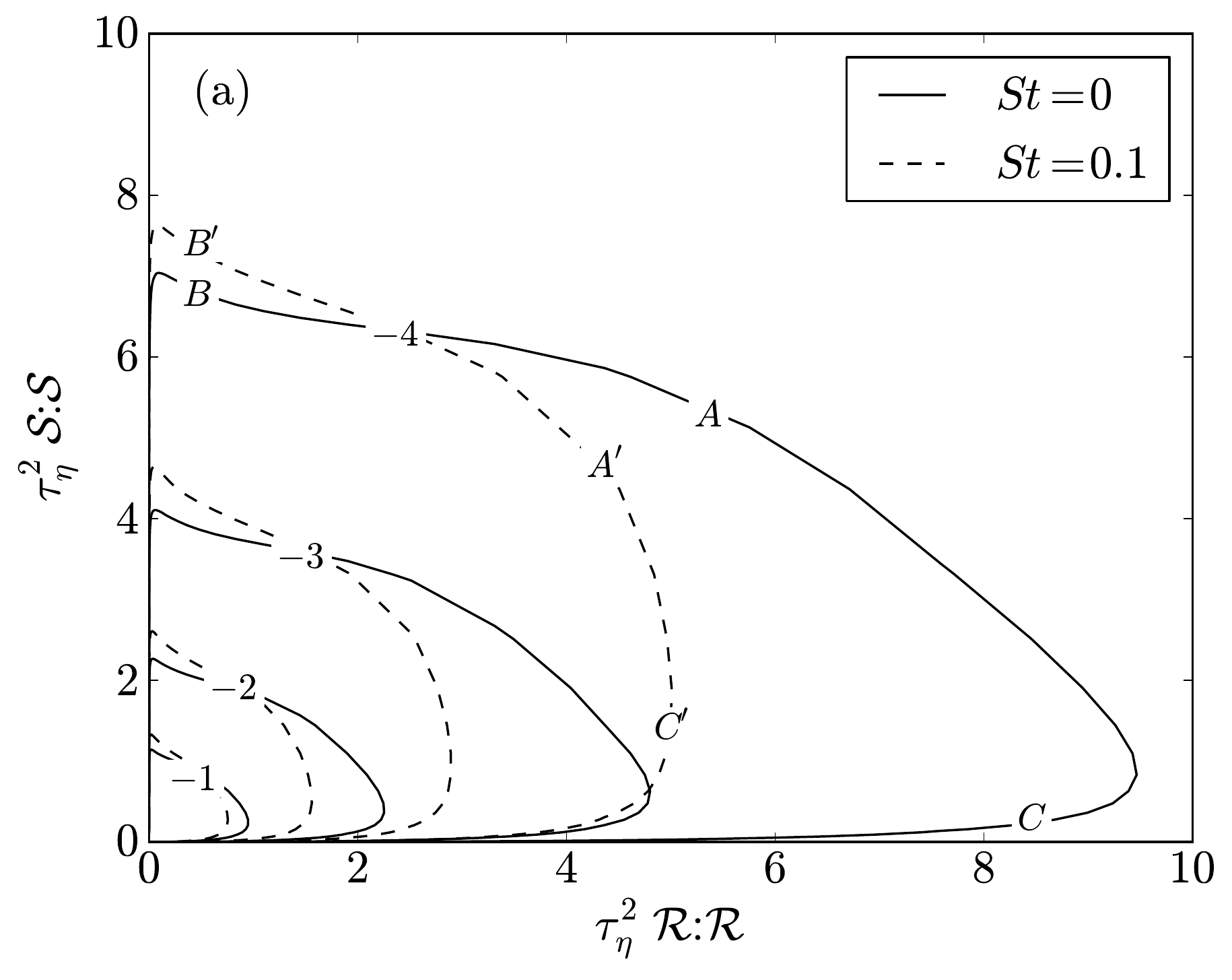}
 \includegraphics[width=2.6in]{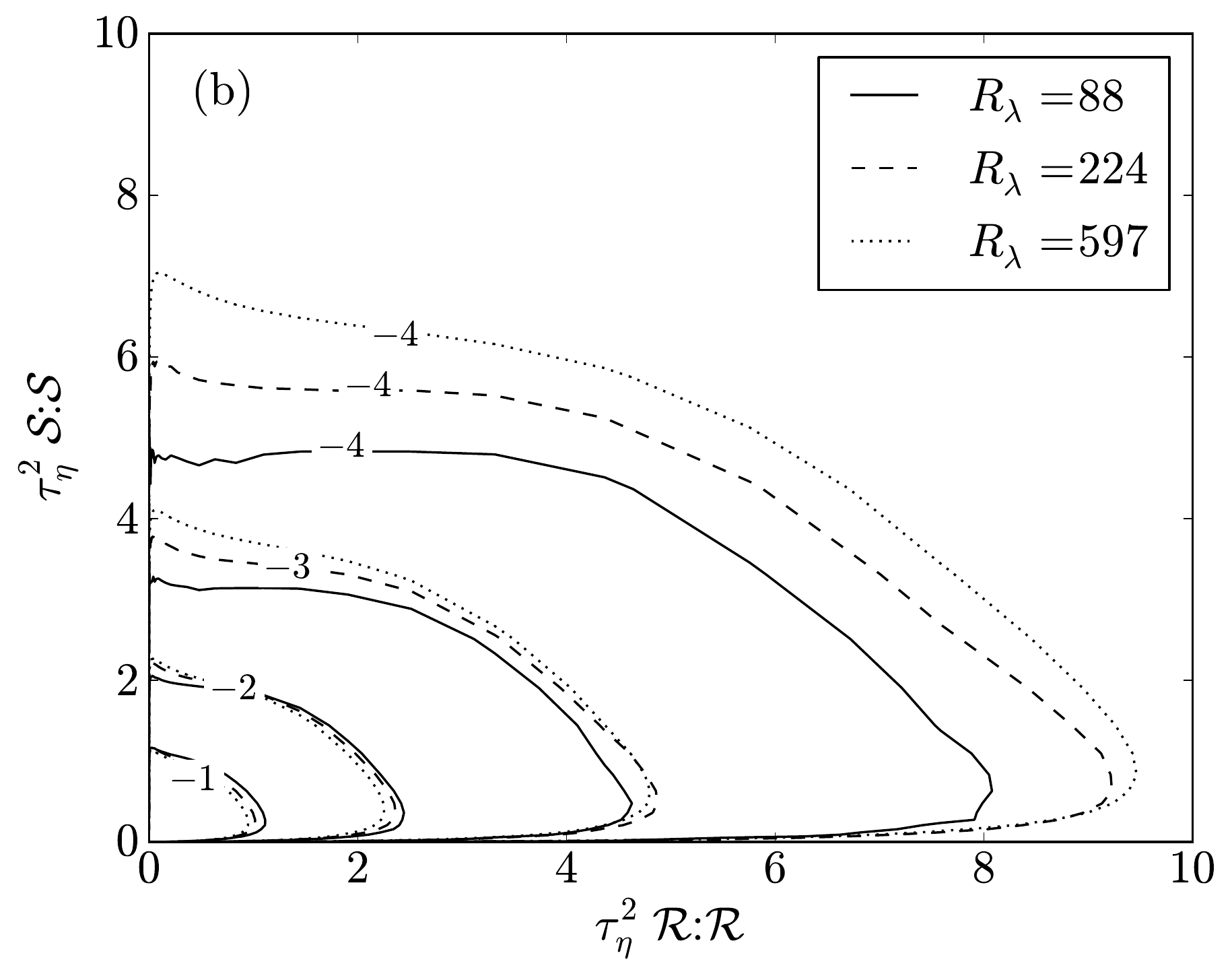}
 \caption{(a) Joint PDFs of $\tau_\eta^2 \boldsymbol{\mathcal{S}}\colon\boldsymbol{\mathcal{S}}$ and 
 $\tau_\eta^2 \boldsymbol{\mathcal{R}}\colon\boldsymbol{\mathcal{R}}$ for $R_\lambda = 597$
 for $St = 0$ and $St = 0.1$ particles. 
 Certain regions of the flow are labeled to aid in the discussion of the trends.
 (b) Joint PDFs of $\tau_\eta^2 \boldsymbol{\mathcal{S}}\colon\boldsymbol{\mathcal{S}}$ and 
 $\tau_\eta^2 \boldsymbol{\mathcal{R}}\colon\boldsymbol{\mathcal{R}}$ for different $R_\lambda$ for $St = 0$ particles.
 In both plots, the exponents of the decade are indicated on the contour lines.}
 \label{fig:S2_R2_jpdf}
\end{figure}

Despite the success of the model of \cite{chun05} in reproducing the trends in the DNS,
the physical explanation for the changes in the mean strain and rotation rates remains unclear. 
In figure~\ref{fig:S2_R2_jpdf}(a), we plot joint PDFs of 
the strain and rotation rates sampled by both $St=0$ and $St=0.1$ particles 
to better understand the specific topological features of the
regions of the flow contributing to these changes.
Following the designations given in \cite{soria94},
we refer to regions with high strain and high rotation 
(indicated by `$A$' in figure~\ref{fig:S2_R2_jpdf}(a))
as `vortex sheets,' regions of low rotation and high strain
(indicated by `$B$') as `irrotational dissipation' areas,
and regions of high rotation and low strain (indicated by `$C$')
as `vortex tubes.'

Our results show three main trends in the particle concentrations.
First, inertial particles are ejected from vortex sheets ($A$) 
into regions of moderate rotation and moderate strain ($A'$).
This ejection from vortex sheets has only recently been discussed in the literature \citep{salazar12b}.
Second, they move from irrotational dissipation regions ($B$) 
to regions of comparable rotation and even higher strain ($B'$).
Third, the particles move out of vortex tubes ($C$)
into regions of lower rotation and higher strain ($C'$).
Evidently, this first effect is primarily responsible for
the decrease in $\tau_\eta^2 \langle \mathcal{S}^2 \rangle^p$ at small $St$,
as suggested in \cite{salazar12b}, and the first and third effects
both contribute to the decrease in $\tau_\eta^2 \langle \mathcal{R}^2 \rangle^p$.
We will revisit these three trends in relation to the 
particle kinetic energies (\textsection \ref{sec:particle_kinetic_energy})
and the particle accelerations (\textsection \ref{sec:particle_acceleration}).

Figure~\ref{fig:S2_R2_jpdf}(b) shows the PDF map for fluid
particles at three values of the Reynolds number.
Notice that as $R_\lambda$ increases,
the probability of encountering a vortex sheet (overlapping high strain and high rotation) increases.
This finding is consistent with the results of \cite{yeung12},
who observed that high strain and rotation events
increasingly overlap in isotropic turbulence as the Reynolds number increases.
It is thus likely that with increasing Reynolds number,
rotation and strain events become increasingly intense,
and the resulting vortex sheets 
become increasingly efficient at expelling particles,
causing both $\tau_\eta^2 \langle \mathcal{S}^2 \rangle^p$ 
and $\tau_\eta^2 \langle \mathcal{R}^2 \rangle^p$ to decrease (cf. figure~\ref{fig:S2_R2}).

\begin{figure}
 \centering
 \includegraphics[height=2.5in]{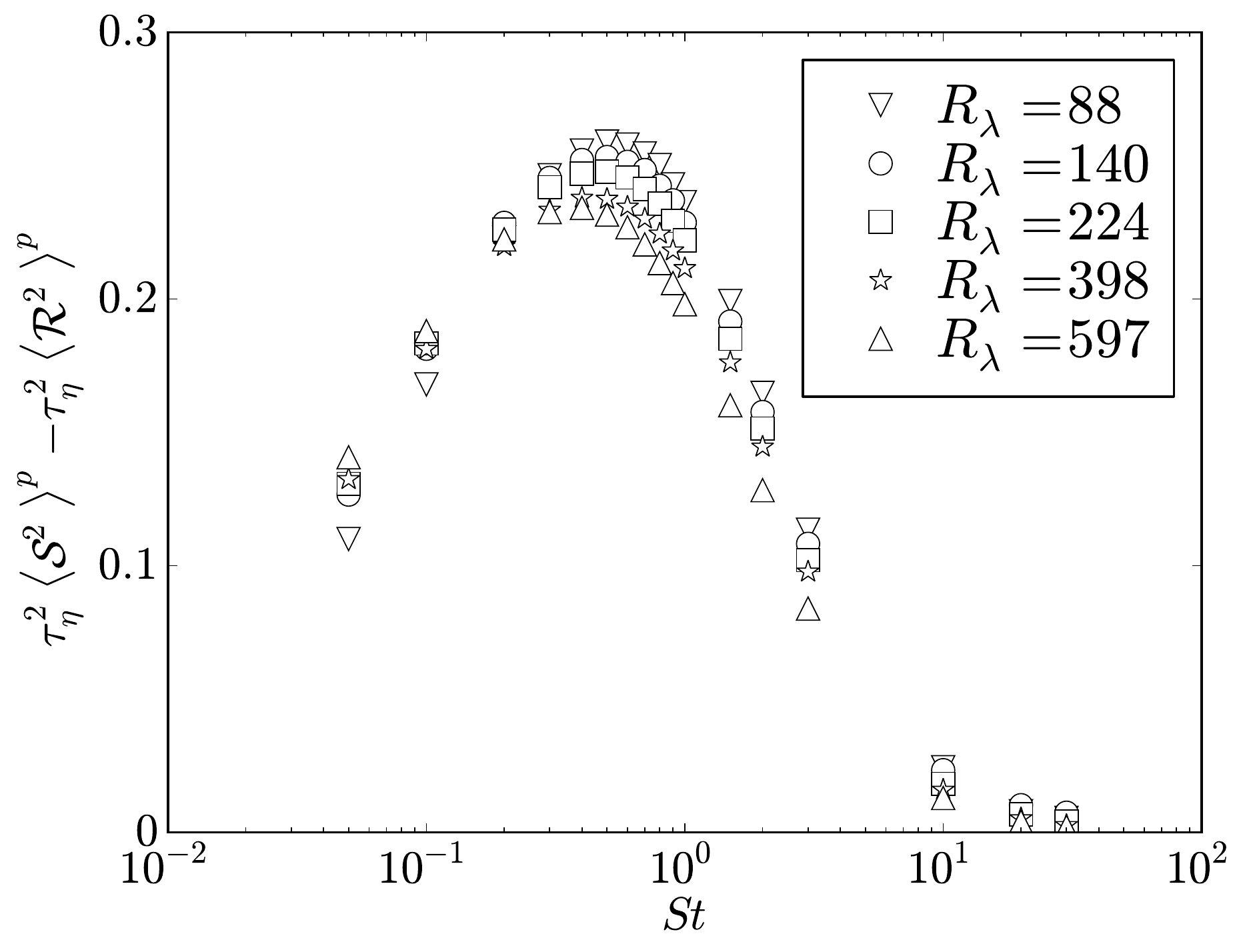}
 \caption{The difference between the mean rates of strain and rotation sampled by the particles
 as a function of $St$ for different values of $R_\lambda$.}
 \label{fig:S2-R2}
\end{figure}

\cite{maxey87b} noted that at low $St$, the compressibility of the particle field (and hence the degree of particle clustering)
is directly related to the difference between the rates of strain and rotation sampled by the particles,
$\tau_\eta^2 \langle \mathcal{S}^2 \rangle^p - \tau_\eta^2 \langle \mathcal{R}^2 \rangle^p$. From figure~\ref{fig:S2-R2}, we see that at low $St$, 
$\tau_\eta^2 \langle \mathcal{S}^2 \rangle^p - \tau_\eta^2 \langle \mathcal{R}^2 \rangle^p$ increases with increasing $R_\lambda$, 
suggesting that the degree of clustering may also increase here. 
We will test this hypothesis in \textsection \ref{sec:particle_clustering}
when we directly measure particle clustering at different values of $St$ and $R_\lambda$.

We finally consider the Lagrangian strain and rotation timescales, which will 
be useful for understanding the trends in particle clustering in \textsection \ref{sec:particle_clustering}.
Since the fluid and particle phases are isotropic, we will have nine statistically equivalent strain timescales:
$T^p_{\mathcal{S}_{11} \mathcal{S}_{11}}$, $T^p_{\mathcal{S}_{11} \mathcal{S}_{22}}$, 
$T^p_{\mathcal{S}_{11} \mathcal{S}_{33}}$, $T^p_{\mathcal{S}_{12} \mathcal{S}_{12}}$, 
$T^p_{\mathcal{S}_{13} \mathcal{S}_{13}}$, $T^p_{\mathcal{S}_{22} \mathcal{S}_{22}}$, 
$T^p_{\mathcal{S}_{22} \mathcal{S}_{33}}$, $T^p_{\mathcal{S}_{23} \mathcal{S}_{23}}$, 
and $T^p_{\mathcal{S}_{33} \mathcal{S}_{33}}$. We take the strain timescale $T^p_{\mathcal{SS}}$ to be
the average of these nine components. 
We similarly take the rotation timescale $T^p_{\mathcal{R} \mathcal{R}}$
to be the average of three statistically equivalent components: 
$T^p_{\mathcal{R}_{12} \mathcal{R}_{12}}$, $T^p_{\mathcal{R}_{13} \mathcal{R}_{13}}$, and $T^p_{\mathcal{R}_{23} \mathcal{R}_{23}}$.

\begin{figure}
 \centering
 \includegraphics[width=2.6in]{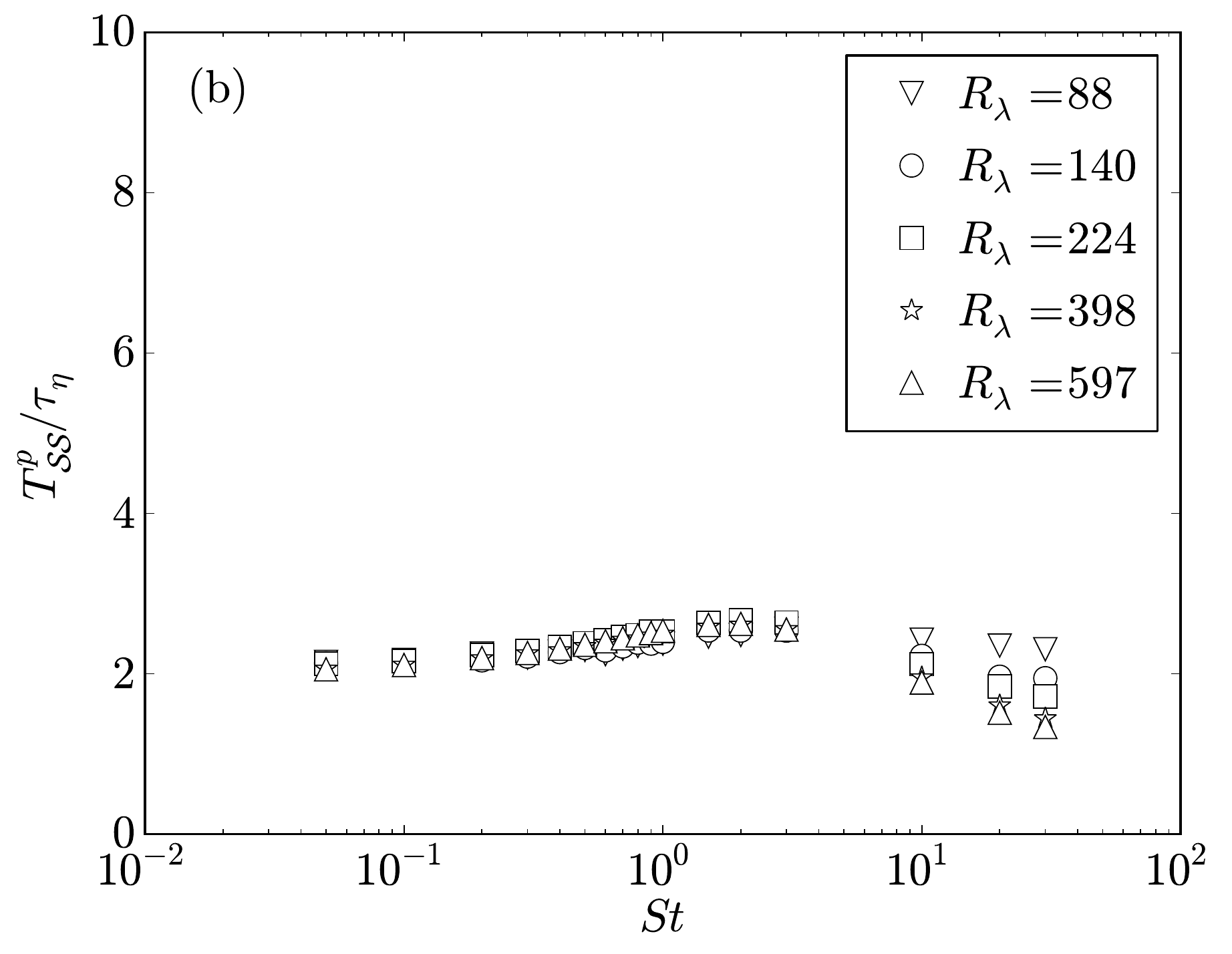}
 \includegraphics[width=2.6in]{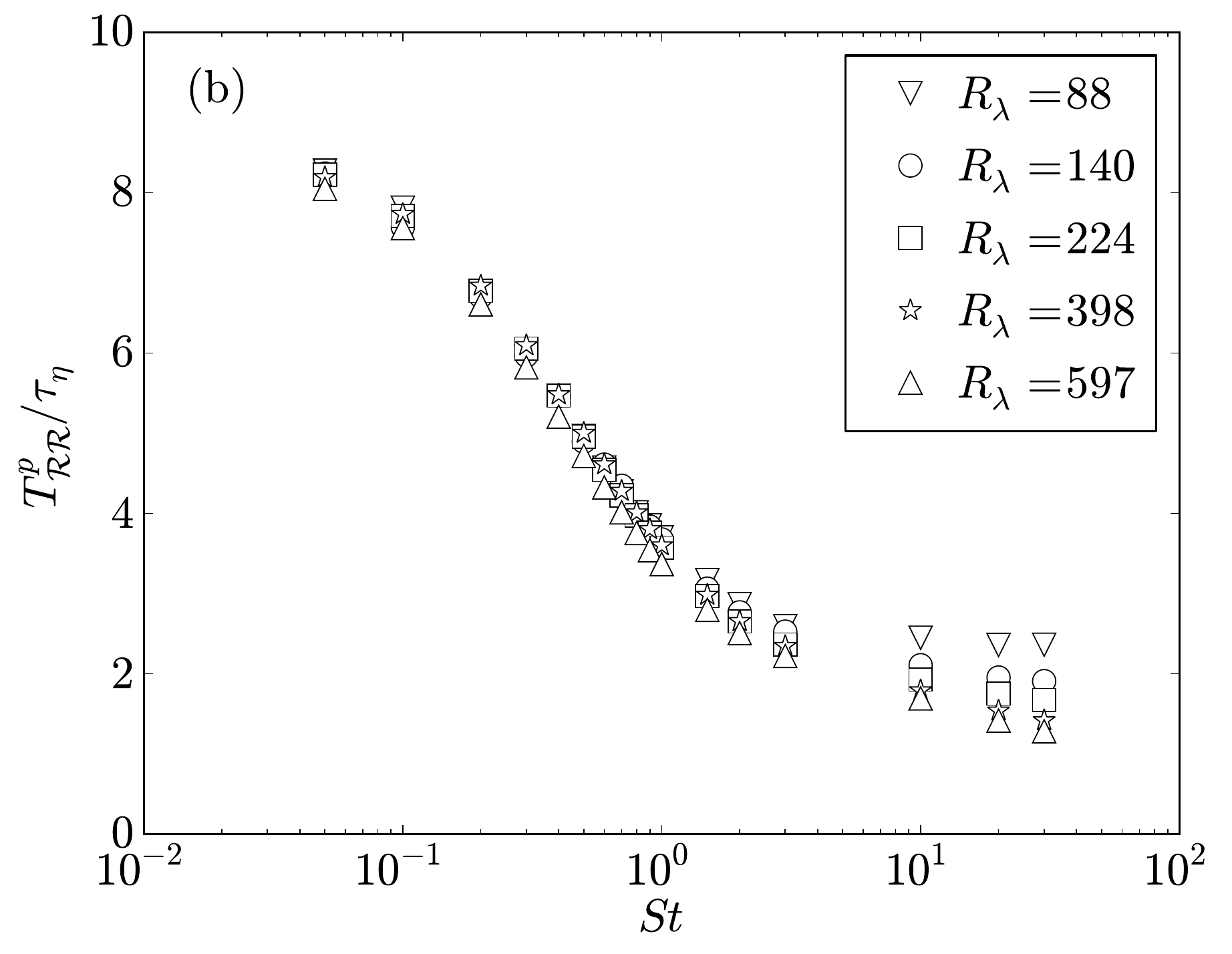}
 \caption{Lagrangian timescales of a single component
 of the strain rate (a) and rotation rate (b) tensors,
 plotted as a function of $St$ for different values of $R_\lambda$.}
 \label{fig:strain_rotation_timescales}
\end{figure}

We see that $T^p_{\mathcal{S} \mathcal{S}}/\tau_\eta$ is independent of $R_\lambda$ for $St < 10$,
and decreases weakly with increasing $R_\lambda$ for $St \geq 10$.
On the other hand, $T^p_{\mathcal{R} \mathcal{R}}/\tau_\eta$ tends to decrease with increasing $R_\lambda$
for all values of $St$, and this decrease becomes more pronounced as $St$ increases.
We also see that $T^p_{\mathcal{R} \mathcal{R}}$ is much 
more sensitive to changes in $St$ than $T^p_{\mathcal{S} \mathcal{S}}$, suggesting that the dominant effect
of inertia is to cause particles to spend less time in strongly rotating regions.
As a result, the particles will generally have less time to respond to fluctuations
in the rotation rate, causing $\langle \mathcal{R}^2 \rangle^p$ to be strongly reduced
with increasing $St$, as was seen above.

\subsection{Particle kinetic energy}
\label{sec:particle_kinetic_energy}
\begin{figure}
 \centering
 \includegraphics[height=1.9in]{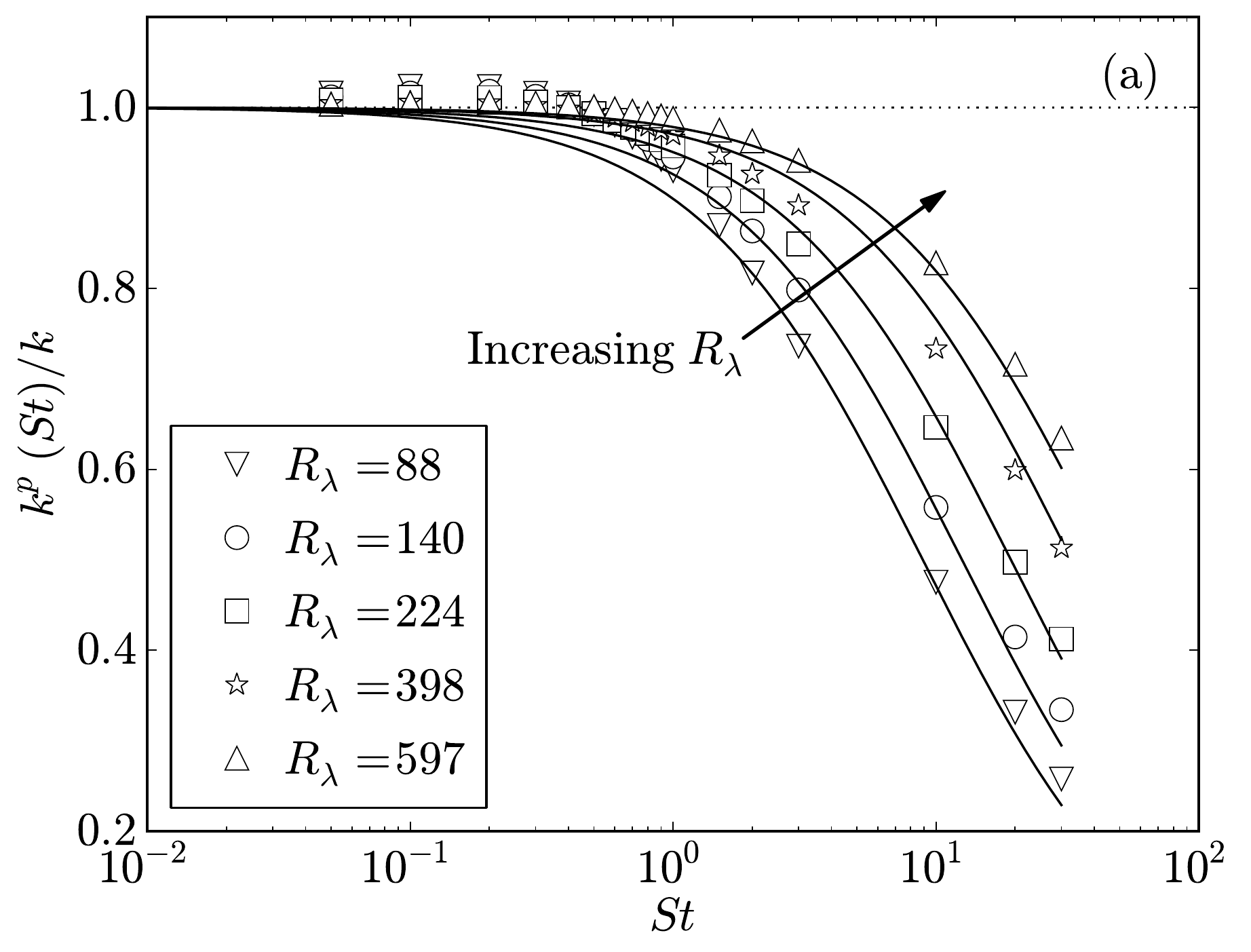}
 \includegraphics[height=1.9in]{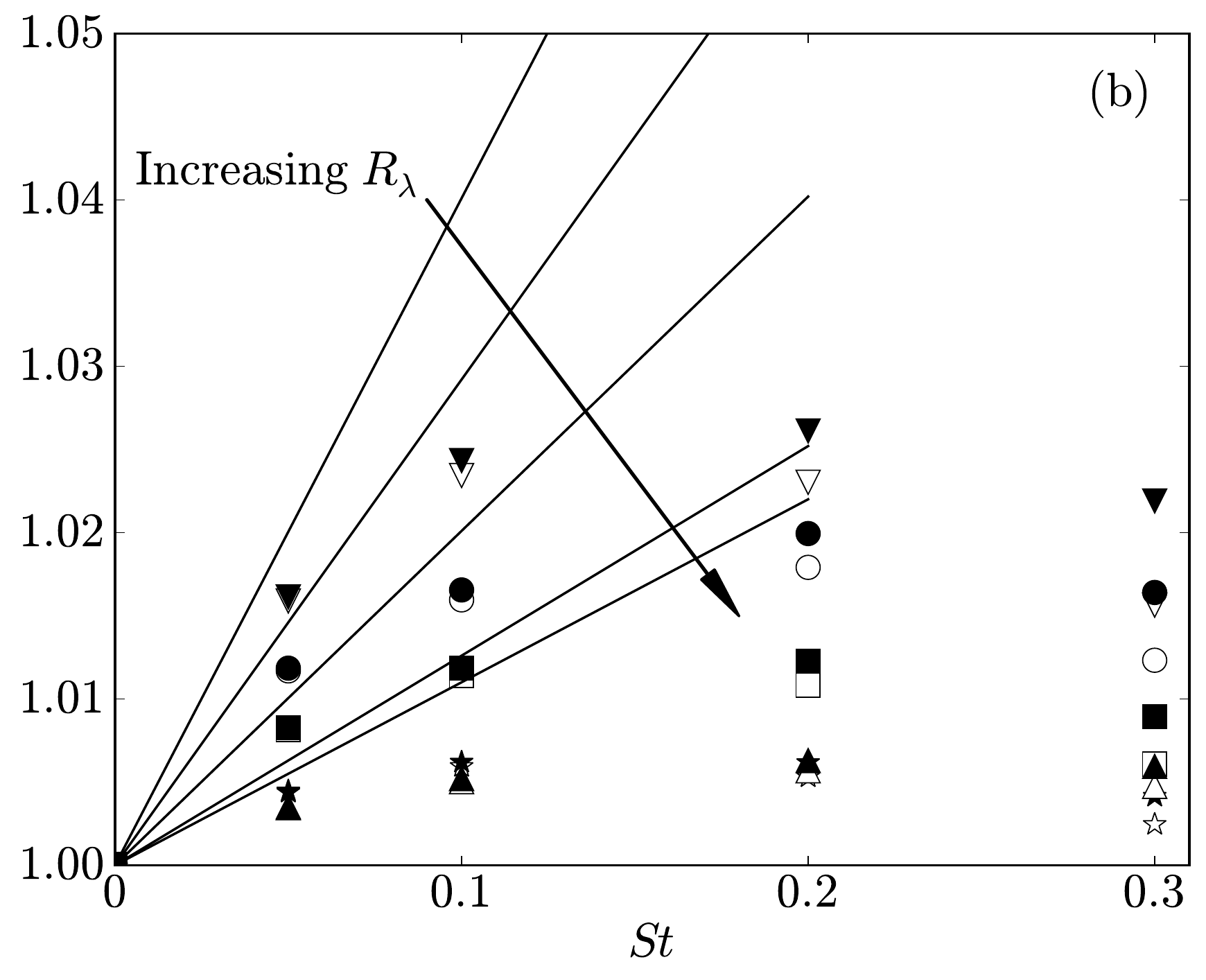}
 \caption{(a) The ratio between the average particle kinetic energy $k^p(St)$ 
 and the average fluid kinetic energy $k$ for different values of $R_\lambda$.
 DNS data are shown with symbols, 
 and the predictions of the filtering model in 
 (\ref{eq:filtering_theory}) are shown with solid lines.
 (b) The ratio between $k^p(St)$ and $k$ (open symbols),
 and the ratio between the average fluid kinetic energy at the particle locations $k^{fp}(St)$ 
 and $k$ (filled symbols), shown at low $St$ to highlight the effects of preferential sampling.
 Also shown is the prediction from the preferential sampling model given in (\ref{eq:chun_theory}) (solid lines).}
 \label{fig:particle_kinetic_energy}
\end{figure}

We now move from small-scale velocity statistics to large-scale velocity statistics.
Figure~\ref{fig:particle_kinetic_energy} shows the average particle kinetic energy 
$k^p(St) \equiv \frac{1}{2} \langle \bm{v}^p(t) \cdot \bm{v}^p(t) \rangle$ 
(normalized by the average fluid kinetic energy $k$) 
for different values of $R_\lambda$.

We first consider the effect of inertial filtering on this statistic,
and then examine the effect of preferential sampling.
It is well-known that filtering leads to a reduction in the particle turbulent 
kinetic energy for large values of $St$.
This reduction is the strongest (weakest) for the lowest (highest) Reynolds numbers,
as seen in figure~\ref{fig:particle_kinetic_energy}(a).
These trends are captured by the model in \cite{abrahamson75}, 
which assumes an exponential decorrelation of the Lagrangian fluid velocity.
Under this assumption, the ratio between the particle and fluid kinetic energies
can be expressed as
\begin{equation}
\label{eq:filtering_theory}
 \frac{k^p(St)}{k} \approx \frac{1}{1 + \tau_p/\tau_\ell} = 
 \frac{1}{1 + St \left(\tau_\eta/\tau_\ell \right)} \mathrm{,}
\end{equation}
where $\tau_\ell$ is the Lagrangian correlation time of the fluid,
which we approximate using the relation given in \cite{zaichik03a}.
The model predictions of $k^p(St)/k$ are included in figure~\ref{fig:particle_kinetic_energy}(a)
and are in good agreement with the DNS at large $St$, 
where filtering is dominant. The trends with $R_\lambda$ are also reproduced well.

We thus have the following physical explanation of inertial filtering on the particle kinetic energies:
for low-Reynolds-number flows, the response time of the largest particles
exceeds the timescales of many large-scale flow features.
The result is a filtered response to the large-scale turbulence 
and an overall reduction in the particle kinetic energy.
As the Reynolds number is increased 
(and the particle response time is fixed with respect to the small-scale turbulence),
more flow features are present with timescales that exceed the particle response time,
and hence the effect of inertial filtering is diminished with increasing $R_\lambda$,
as predicted by (\ref{eq:filtering_theory}).

To highlight the effect of preferential sampling on the particle kinetic energy,
figure~\ref{fig:particle_kinetic_energy}(b) shows both the average particle kinetic energy $k^p(St)$
and the average kinetic energy of the fluid sampled along an inertial particle trajectory,
$k^{fp}(St) \equiv \frac{1}{2} \langle \bm{u}(\bm{x}^p(t),t) \cdot \bm{u}(\bm{x}^p(t),t) \rangle$.
As is evident in figure~\ref{fig:particle_kinetic_energy}(b), the particle kinetic energy
exceeds $k$ for low values of $St$.
By comparing $k^p$ to $k^{fp}$,
we see that the increased kinetic energy of the smallest particles is due
almost entirely to preferential sampling of the flow field.
While \cite{salazar12a} were the first to show an increase in $k^p(St)/k$ for low $St$
(which they attributed to preferential sampling),
this trend is also suggested by the early study of \cite{squires91a},
in which the authors observed that small inertial particles 
preferentially sample certain high kinetic energy regions
they referred to as `streaming zones.'
Figure~\ref{fig:particle_kinetic_energy}(b) also shows that
at small values of $St$, $k^p(St)/k$ decreases with increasing Reynolds number.

The solid lines in figure~\ref{fig:particle_kinetic_energy}(b) show the predictions of the particle kinetic
energy from (\ref{eq:chun_theory}). In the limit of small $St$, the model of \cite{chun05} is able to
capture qualitatively both the increase in $k^p(St)/k$ with increasing $St$ and 
the decrease in $k^p(St)/k$ with increasing $R_\lambda$.

\begin{figure}
 \centering
 \includegraphics[height=2.2in]{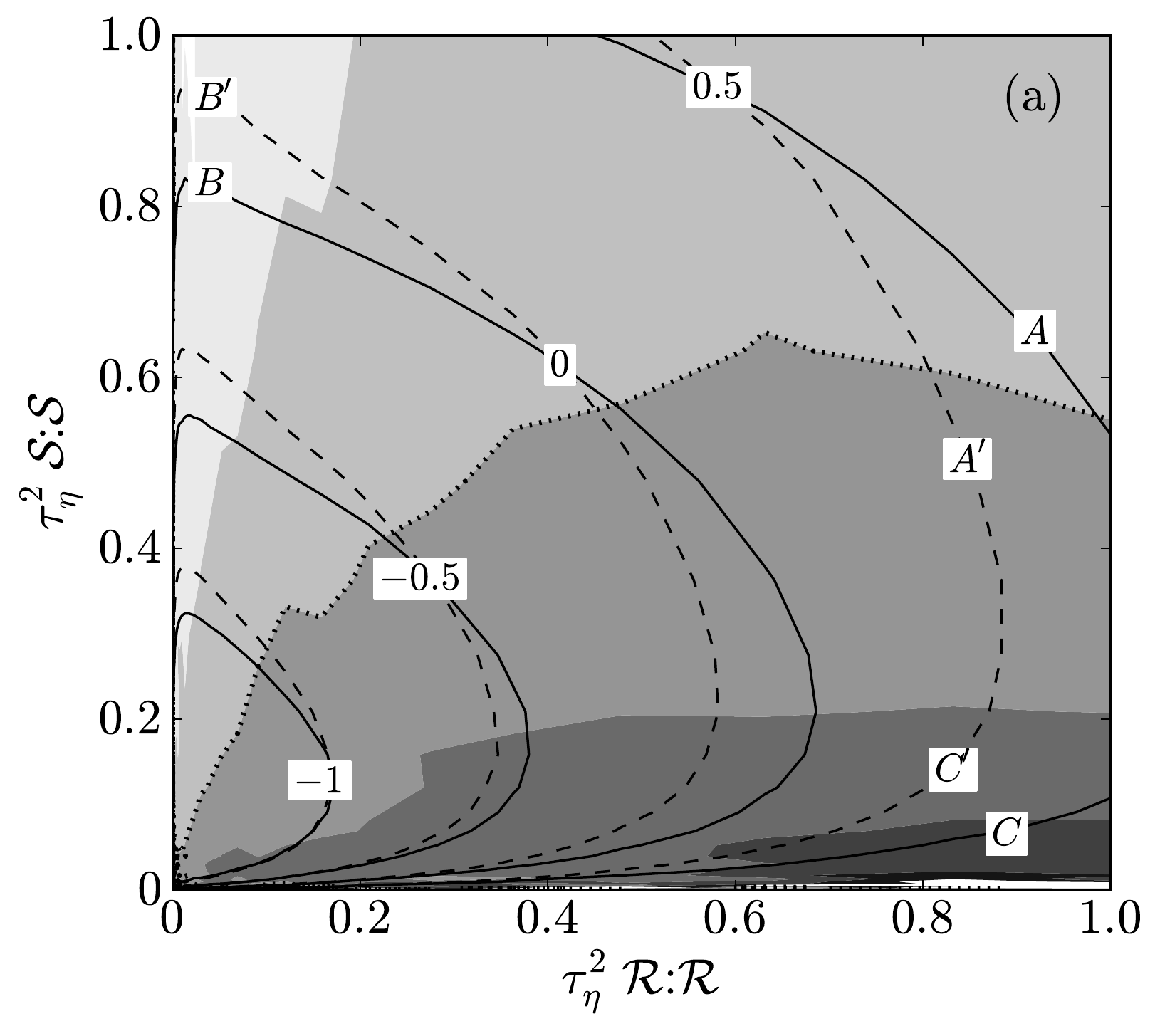}
 \includegraphics[height=2.2in]{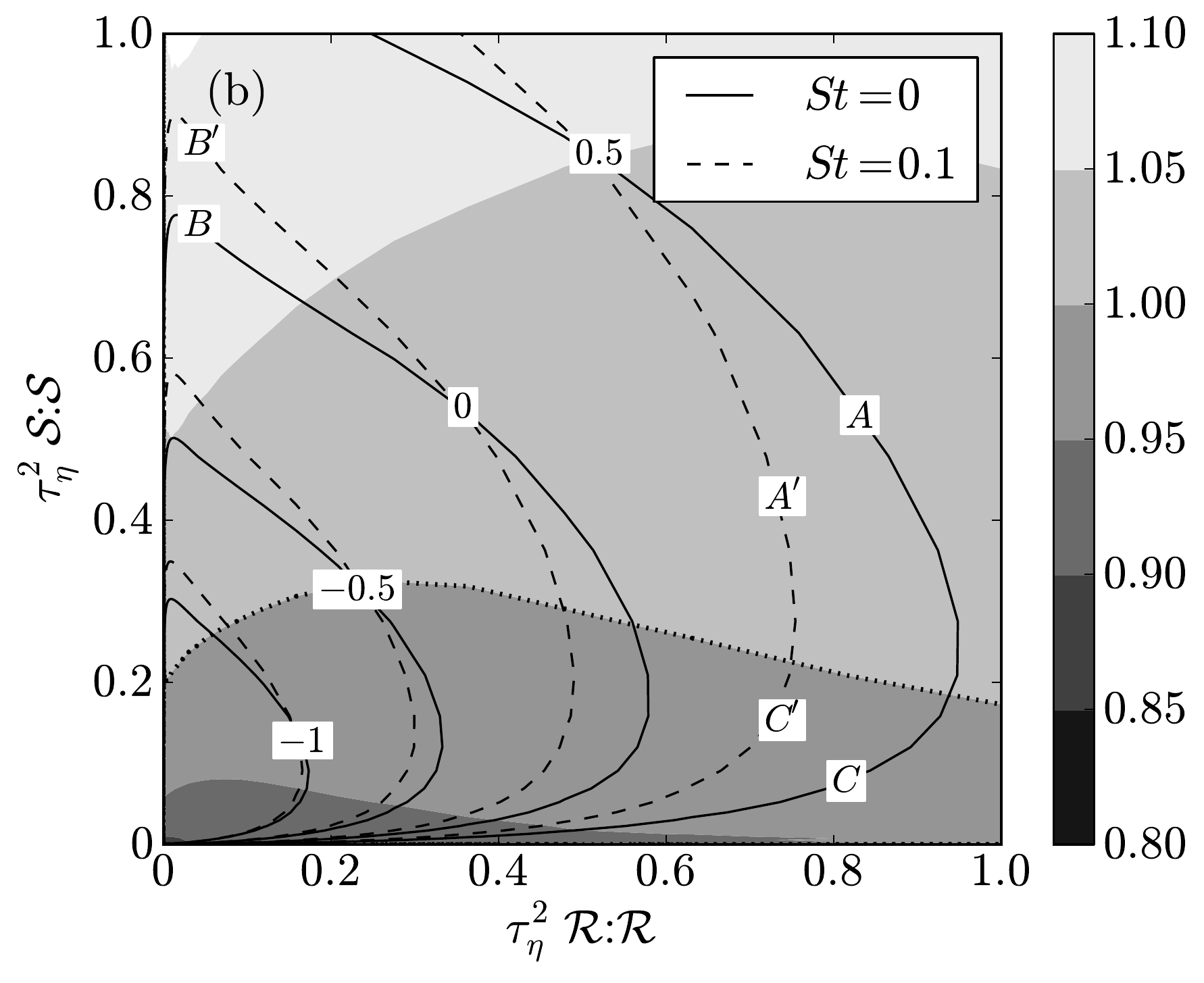}
 \caption{Filled contours of the fluid kinetic energy conditioned on $\mathcal{S}^2$ and $\mathcal{R}^2$,
 $k_{\mathcal{S}^2,\mathcal{R}^2}$, normalized by the unconditioned mean fluid kinetic energy $k$
 at (a) the lowest Reynolds number and (b) the highest Reynolds number.
 The dotted contour lines indicate $k_{\mathcal{S}^2,\mathcal{R}^2}/k = 1$.
 Isocontours of particle concentration for $St=0$ and $St=0.1$ particles are included
 for reference, with the exponents of the decade indicated on the contour lines.
 Certain regions of the flow are labeled to aid in the discussion of the trends.}
 \label{fig:conditional_ke}
\end{figure}

To further elucidate the physical mechanisms leading to these trends,
we plot the mean kinetic energy of the fluid conditioned on $\mathcal{S}^2$ and $\mathcal{R}^2$,
$k_{\mathcal{S}^2,\mathcal{R}^2}$, in figure~\ref{fig:conditional_ke}.
Isocontours of the concentrations of $St=0$ and $St=0.1$ particles
are shown for comparison.
While the data contain considerable statistical noise,
we can draw a few conclusions about the qualitative trends.

From figure~\ref{fig:conditional_ke}(a), 
we see that the change in kinetic energy at $R_\lambda = 88$ can be divided into the
three mechanisms discussed in \textsection \ref{sec:topology}.
First, particles are ejected from vortex sheets ($A$)
into moderate rotation and moderate strain regions ($A'$), 
which generally tends to decrease the particle kinetic energy.
Second, as $St$ increases, particles in irrotational straining regions ($B$) 
travel into regions of higher strain ($B'$),
which are characterized by higher kinetic energy. 
Third, some inertial particles are ejected from vortex tubes ($C$),
which are characterized by lower kinetic energies,
and travel into lower rotation and higher strain regions ($C'$), which have
higher kinetic energies. The observed increase in $k^p(St)/k$ must therefore be due to the
second and third mechanisms.

At high Reynolds numbers (figure~\ref{fig:conditional_ke}(b)), however, 
a larger portion of the flow is occupied by regions of overlapping high strain and high rotation
from which particles are ejected (see \textsection \ref{sec:topology}).
The first mechanism (which tends to decrease the kinetic energy) therefore plays a larger role.
Also, at $R_\lambda = 597$, high rotation and low strain regions ($C$) 
are no longer associated with very low kinetic energies,
causing the third mechanism to be less effective at increasing the particle kinetic energy.
The overall result is a decrease in $k^p(St)/k$ with increasing Reynolds number at small values of $St$.

\subsection{Particle accelerations}
\label{sec:particle_acceleration}

\begin{figure}
 \centering
 \includegraphics[width=2.6in]{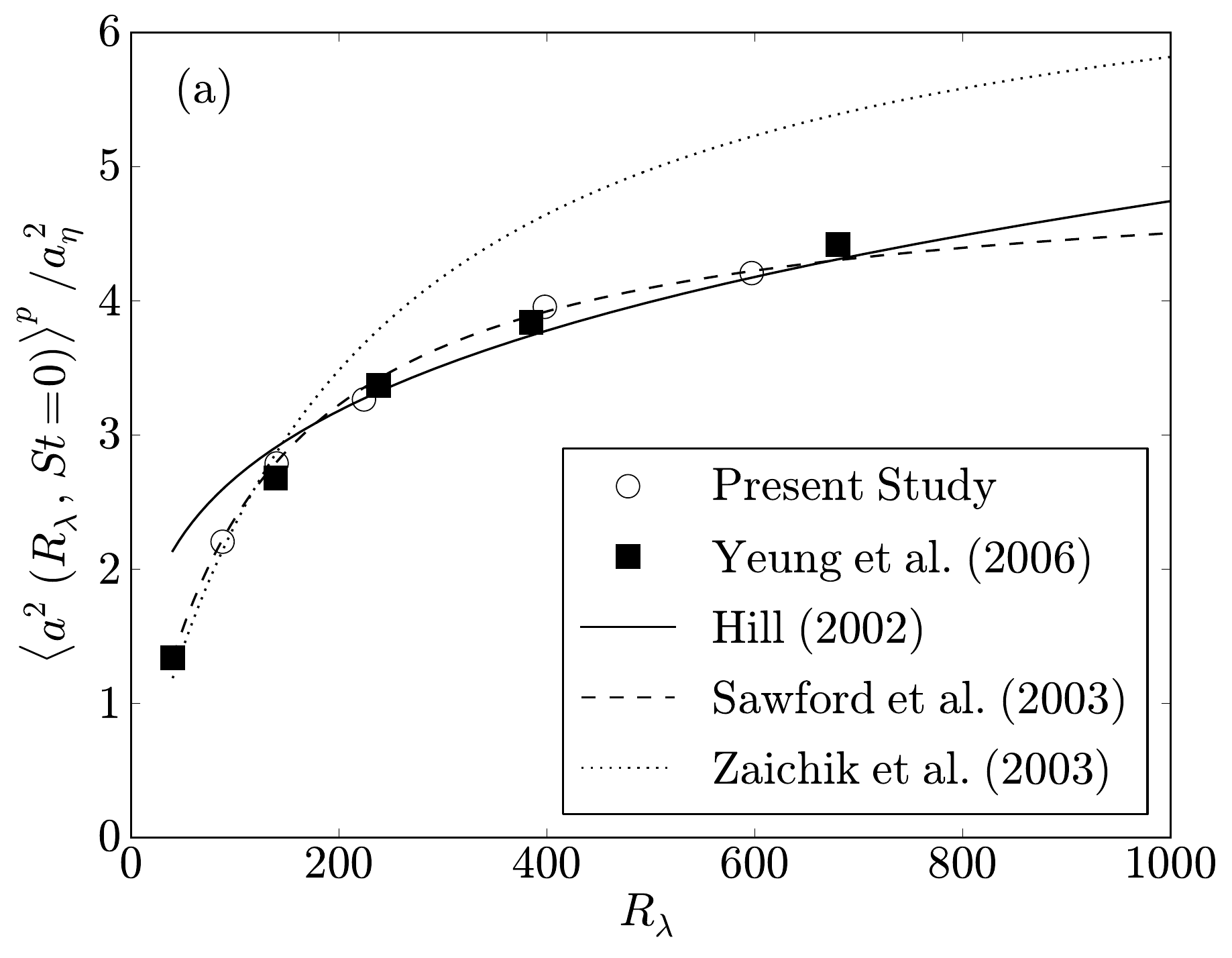}
 \includegraphics[width=2.6in]{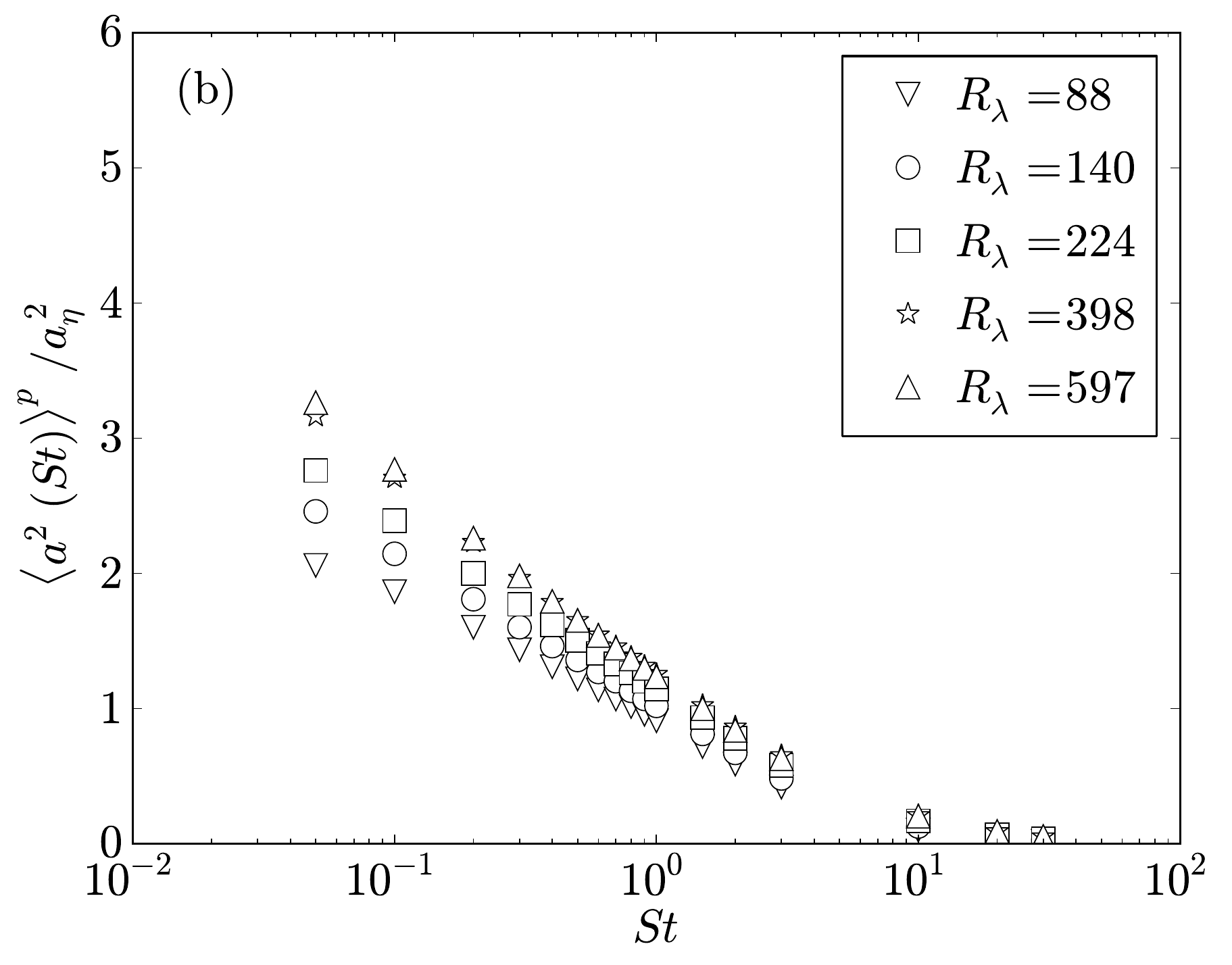}
 \caption{(a) The acceleration variance of Lagrangian fluid particles as a function of $R_\lambda$.
 The results from the present study (open circles) are compared to DNS data
 from \cite{yeung06b} (filled squares) and several theoretical predictions (lines).
 (b) The acceleration variance of inertial particles as a function of $St$ for different
 values of $R_\lambda$.}
 \label{fig:acceleration_aeta}
\end{figure}

In this section, we analyze fluid and inertial particle accelerations 
$\bm{a}^p(t) \equiv d\bm{v}^p(t)/dt$.
Fluid particle accelerations are known to be strongly intermittent
\citep[e.g., see][]{voth02,ishihara07},
with the probability of intense acceleration events 
increasing with the Reynolds number.
Before accounting for inertial effects,
we consider the effect of $R_\lambda$ on the 
acceleration variance 
$\langle a^2 \rangle^p \equiv \langle\bm{a}^p(t)\bm{\cdot}\bm{a}^p(t) \rangle/3$
of Lagrangian fluid particles in figure~\ref{fig:acceleration_aeta}(a).
To facilitate comparison between the different Reynolds numbers,
we have normalized $\langle a^2 \rangle^p$ by the Kolmogorov acceleration
variance $a_\eta^2 \equiv \sqrt{\epsilon^3/\nu}$.
The DNS data from \cite{yeung06b} and the theoretical predictions of 
\cite{hill02}, \cite{sawford03}, and \cite{zaichik03a} are shown for comparison.
We see that our DNS data agrees well with \cite{yeung06b}, and that the
model of \cite{sawford03} best reproduces the trends in the DNS.
\cite{hill02} breaks down at low $R_\lambda$, while 
\cite{zaichik03a} fails at high $R_\lambda$.

We turn our attention to inertial particle accelerations in figure~\ref{fig:acceleration_aeta}(b).
The observed trend for inertial particles is analogous to that for fluid particles:
at each value of $St$ considered, the particle acceleration variance (normalized by Kolmogorov units)
monotonically increases with $R_\lambda$ \citep[cf.][]{bec06a}.
As $St$ increases, the acceleration variance decreases,
presumably as a result of both preferential sampling of the flow
field and inertial filtering.

\begin{figure}
 \centering
 \includegraphics[height=1.9in]{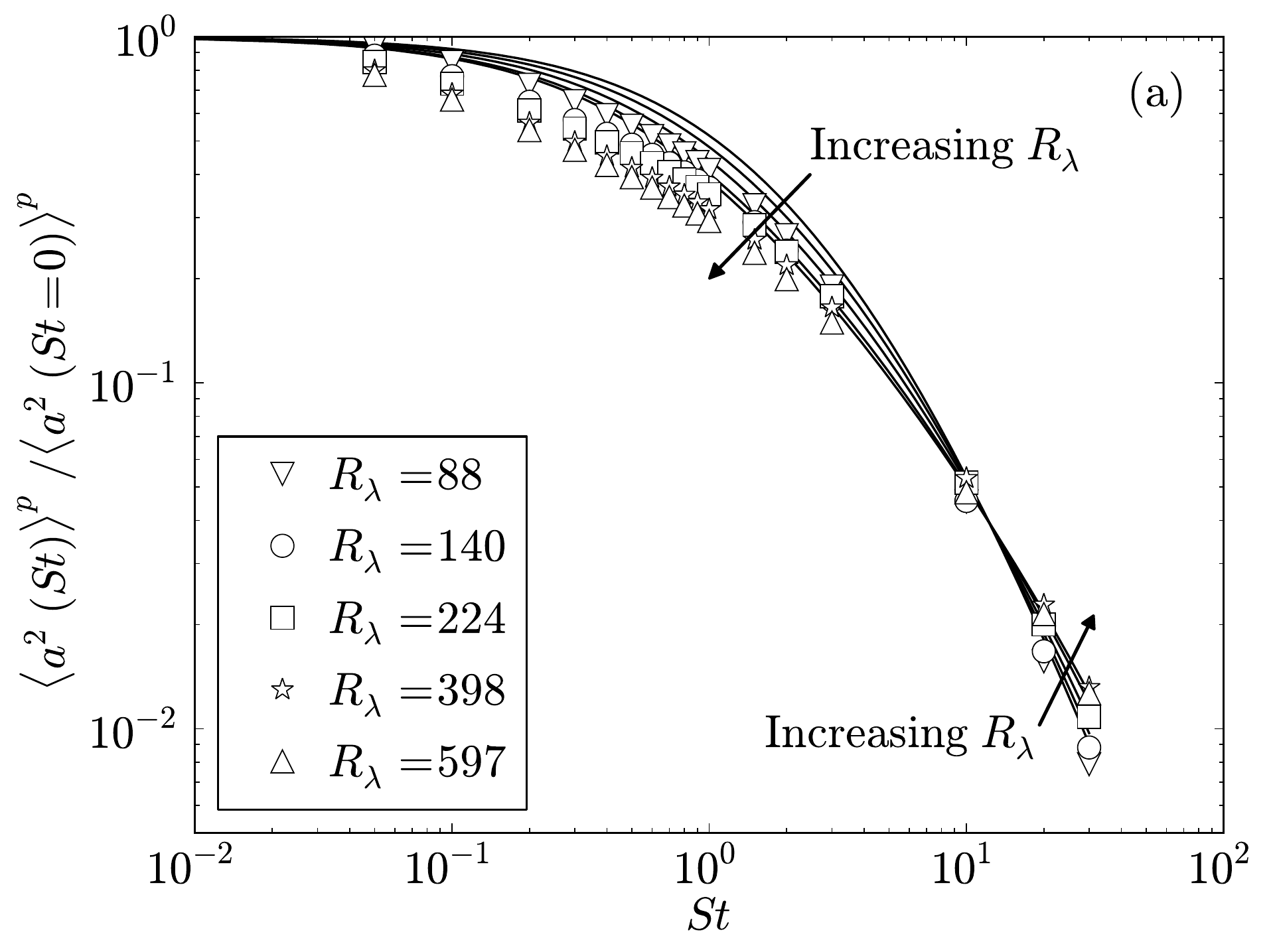}
 \includegraphics[height=1.9in]{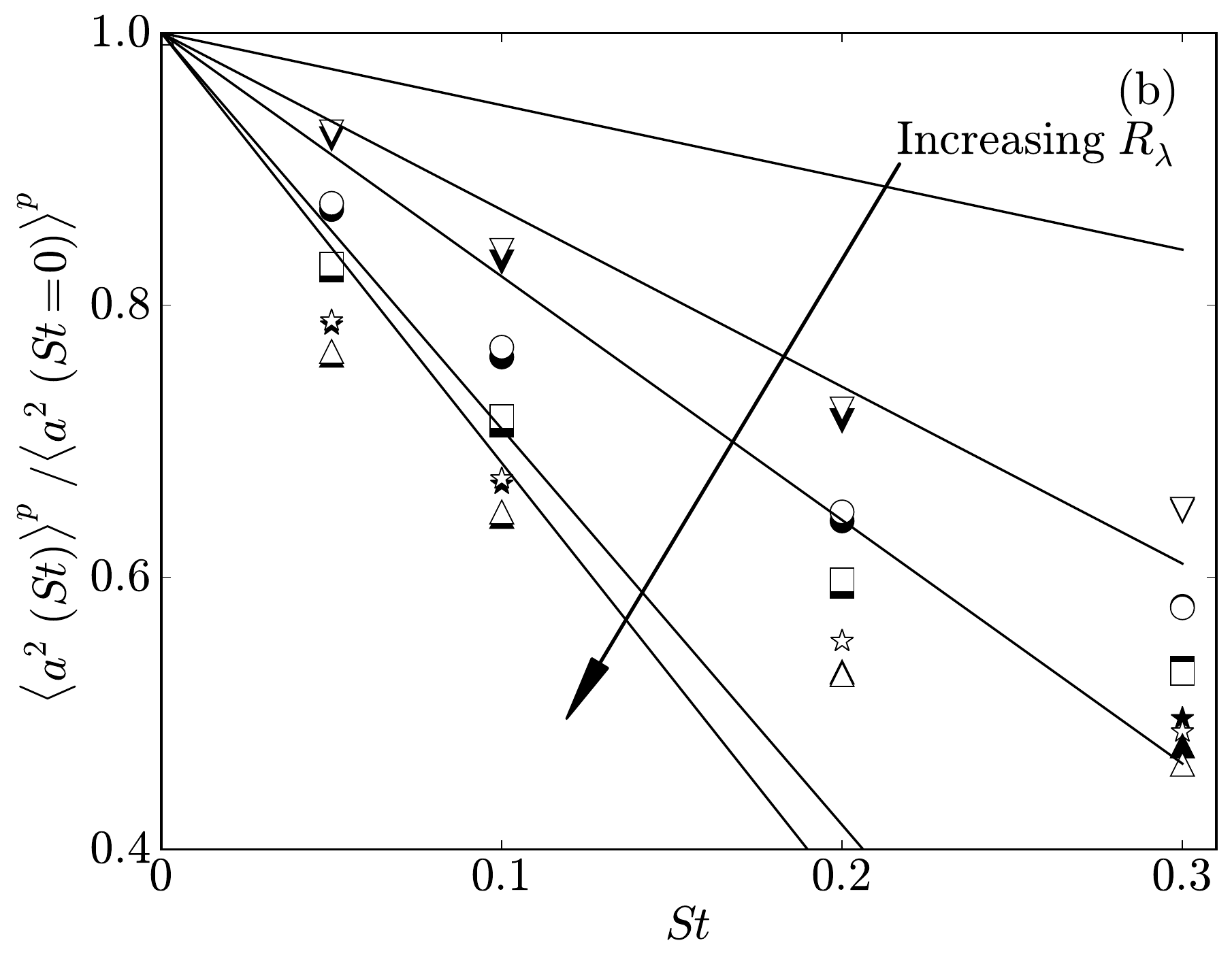}
 \caption{(a) Inertial particle acceleration variances scaled by
 the fluid particle acceleration variance (open symbols). The solid lines
 and arrows indicate the predictions from the filtering model of \cite{zaichik08}.
 (b) The variance of the inertial particle accelerations (open symbols)
 and the fluid velocity accelerations along the particle trajectories (filled symbols),
 shown at low $St$ to highlight the effect of preferential sampling.
 The solid lines indicate the predictions from the preferential sampling model given
 in (\ref{eq:chun_theory}).}
 \label{fig:acceleration_afluid}
\end{figure}

We now seek to understand and model how inertia changes the accelerations of
particles through the filtering and preferential sampling effects.
To do so, we rescale the inertial particle acceleration variance by that of fluid particles
and plot the results in figure~\ref{fig:acceleration_afluid}.
In figure~\ref{fig:acceleration_afluid}(a),
we compare the rescaled acceleration variance to the model of \cite{zaichik08},
which only accounts for inertial filtering of the underlying flow.
The model of \cite{zaichik08} is able to capture all the qualitative trends
in $R_\lambda$ and $St$, and the model predictions provide remarkably good
quantitative agreement with the DNS at the largest values of $St$,
where filtering is the dominant mechanism.
At lower values of $St$, the rescaled particle acceleration variance
decreases with increasing $R_\lambda$. In this case, as $R_\lambda$ increases,
the underlying flow is subjected to increasingly intermittent acceleration events, 
and the inertial particles filter a larger fraction of these events.
At the largest values of $St$, most intermittent accelerations are filtered,
and a particle's acceleration variance is determined by its interaction with the largest
turbulence scales.  Since the range of available large scales increases with $R_\lambda$,
the rescaled particle acceleration variance increases with $R_\lambda$ for the largest values of $St$.

We now consider the effect of preferential sampling on the acceleration variances.
In figure~\ref{fig:acceleration_afluid}(b), we plot the variance of both inertial particle accelerations 
and fluid accelerations along inertial particle trajectories
(scaled by the acceleration variance of $St=0$ particles).
As expected, for $St \ll 1$, where preferential sampling is the dominant mechanism,
inertial particle accelerations are almost equivalent to the accelerations of the underlying flow
sampled along the particle trajectories.
The model of \cite{chun05} (\ref{eq:chun_theory}) 
is able to reproduce all the qualitative trends correctly
in the limit of small $St$.
The scaled variances decrease with increasing $R_\lambda$,
and we expect that this trend is due to the fact that high vorticity regions
are associated with high accelerations \citep{biferale05b} and
become increasingly efficient at ejecting particles (refer to \textsection \ref{sec:topology}).

\begin{figure}
 \centering
 \includegraphics[height=2.2in]{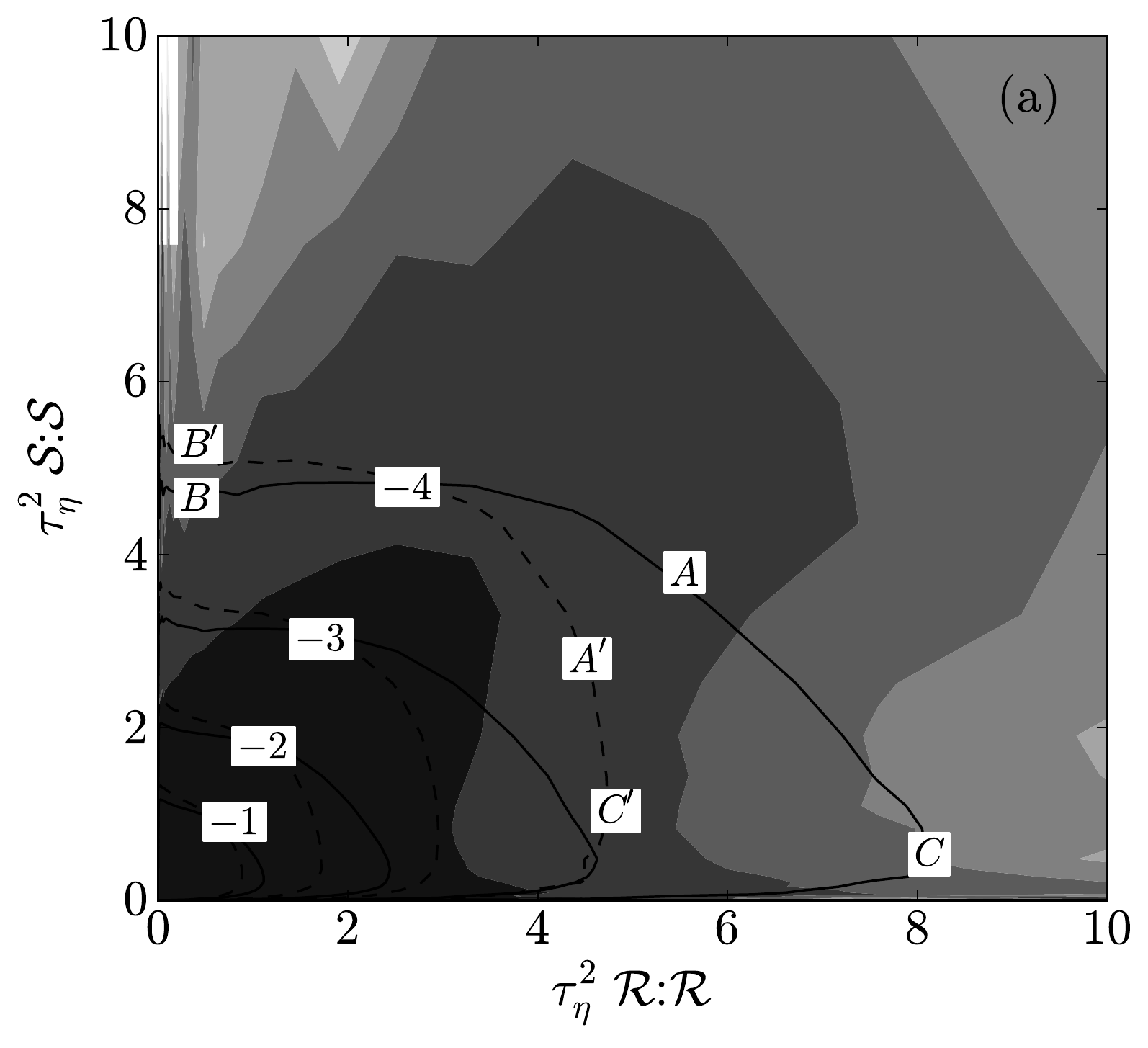}
 \includegraphics[height=2.2in]{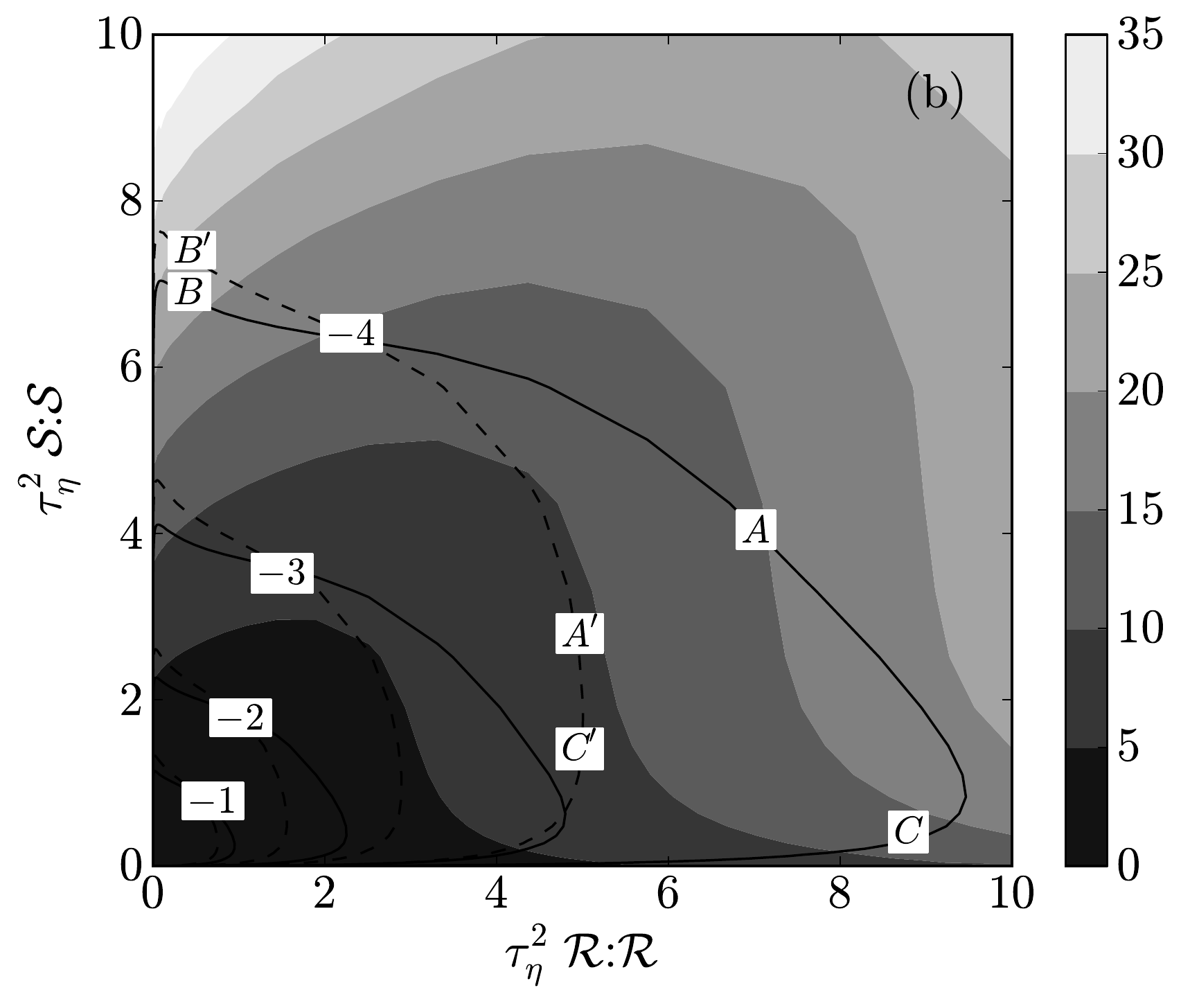}
 \caption{Filled contours of the variance of the fluid particle accelerations conditioned on $\mathcal{S}^2$ and $\mathcal{R}^2$,
 $\langle a^2 \rangle^p_{\mathcal{S}^2,\mathcal{R}^2}$, normalized by the unconditioned 
 fluid particle acceleration variance $\langle a^2 \rangle^p$,
 at (a) $R_\lambda = 88$ and (b) $R_\lambda = 597$.
 Isocontours of particle concentration for $St=0$ and $St=0.1$ particles are included
 for reference, with the exponents of the decade indicated on the contour lines.
 Certain regions of the flow are labeled to aid in the discussion of the trends.}
 \label{fig:conditional_a}
\end{figure}

We test this expectation in figure~\ref{fig:conditional_a} by plotting
the acceleration variance for fluid particles 
conditioned on $\mathcal{S}^2$ and $\mathcal{R}^2$, $\langle a^2 \rangle^p_{\mathcal{S}^2,\mathcal{R}^2}$,
and normalized by the unconditioned variance $\langle a^2 \rangle^p$.
We see that inertial particles are indeed ejected from high vorticity regions
(both vortex sheets and vortex tubes)
into lower vorticity regions (e.g., $A$ into $A'$ and $C$ into $C'$),
and that these high vorticity regions are marked by very large accelerations.
Though some inertial particles experience higher accelerations 
as they move into irrotational straining regions with higher strain rates
(e.g., $B$ into $B'$), this effect is relatively weak,
and the overall trend is a decrease in the particle accelerations
with increasing inertia.

\begin{figure}
 \centering
 \includegraphics[width=2.6in]{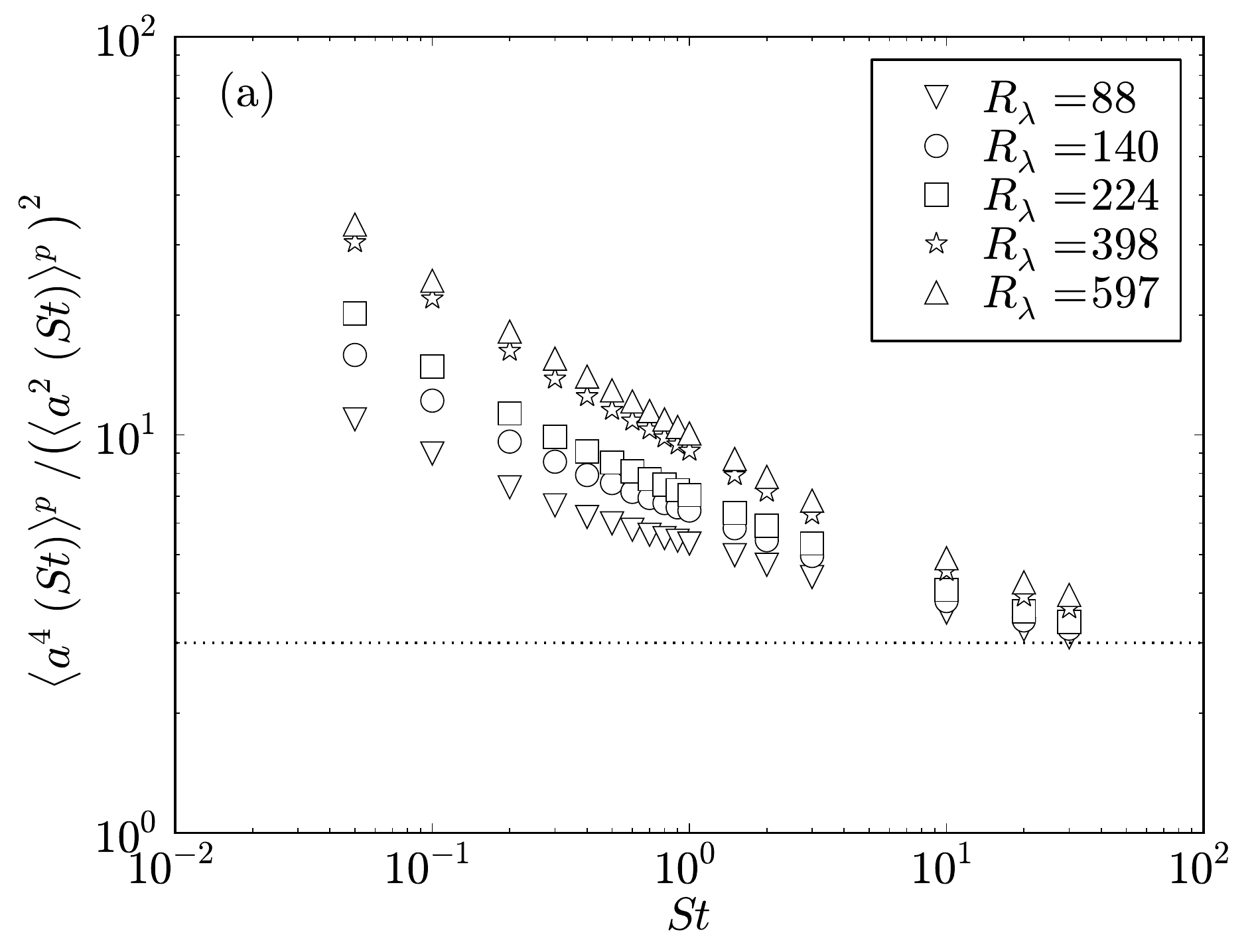}
 \includegraphics[width=2.6in]{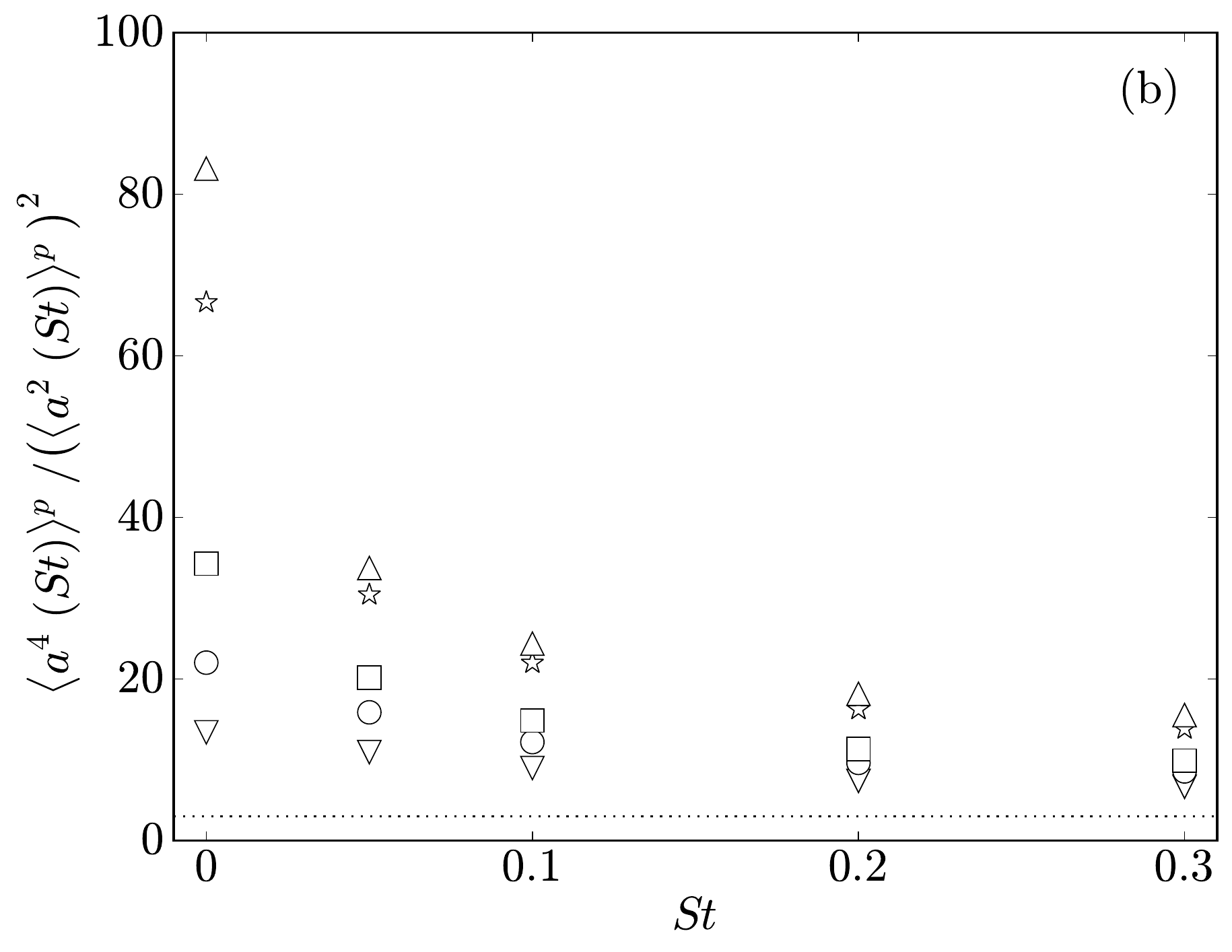}
 \caption{Particle acceleration kurtosis as a function of $St$
 for different values of $R_\lambda$. 
 The dotted line indicates  a kurtosis of $3$, 
 the value for a Gaussian distribution. Values over the whole range
 of non-zero $St$ are shown in (a).
 (b) shows only small-$St$ results on a linear plot
 to emphasize the rapid reduction in kurtosis as $St$ increases from 0.}
 \label{fig:acceleration_hom}
\end{figure}

To investigate the intermittency of inertial particle accelerations,
we plot the kurtosis of the particle accelerations,
$\langle a^4 \rangle^p/{(\langle a^2 \rangle^p)^2}$, in figure~\ref{fig:acceleration_hom},
where $\langle a^4 \rangle^p \equiv \langle a^p_1(t)^4 + a^p_2(t)^4 + a^p_3(t)^4\rangle^p/3$.
(Note that a Gaussian distribution has a kurtosis of $3$,
as indicated in figure~\ref{fig:acceleration_hom} by a dotted line.)
As expected, the particle accelerations are highly intermittent,
with the degree of intermittency increasing with increasing $R_\lambda$.
The kurtosis decreases very rapidly as $St$ increases.
Figure~\ref{fig:acceleration_hom}(b) indicates that
the kurtosis of very small particles ($St=0.05$)
at the highest value of $R_\lambda$
is over a factor of two smaller than that of fluid particles.
The largest-$St$ particles have kurtosis values 
approaching those of a Gaussian distribution.
These trends can be explained by the fact that
both preferential sampling and inertial filtering decrease the
probability of high-intensity acceleration events.
Standardized moments of up to order $10$ (not shown) were also analyzed
and found to exhibit the same trends.

We should note that the grid resolution study in \cite{yeung06b}
suggests that the acceleration moments from our DNS may be under-predicted.
\cite{yeung06b} showed that at $R_\lambda \approx 140$,
increasing the grid resolution $k_\mathrm{max} \eta$ from $1.5$ to $12$
led to a $10\%$ increase in the fluid acceleration variance
and a $30\%$ increase in the fluid acceleration kurtosis.
It is unclear how these trends will change at higher $R_\lambda$,
but it suggests that the quantitative results reported here should be interpreted with caution.
(The velocity gradients presented earlier are likely reliable, however,
since \cite{yeung06b} found that such statistics are less dependent on the grid resolution.)
\section{Two-particle statistics}
\label{sec:two_particle_stats}
We now consider two-particle statistics relevant for predicting inertial particle collisions.
We analyze particle relative velocities in \textsection \ref{sec:relative_velocities},
clustering in \textsection \ref{sec:particle_clustering},
and use these data to compute the collision kernel in \textsection \ref{sec:collision_kernel}.
(The mean-squared separation of inertial particle pairs
was also studied from these data and is the topic of a separate publication \citep{bragg15}.)

\subsection{Particle relative velocities}
\label{sec:relative_velocities}

We study particle relative velocities as a function of both $St$ and $R_\lambda$.
The relative velocities for inertial particles are defined by the relation
\begin{equation}
 w^p_{\parallel,\perp}(t) \equiv 
 \left[ \bm{v}^p_2(t) - \bm{v}^p_1(t) \right] \cdot \bm{e}^p_{\parallel,\perp}(t)
 \mathrm{.}
\end{equation}
Here, $\bm{v}^p_1$ and $\bm{v}^p_2$ indicate the velocities of particles 1 and 2, respectively,
which are separated from each other by a distance $r^p(t)=\vert\bm{r}^p(t)\vert$.
The subscripts $\parallel$ and $\perp$ 
indicate directions parallel (longitudinal) to the separation vector
or perpendicular (transverse) to the separation vector, respectively,
and $\bm{e}^p_{\parallel,\perp}$ denotes the unit vector in the corresponding direction.
(We use the method discussed in \cite{pan13} to compute
the transverse components.)

We will also examine the velocity differences of the fluid at the particle locations, defined as
\begin{equation}
 \Delta u^p_{\parallel,\perp}(t) \equiv 
 \left[ \bm{u}^p_2(t) - \bm{u}^p_1(t) \right] \cdot \bm{e}^p_{\parallel,\perp}(t)
 \mathrm{,}
\end{equation}
where $\bm{u}^p_1$ and $\bm{u}^p_2$ are the velocities of the fluid underlying particles 1 and 2, respectively.
Note that for uniformly-distributed fluid ($St=0$) particles, 
the particle velocity statistics are equivalent to the underlying fluid velocity statistics.

Following the nomenclature in \cite{bragg14,bragg14b}, we denote particle relative velocity moments of order $n$ as
\begin{equation}
 S^p_{n\parallel}(r) \equiv \Big\langle \left[w_\parallel^p(t)\right]^n \Big\rangle_r \mathrm{,}
\end{equation}
for the components parallel to the separation vector, and as
\begin{equation}
 S^p_{n\perp}(r) \equiv \Big\langle \left[w_\perp^p(t)\right]^n \Big\rangle_r \mathrm{,}
\end{equation}
for components perpendicular to the separation vector.  
In these expressions $\langle \cdot \rangle_r$ denotes an ensemble average conditioned on 
${r}^p(t) = r$.

For the purposes of computing the collision kernel (see \textsection \ref{sec:collision_kernel}),
we are also interested in the mean inward relative velocity parallel to the separation vector,
defined as
\begin{equation}
S^p_{-\parallel}(r) \equiv 
-\int_{-\infty}^{0} w_\parallel p(w_\parallel \vert r) d w_\parallel \mathrm{,}
\end{equation}
where $p(w_\parallel | r)=\langle\delta(w^p_\parallel(t)-w_\parallel)\rangle_r$ is the 
PDF for the longitudinal particle relative velocity conditioned on ${r}^p(t) = r$.

Finally, in some cases we are also interested in moments of the fluid velocity differences.
We use a superscript $fp$ to denote the moments of fluid velocity differences at the particle locations,
and a superscript $f$ to denote the moments of fluid velocity differences at fixed points with separation $r$.
We therefore have
\begin{equation}
 S^{fp}_{n\parallel}(r) \equiv \Big\langle \left[\Delta u_\parallel({r}^p(t),t)\right]^n \Big\rangle_r \mathrm{,}
\end{equation}
and
\begin{equation}
 S^{f}_{n\parallel}(r) \equiv \Big\langle \left[\Delta u_\parallel(r,t) \right]^n \Big\rangle \mathrm{.}
\end{equation}
The components perpendicular to the separation vector are defined analogously.

We consider dissipation-range statistics in \textsection \ref{sec:wr_dissipation}
and inertial-range statistics in \textsection \ref{sec:wr_inertial}.

\subsubsection{Dissipation range relative velocity statistics}
\label{sec:wr_dissipation}

\begin{figure}
 \centering
 \includegraphics[width=2.6in]{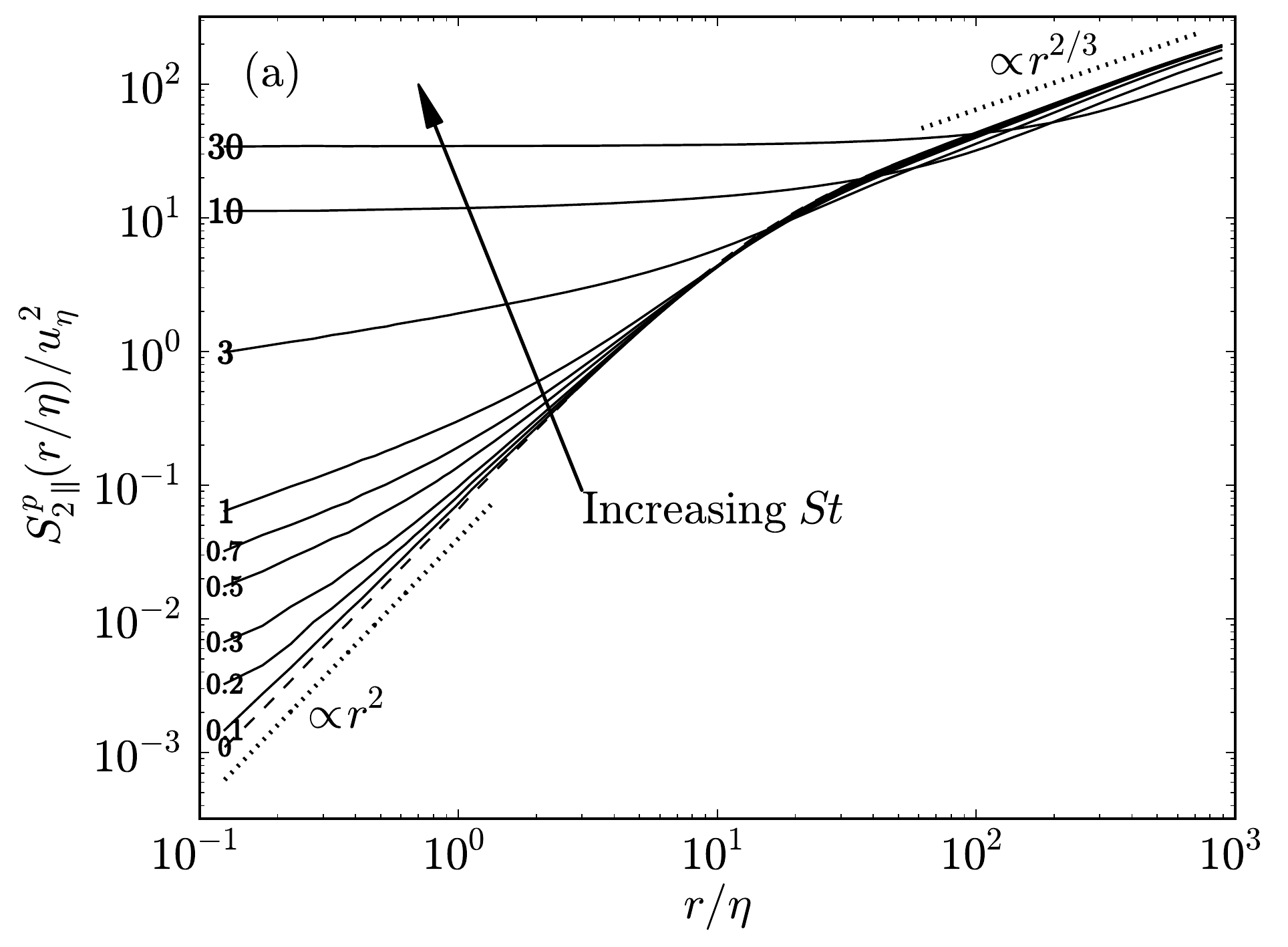}
 \includegraphics[width=2.6in]{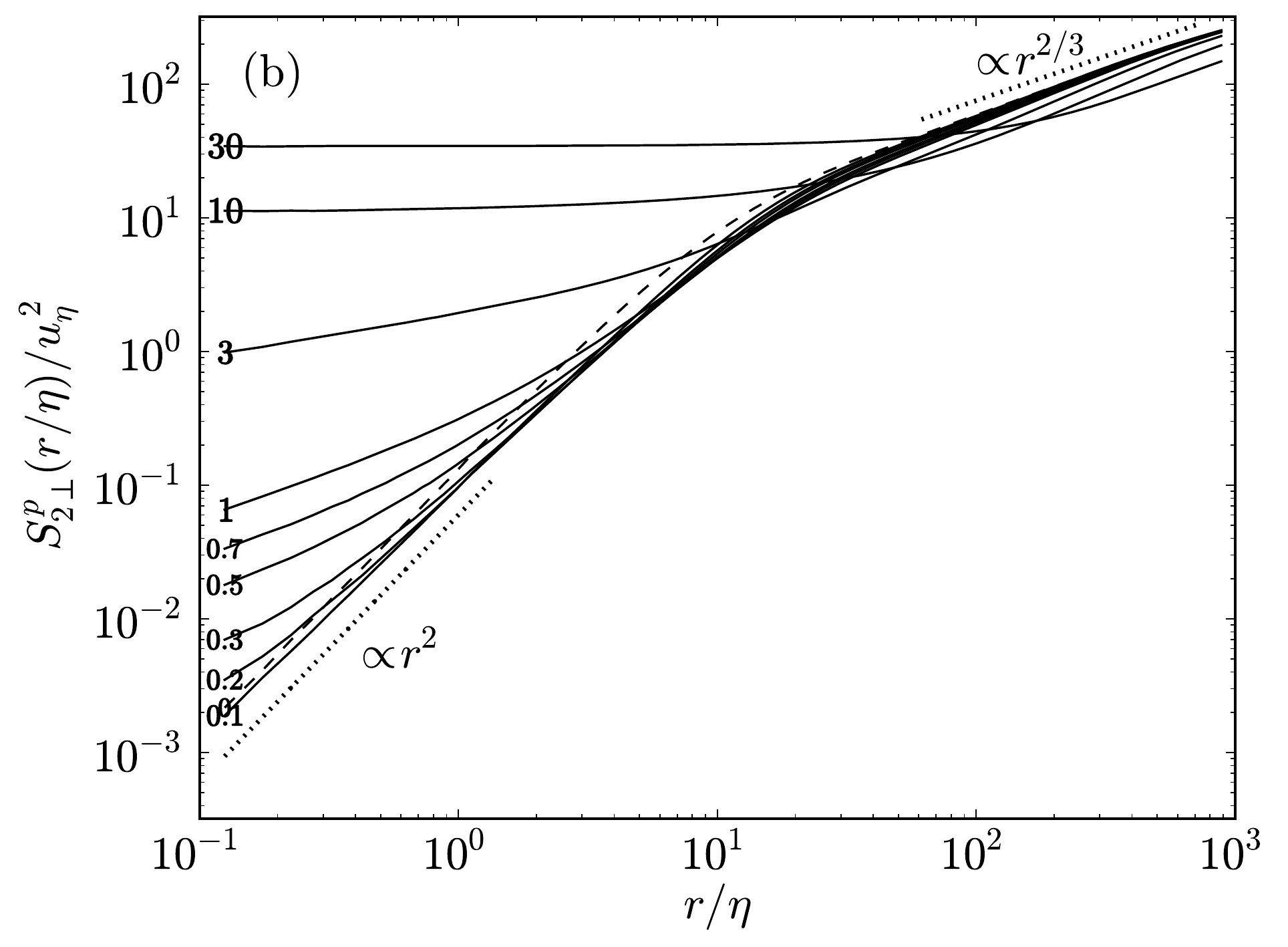}
 \caption{The particle relative velocity variances parallel to the separation vector (a)
 and perpendicular to the separation vector (b),
 plotted as a function of the separation $r/\eta$ for $R_\lambda = 597$.
 The Stokes numbers are indicated by the line labels, and the $St=0$ curves are shown
 with dashed lines for clarity.  The expected dissipation and inertial range scalings 
 \citep[based on][]{kolmogorov41a} are included for reference.}
 \label{fig:wr_lom}
\end{figure}

In figure~\ref{fig:wr_lom}, we plot the relative velocity variances
$S^p_{2 \parallel}$ and $S^p_{2 \perp}$ versus $r/\eta$ at $R_\lambda = 597$. 
The mean inward relative velocity (not shown) has the same qualitative trends,
and will be considered later in this section.
For the purposes of the following discussion,
we define the dissipation range as the region over which the fluid velocity variances
follow $r^2$-scaling, which is seen to be $0 \leq r/\eta \lesssim 10$ in figure~\ref{fig:wr_lom},
in agreement with \cite{ishihara09}.

At small separations, the relative velocity variances parallel to the separation vector
(figure~\ref{fig:wr_lom}(a)) increase monotonically
with $St$ and deviate from $r^2$-scaling,
while the relative velocity variances perpendicular to the separation vector decrease for $St \lesssim0.1$
and then increase monotonically with $St$ for $St \gtrsim 0.1$ (figure~\ref{fig:wr_lom}(b)).
We expect that the trends at small separations and small $St$ are primarily 
due to preferential sampling of the underlying flow,
which also dictates much of the single-particle dynamics for small $St$
(refer to \textsection \ref{sec:single_particle}).

\begin{figure}
 \centering
 \includegraphics[width=2.6in]{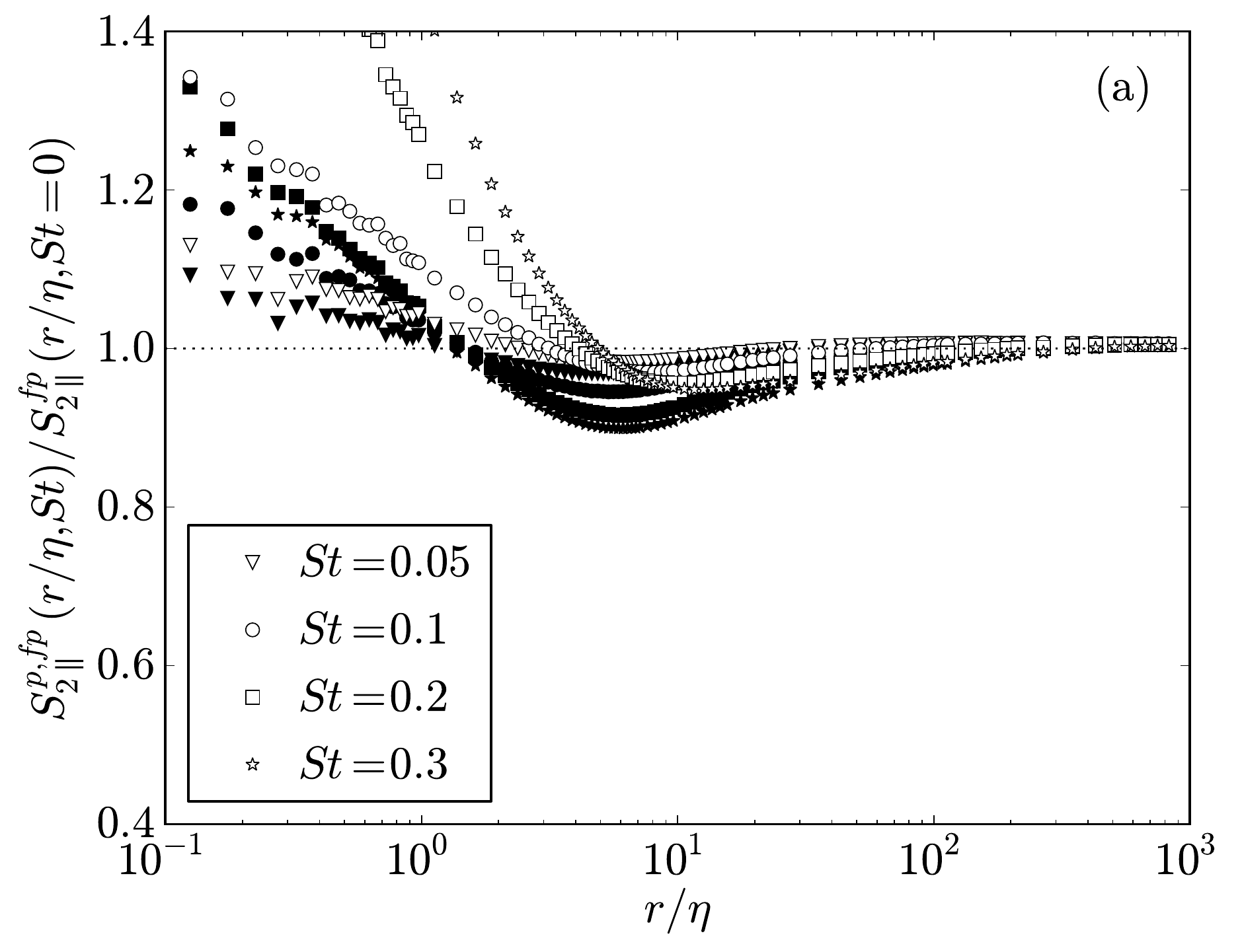}
 \includegraphics[width=2.6in]{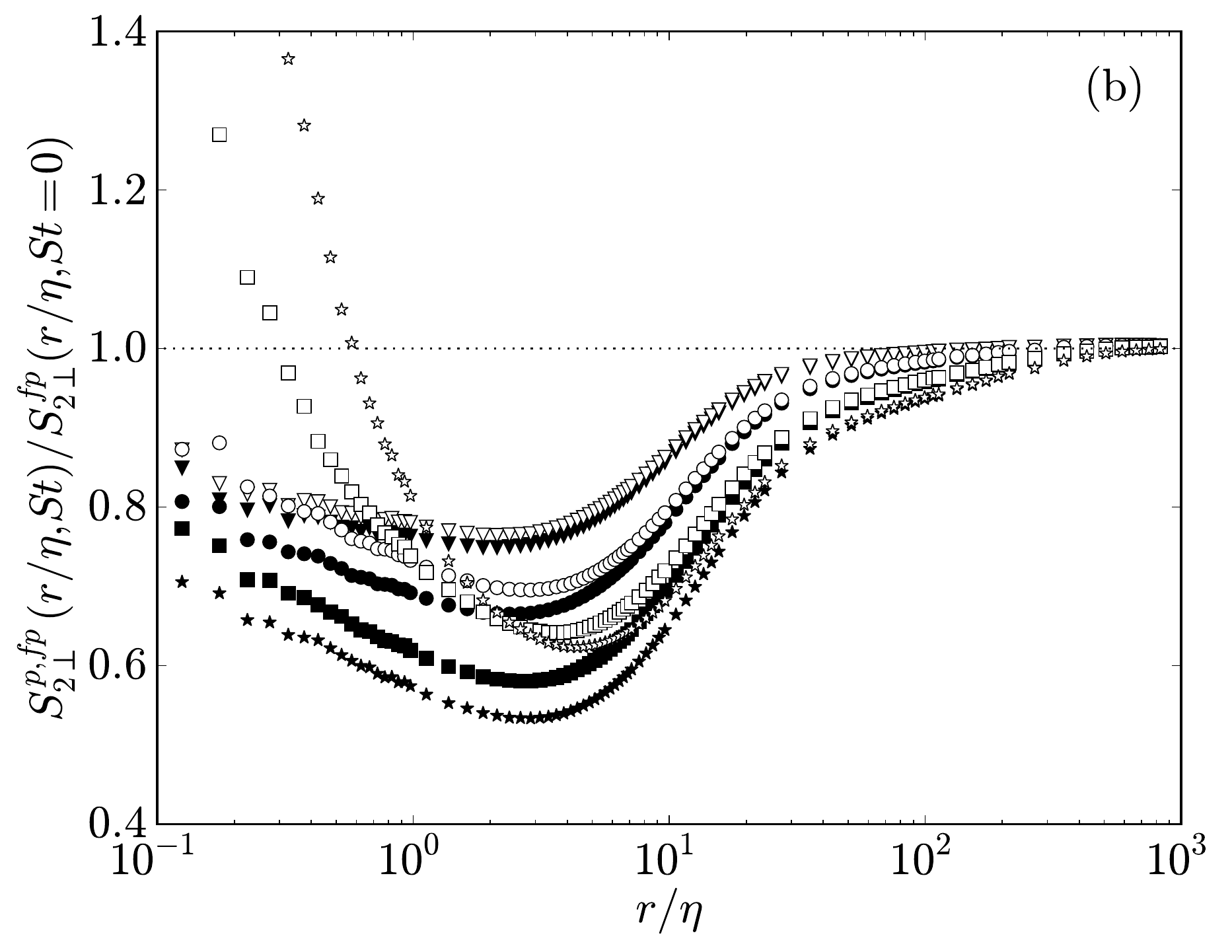}
 \caption{The parallel (a) and perpendicular (b) relative velocity variances of 
 inertial particles ($S^p_{2\parallel}$ and $S^p_{2\perp}$, open symbols) and of
 the fluid at inertial particle positions ($S^{fp}_{2\parallel}$ and $S^{fp}_{2\perp}$, filled symbols)
 for $R_\lambda=597$. All quantities are normalized by the relative velocity
 variances of $St=0$ particles.}
 \label{fig:delu_particle_fluid}
\end{figure}

To test this expectation, we compare the particle relative velocity variances to those of 
the fluid sampled by the particles in figure~\ref{fig:delu_particle_fluid}.
In all cases, the velocity variances are normalized by those of $St=0$ particles.
At $St=0.05$ and $St = 0.1$, the effect of preferential sampling is dominant at all separations,
as evidenced by the fact that $S^{fp}_{2\parallel}$ and $S^{fp}_{2\perp}$
are close to $S^p_{2\parallel}$ and $S^p_{2\perp}$, respectively.
We note that for small $St$ and small $r/\eta$, preferential sampling leads to an increase in
$S^{fp}_{2\parallel}$ with increasing $St$ and to a decrease in $S^{fp}_{2\perp}$ with increasing $St$.
This is consistent with the trends observed in figure~\ref{fig:wr_lom} and
with our argument (\textsection \ref{sec:topology}) that inertia causes particles to be ejected
from vortex tubes. We expect that two particles which are rotating in a vortex tube
will experience small (large) relative velocities parallel (perpendicular) to the particle separation vector,
and that the parallel (perpendicular) relative velocities will increase (decrease)
as particles are ejected from a vortex tube.

For $St \gtrsim 0.2$, the particle relative velocities 
are much larger than the underlying fluid velocity differences at small separations.
This difference is due to path-history effects \citep[see][]{bragg14,bragg14b}.
That is, as inertial particles approach each other,
they retain a memory of more energetic turbulence scales along their path histories,
leading to relative velocities that exceed the local fluid velocity difference.
These path-history effects imply
that inertial particles can come together from different regions in the flow,
occupy the same position in the flow at the same time, and yet have different
velocities due to their differing path histories.
This effect is referred to as `caustics,' `crossing trajectories,' or `the sling effect,' 
causes a departure from $r^2$-scaling in the second-order structure functions
at small separations,
and can lead to large relative velocities
\citep{yudine59,falkovich02,wilkinson05,wilkinson06,falkovich07}.
(Also note that while caustics are instantaneous events, 
the statistical manifestation of caustics is known as
`random, uncorrelated motion' and is discussed in \cite{ijzermans10}.)
Since the timescale over which the particles retain a memory of their interactions with turbulence 
increases with increasing inertia, caustics become more prevalent as $St$ increases.

One effect of caustics is to make the parallel and perpendicular relative
velocity components nearly the same in the dissipation range, 
as can be seen in figure~\ref{fig:wr_lom} for $St \gtrsim 0.3$. (Note that fluid particles
do not experience caustics and have $2S^p_{2\parallel} = S^p_{2\perp}$ for $r/\eta \ll 1$
as a result of continuity \citep[e.g., see][]{pope}.)
For $St \geq 10$, the relative velocities are almost 
unaffected by the underlying turbulence in the dissipation range.
As a result, the relative velocities are nearly independent of $r/\eta$ in this range.

\begin{figure}
 \centering
 \includegraphics[width=2.6in]{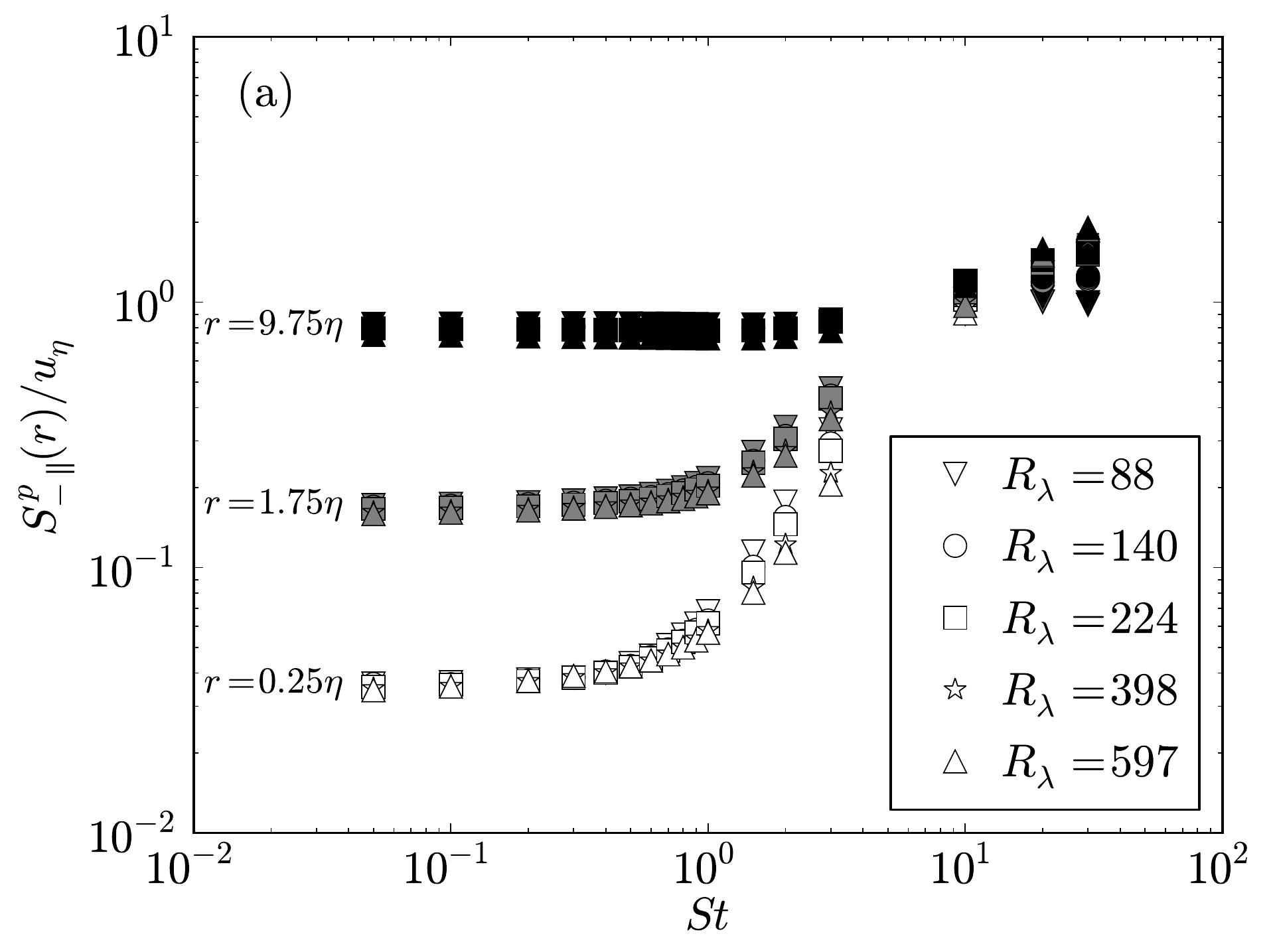}
 \includegraphics[width=2.6in]{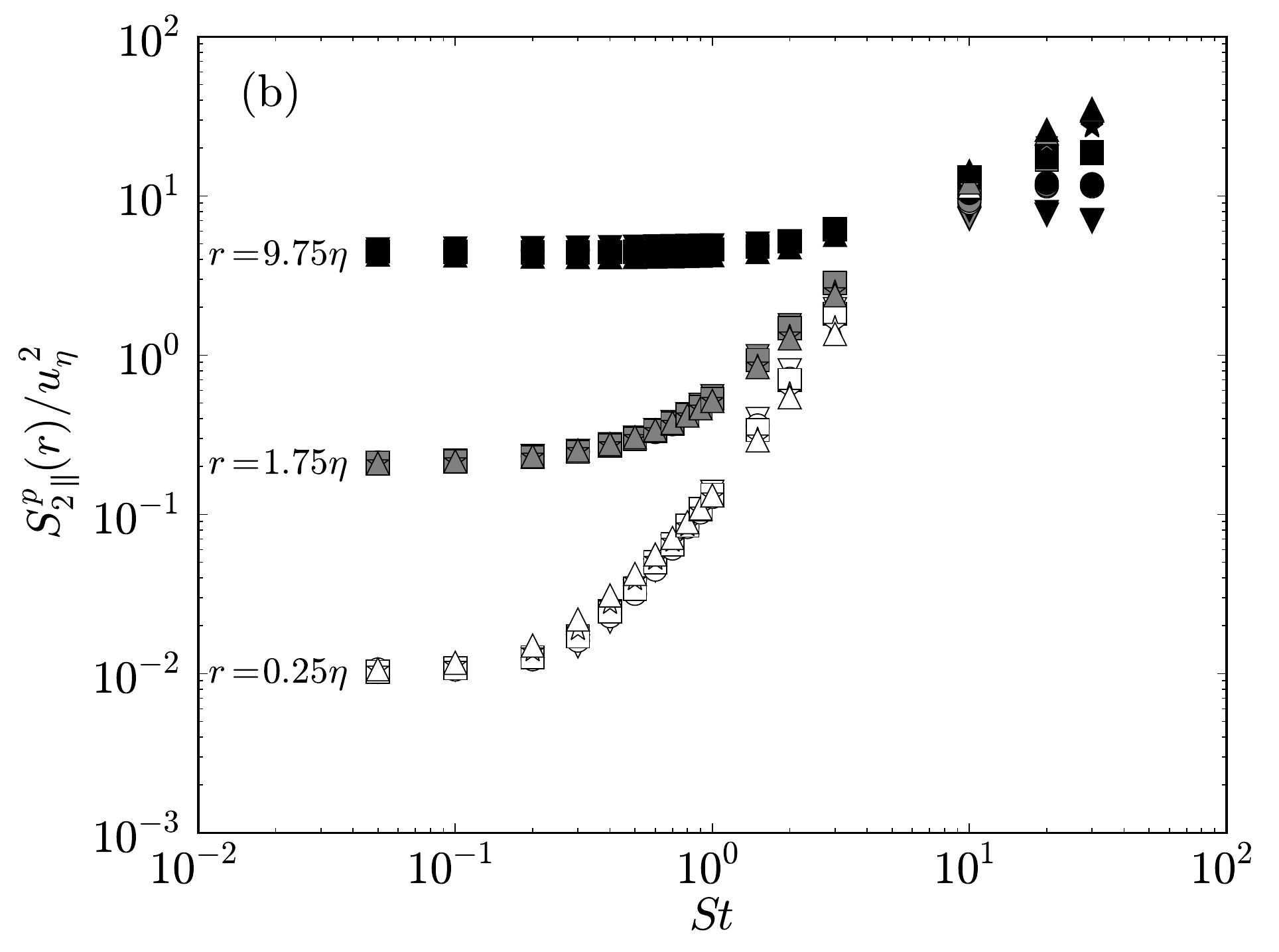}
 \includegraphics[width=2.6in]{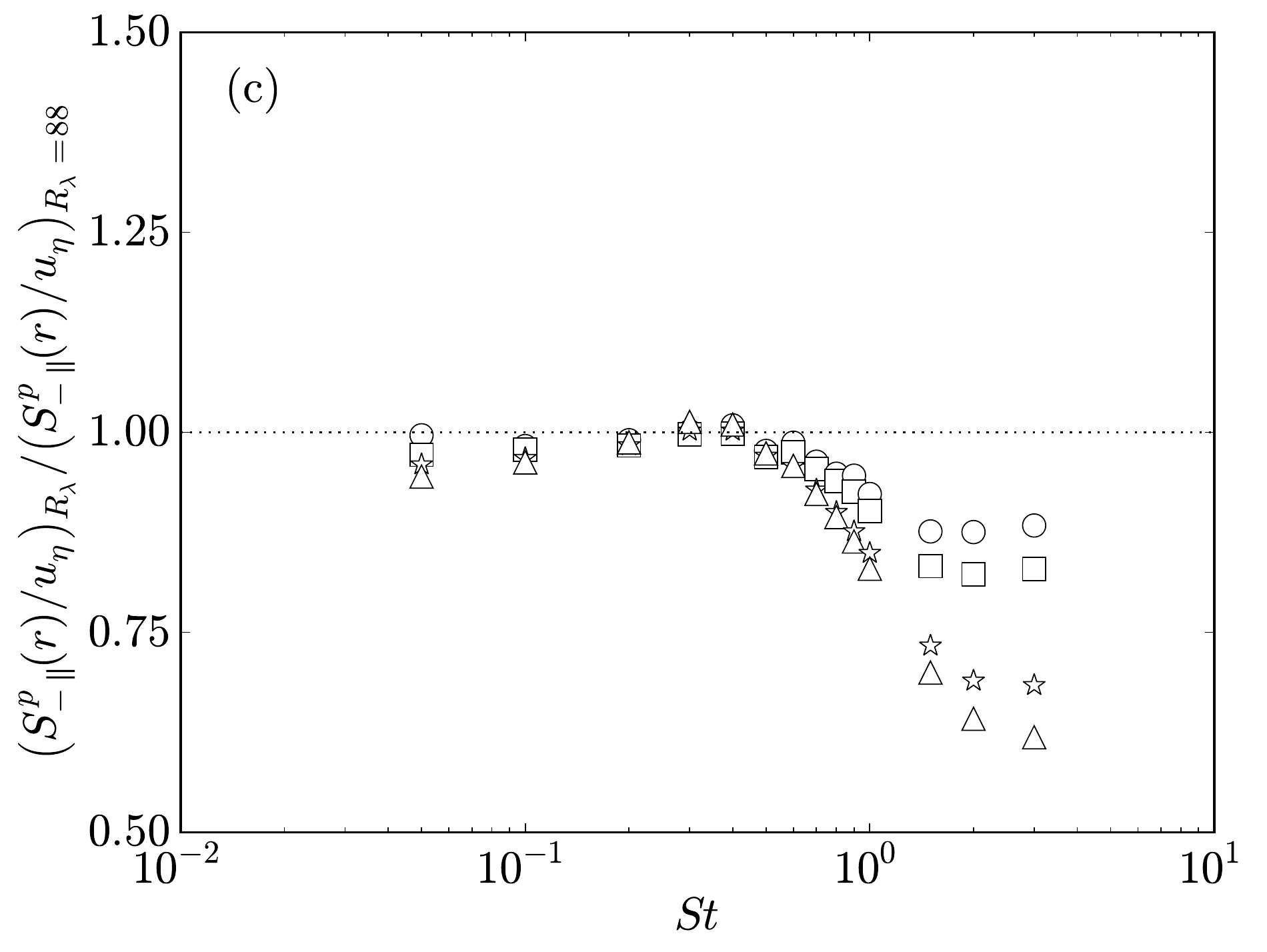}
 \includegraphics[width=2.6in]{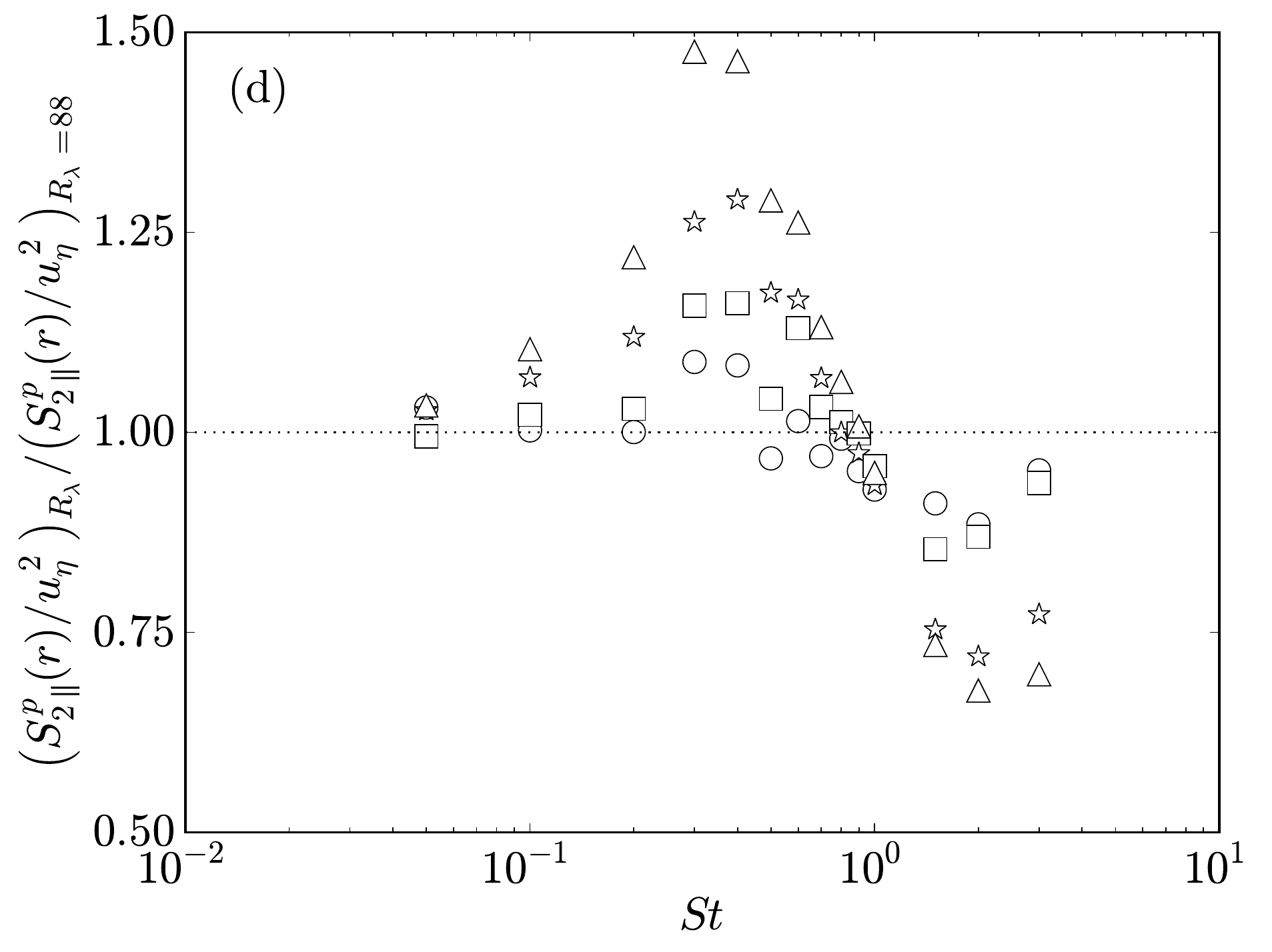}
 \caption{(a) The mean inward relative velocities
 and (b) the relative velocity variances, plotted as a function
 of $St$ for small separations and different values of $R_\lambda$.
 Open symbols denote $r = 0.25\eta$, gray filled symbols
 denote $r = 1.75\eta$, and black filled symbols denote $r = 9.75\eta$.
 To emphasize any Reynolds-number dependencies for $St \leq 3$, we also
 plot (c,d) the ratio between the value of these quantities at a given Reynolds number
 to their value at $R_\lambda = 88$ at separation $r=0.25 \eta$.}
 \label{fig:wr_lom_re}
\end{figure}

The effect of caustics can also be clearly seen in figure~\ref{fig:wr_lom_re}(a,b),
where we plot the parallel relative velocities at a given separation as a function of $St$.
From this figure, it is evident that the particle
relative velocities at the smallest separation sharply increase as $St$ exceeds about 0.2.
The rapid increase in the particle relative velocities with $St$
is consistent with the notion that caustics take an activated form \citep{wilkinson06}
and that they are negligible below a critical value of $St$ \citep{salazar12a,ijzermans10}.
Our data suggest a critical Stokes number for caustics of about $0.2$ to $0.3$,
in agreement with \cite{falkovich07} and \cite{salazar12a}.
The increase in the relative velocities occurs at higher values of $St$ as the separation increases.
In this case, the particles are subjected to larger-scale turbulence,
and hence the particles must have more inertia for their motion to deviate 
significantly from that of the underlying flow.

We now examine the Reynolds-number dependence of the relative velocities, 
restricting our attention to the component parallel to the separation vector.
The relative velocities of the largest particles ($St \gtrsim 10$)
increase strongly with increasing $R_\lambda$ in figure~\ref{fig:wr_lom_re}(a,b).
There are two reasons for this trend. The first is that the effect
of filtering on the larger turbulence scales decreases 
as $R_\lambda$ is increased (see \textsection \ref{sec:particle_kinetic_energy}).
The second is that $u^\prime/u_\eta$ increases with increasing $R_\lambda$, 
indicating that large-$St$ particles in the dissipation range 
carry a memory of increasingly energetic turbulence 
(relative to the Kolmogorov scales) in their path history as $R_\lambda$ is increased.

For smaller values of $St$ ($St \leq 3$), the relative velocities in figure~\ref{fig:wr_lom_re}(a,b)
are only weakly dependent on $R_\lambda$, 
in agreement with previous DNS studies \citep{wwz00,bec10a,rosa13,onishi13,onishi14} and the model of \cite{pan10}.
To highlight any small Reynolds-number effects in this range, we therefore
divide the relative velocities at $r=0.25\eta$ and a certain $R_\lambda$ by their
value at $R_\lambda = 88$ and plot the results in figure~\ref{fig:wr_lom_re}(c,d).

For $St \lesssim 1$, the relative velocity variances
increase weakly with increasing $R_\lambda$ (figure~\ref{fig:wr_lom_re}(d)).
As $R_\lambda$ increases, the range of velocity scales $u'/u_\eta$ increases,
allowing some particle pairs in this Stokes-number range to sample more energetic
turbulence as they converge to small separations, and causing the the relative
velocity variances to increase with increasing Reynolds number.
Furthermore, turbulence intermittency, which also increases with increasing Reynolds number,
may also contribute to the trend in the relative velocity statistics.
We note that the mean inward velocities (figure~\ref{fig:wr_lom_re}(c))
are less affected by changes in Reynolds number, presumably
because the mean inward velocity is a lower-order statistic that is less
influenced by the relatively rare events described above.

For $1 \lesssim St \lesssim 3$, we also expect 
the increased scale separation, the increased intermittency of the turbulence, or both
to act to increase
the relative velocities. However, we observe an overall \textit{decrease} in the relative
velocities with increasing $R_\lambda$ here, in agreement
with \cite{bec10a,rosa13}. These reduced relative velocities
are likely linked to the decrease in the Lagrangian rotation timescales $T^p_{\mathcal{R} \mathcal{R}}/\tau_\eta$ 
with increasing $R_\lambda$ observed in \textsection \ref{sec:topology}. That is, as 
$T^p_{\mathcal{R} \mathcal{R}}/\tau_\eta$ decreases with increasing $R_\lambda$, the 
particles have a shorter memory of fluid velocity differences along their path histories,
which in turn causes the relative velocities to decrease.

We now examine the behavior of the scaling exponents
of $S^p_{-\parallel} \propto r^{\zeta_{\parallel}^-}$ and $S^p_{2\parallel} \propto r^{\zeta_{\parallel}^2}$
at small separations. (These scaling exponents will also be used in
\textsection \ref{sec:particle_clustering} to understand and predict the trends in the particle clustering.)
We compute $\zeta_{\parallel}^-$ and $\zeta_{\parallel}^2$ using a linear least-squares
regression for $0.75 \leq r/\eta \leq 2.75$ at different values of $St$ and $R_\lambda$.
Note that while using such a large range of $r/\eta$ will necessarily introduce
finite-separation effects, there is generally too much noise in the data to accurately
compute the scaling exponents over smaller separations.

\begin{figure}
 \centering
 \includegraphics[width=2.6in]{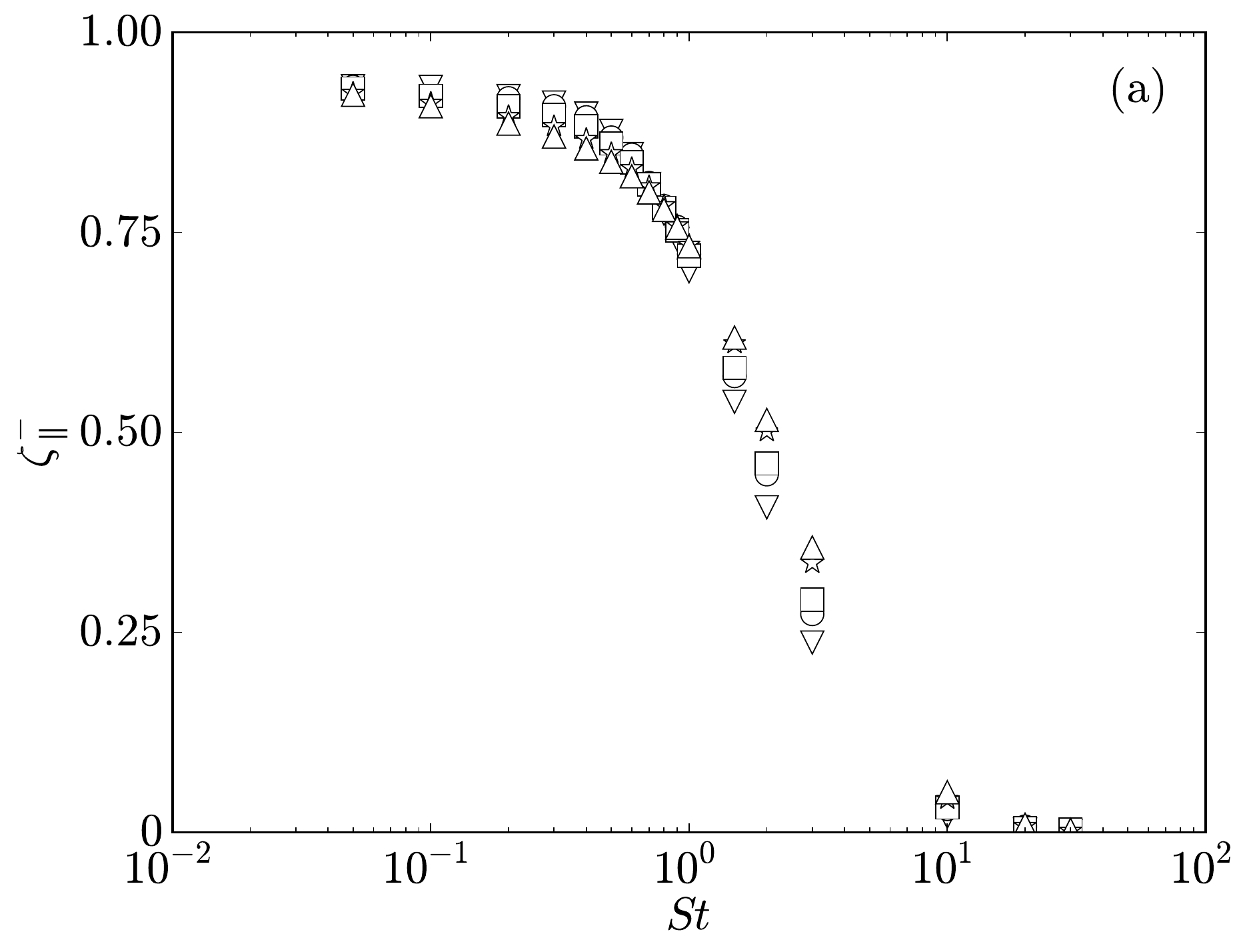}
 \includegraphics[width=2.6in]{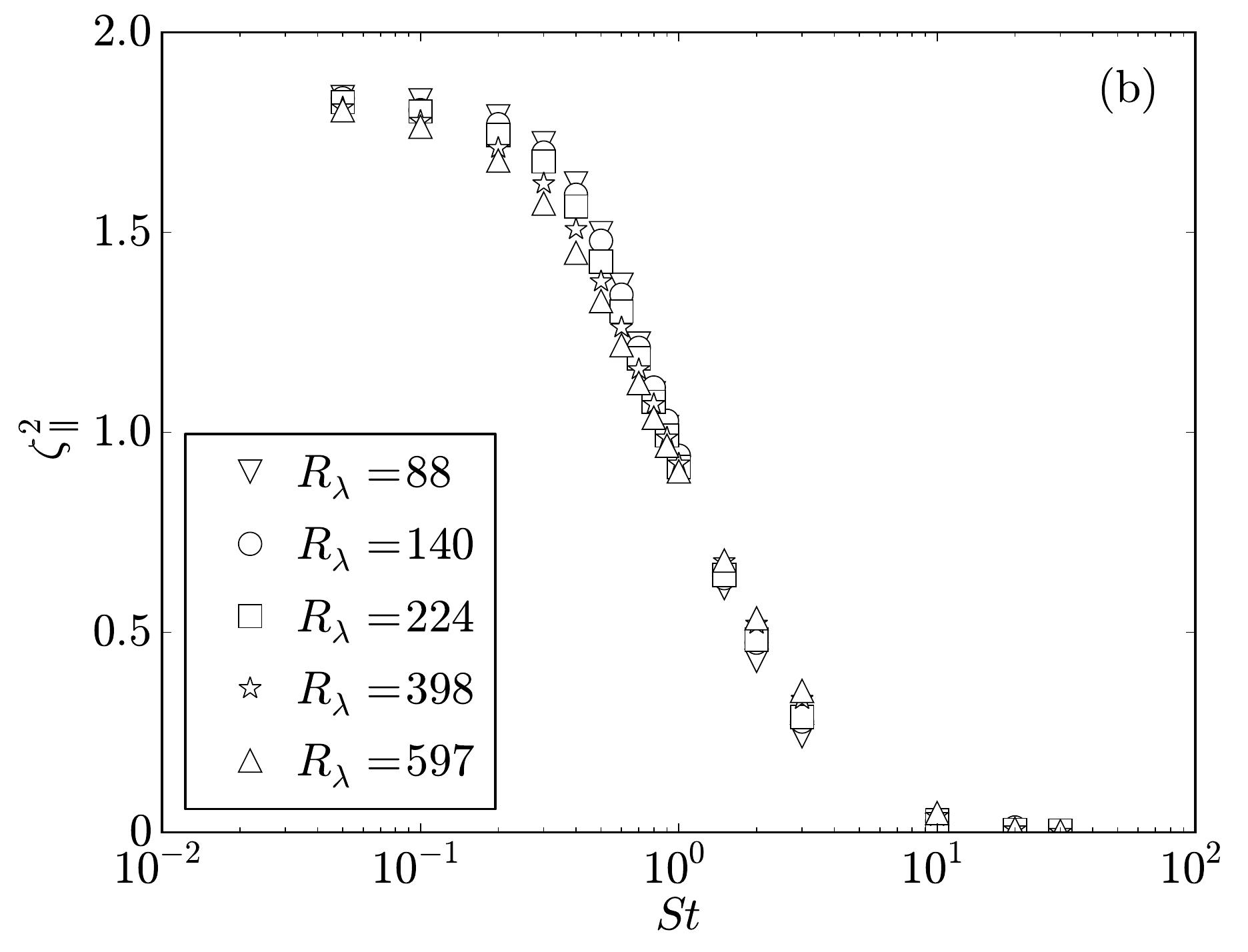}
  \caption{Dissipation-range scaling exponents for $S^p_{-\parallel}$ (a) 
  and $S^p_{2\parallel}$ for various values of $St$ and $R_\lambda$.
  The exponents are computed from linear least-squares regression for $0.75 \leq r/\eta \leq 2.75$.}
 \label{fig:sf_scaling_dissipation}
\end{figure}

The scaling exponents are plotted in figure~\ref{fig:sf_scaling_dissipation}.
We note that the scaling exponents are below those predicted by \cite{kolmogorov41a} (hereafter `K41') for fluid ($St=0$) particles
($\zeta^-_{\parallel} = 1$ and $\zeta^2_{\parallel} = 2$)
and, like the relative velocities themselves, vary only slightly as $R_\lambda$ changes.

For $St \geq 10$, the scaling exponents are about zero, indicating that the 
relative velocities are generally independent of $r$, as explained above.
The scaling exponents for $1 \lesssim St \lesssim 3$ generally increase with increasing $R_\lambda$,
since path-history interactions (which generally decrease the scaling exponents)
become less important, as explained above.
Finally, we note that $\zeta^2_\parallel$ decreases with increasing $R_\lambda$ for $St \lesssim 1$, since
intermittent path-history effects are expected to be more important here.

\begin{figure}
 \centering
 \includegraphics[width=2.6in]{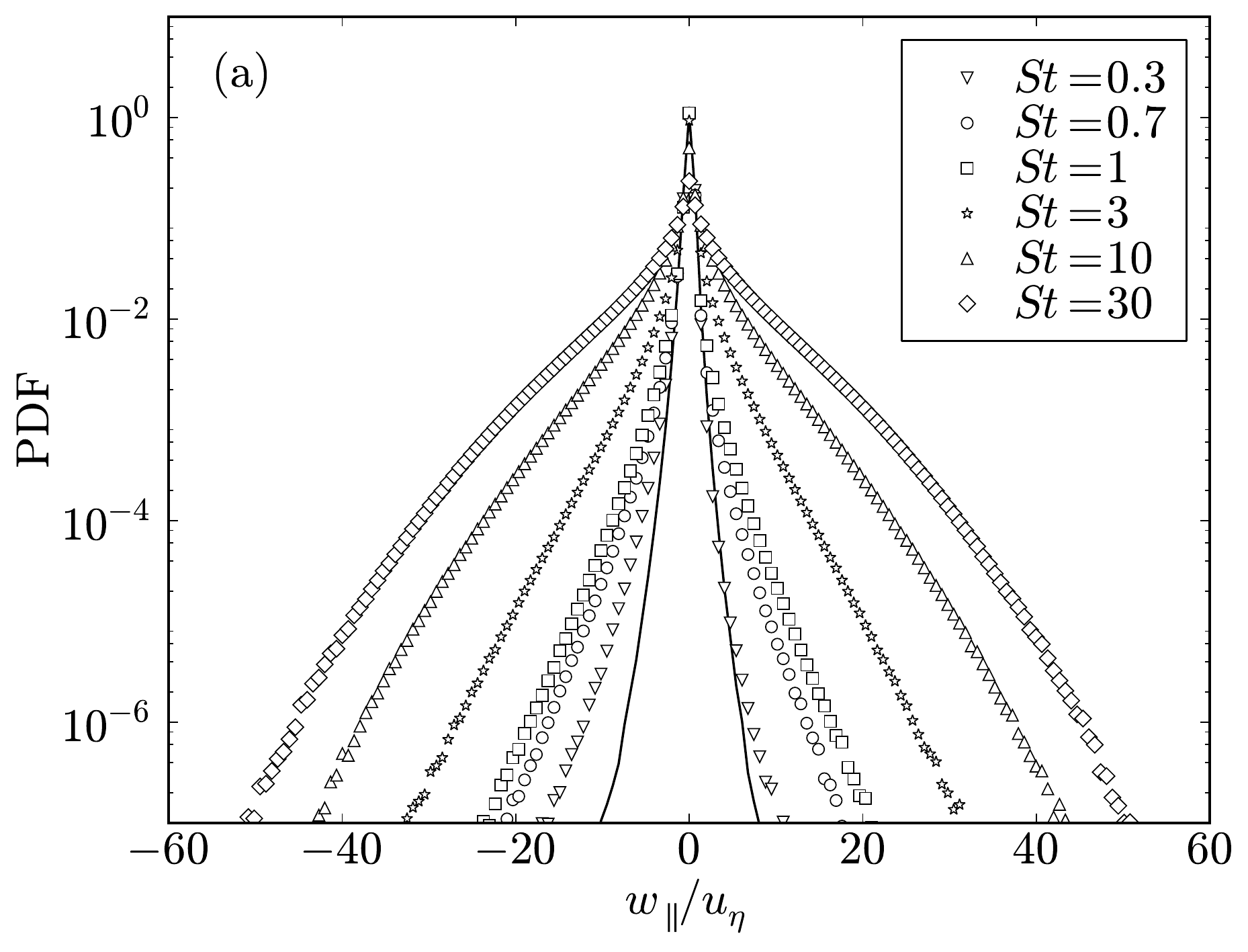}
 \includegraphics[width=2.6in]{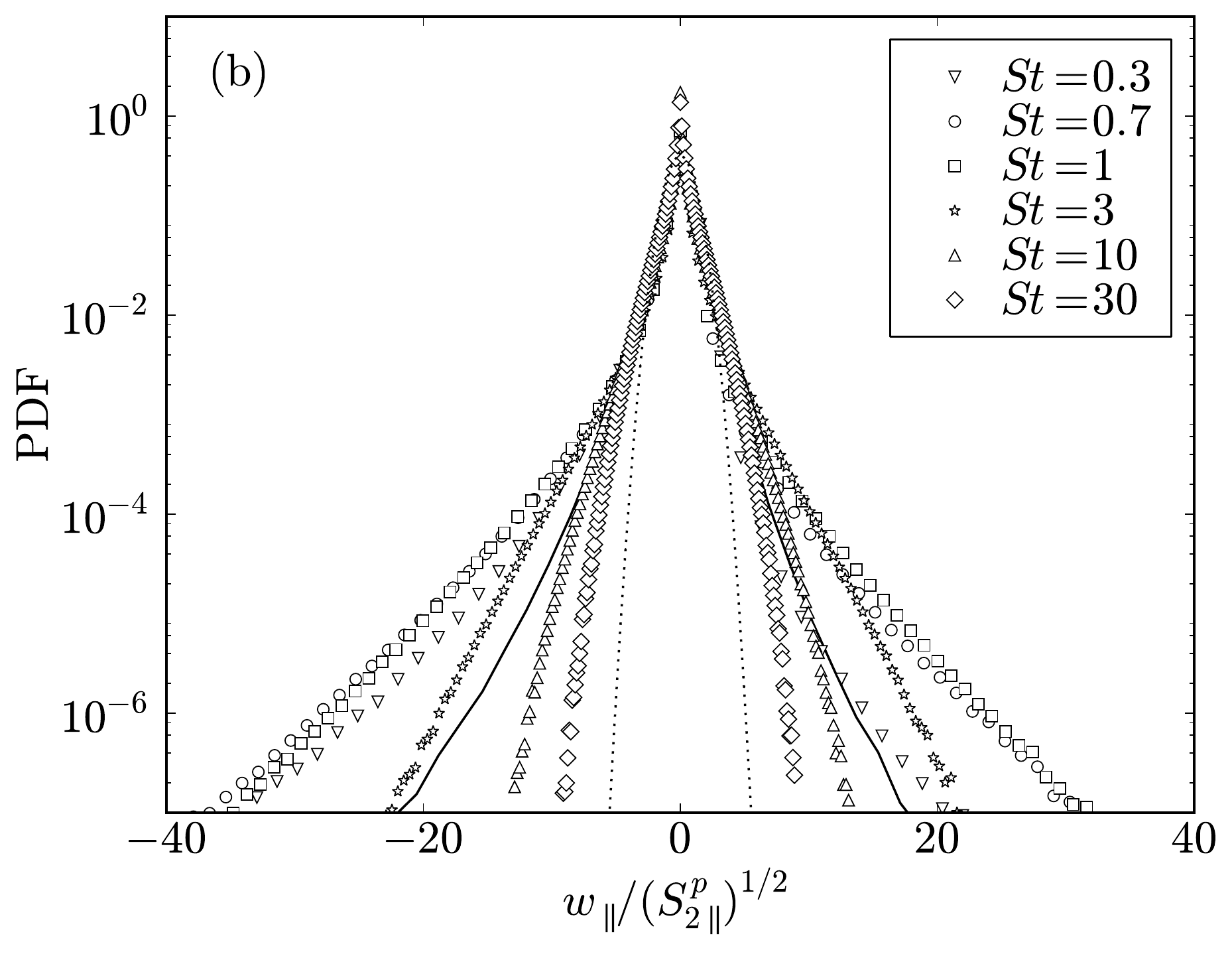}
  \caption{PDFs of the particle relative velocities $w^p_\parallel$ 
  for separations $0 \leq r/\eta \leq 2$ and $R_\lambda = 597$.
  The relative velocities are normalized by both $u_\eta$ (a) and 
  $(S^p_{2\parallel})^{1/2}$ (b). The solid lines denote the relative velocity PDFs for $St=0$ particles,
  and the dotted line in (b) indicates a standard normal distribution.}
 \label{fig:wr_diss_pdf}
\end{figure}

We next consider the PDFs of the relative velocities in the dissipation range.
Figure~\ref{fig:wr_diss_pdf} shows the PDFs for $0 \leq r/\eta \leq 2$ and $R_\lambda = 597$.
In figure~\ref{fig:wr_diss_pdf}(a), we see that as $St$ increases, 
the tails of the PDF of $w^p_\parallel/u_\eta$ become more pronounced,
indicating that larger relative velocities become more frequent, in agreement with our observations above.

We show PDFs in standardized form in figure~\ref{fig:wr_diss_pdf}(b)
to analyze the extent to which they deviate from that of a Gaussian distribution.
It is evident that the degree of non-Gaussianity peaks for $St \sim 1$
and becomes smaller as $St$ increases.
The physical explanation for this intermittency at $St \sim 1$
is that the motion of these particles is affected by both the small-scale underlying turbulence
and by the particles' memory of large-scale turbulent events in their path histories.
This combination of contributions from both large- and small-scale events leads to strong intermittency.
We also see that the underlying fluid is itself quite intermittent at this small separation,
as expected \citep[e.g., see][]{gotoh02}.

We now use three statistical measures to quantify the shape of the PDFs.
The first is the ratio between the mean inward relative velocities and the standard deviation of the relative
velocities, $S^p_{-\parallel} / (S^p_{2\parallel})^{1/2}$;
the second is the skewness of the relative velocities, 
$S^p_{3\parallel} / (S^p_{2\parallel})^{3/2}$;
and the third is the kurtosis of the relative velocities, 
$S^p_{4\parallel} / (S^p_{2\parallel})^{2}$.
(Due to insufficient statistics, we will not consider data from these latter two
quantities for $r/\eta < 1.75$.)

\begin{figure}
 \centering
 \includegraphics[height=2.5in]{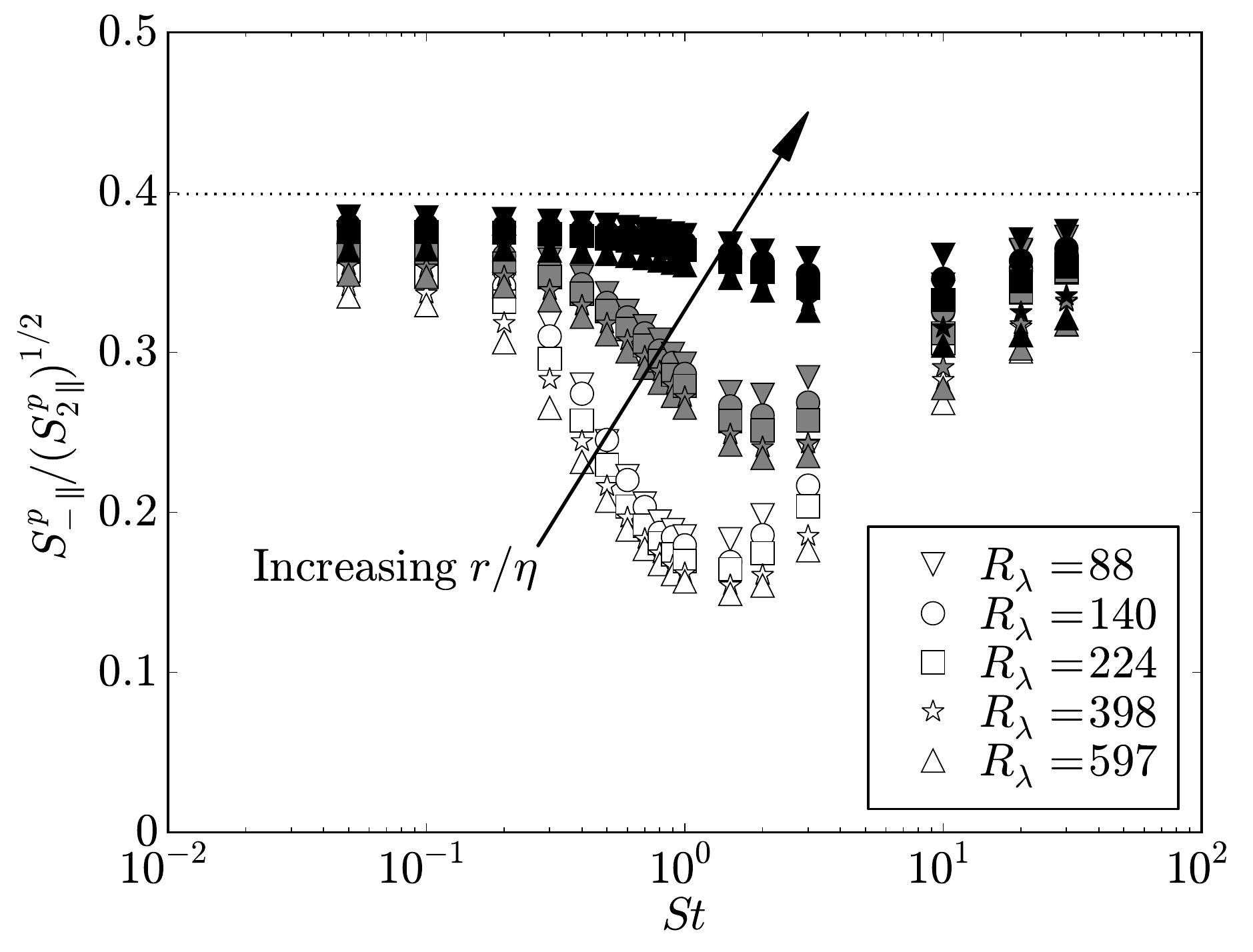}
  \caption{The ratio between mean inward relative velocities 
 and the standard deviation of the relative velocities as a function
 of $St$ for small separations and different values of $R_\lambda$.
 Open symbols denote $r = 0.25\eta$, gray filled symbols
 denote $r = 1.75\eta$, and black filled symbols denote $r = 9.75\eta$.
 The horizontal dotted line indicates that value of this quantity for a Gaussian distribution.}
 \label{fig:wr_mean_std_ratio}
\end{figure}

We show the ratio $S^p_{-\parallel} / (S^p_{2\parallel})^{1/2}$ in figure~\ref{fig:wr_mean_std_ratio}.
One motivation for looking at this ratio is that existing theories \citep[e.g., see][]{zaichik03a,pan10} 
only predict the relative velocity variance,
and by assuming the relative velocities have a Gaussian distribution,
relate this variance to the mean inward relative velocity.
For a Gaussian distribution, this ratio is approximately 0.4. 
At all values of $St$, $R_\lambda$, and $r/\eta$, our data indicate that the ratio is
below 0.4 and thus that the particle relative velocities are intermittent \citep[see also][]{wwz00,pan13}.
The degree of intermittency peaks for order unity $St$, high $R_\lambda$, and small $r/\eta$, 
and using a Gaussian prediction in this regime would lead to predictions
of the mean inward velocity which are in error by more than a factor of 2.

\begin{figure}
 \centering
 \includegraphics[width=2.6in]{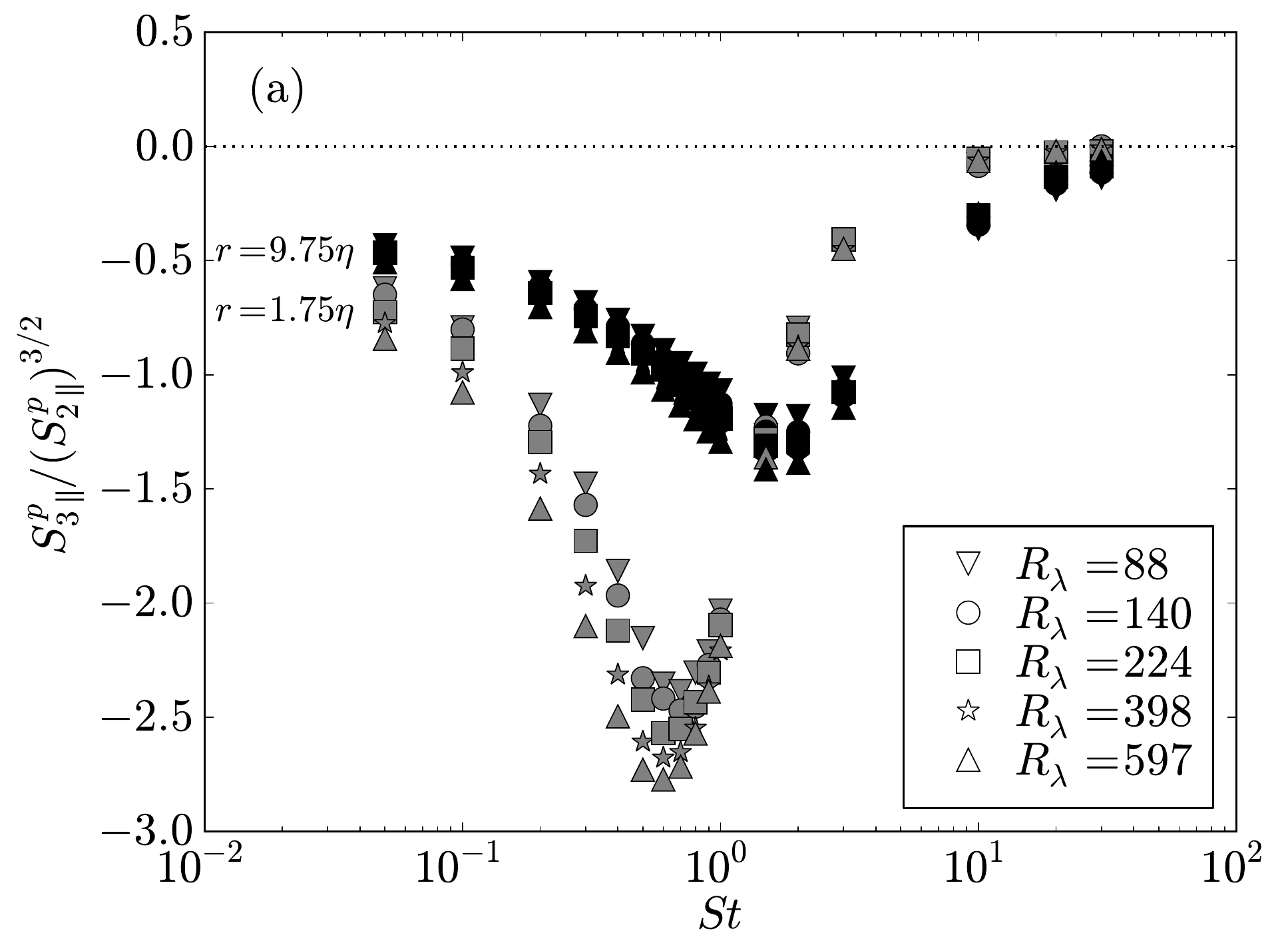}
 \includegraphics[width=2.6in]{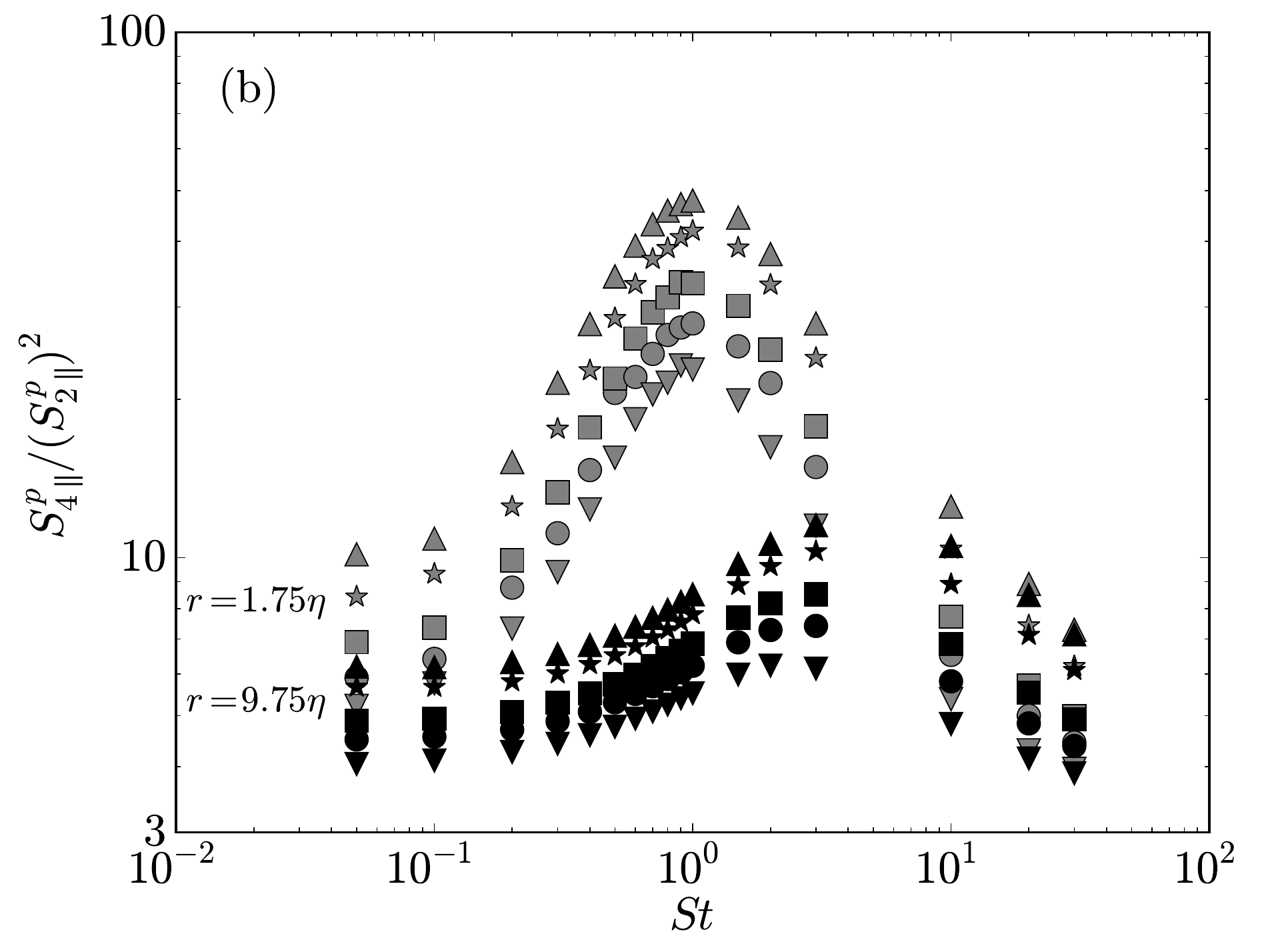}
  \caption{The (a) skewness and (b) kurtosis of the relative velocities
  as a function of $St$ for separations in the dissipation range and different values of $R_\lambda$.
 Gray filled symbols denote $r = 1.75\eta$, and black filled symbols denote $r = 9.75\eta$.}
 \label{fig:vel_skew_kurt}
\end{figure}

We next consider the skewness, $S^p_{3\parallel} / (S^p_{2\parallel})^{3/2}$,
to provide information about the asymmetry of the relative velocities.
Figure~\ref{fig:vel_skew_kurt}(a) indicates that the relative velocities are negatively
skewed \citep{wwz00,ray11}. This skewness is a result of two contributions.
First, the velocity derivatives of the underlying turbulence are negatively skewed,
a consequence of the energy cascade \citep{tavoularis78}.
Second, additional skewness arises from the path-history effect described earlier \citep[see also][]{bragg14b}.  
Figure~\ref{fig:vel_skew_kurt}(a) shows by implication that at $St\sim1$ it is the latter effect 
that dominates the skewness behavior.  At even larger values of $St$, the effect of both mechanisms 
decreases because, with increasing Stokes number, 
the particle velocity dynamics become increasingly decoupled from the small-scale fluid velocity 
field and their motion becomes increasingly ballistic in the dissipation range.

Finally, we consider the kurtosis of the relative velocities, $S^p_{4\parallel} / (S^p_{2\parallel})^{2}$,
in figure~\ref{fig:vel_skew_kurt}(b)
to quantify the contributions from intermittent events in the tails of the PDFs. The trends are
similar to those in $S^p_{-\parallel} / (S^p_{2\parallel})^{1/2}$, as expected,
indicating that contributions from intermittent events become strongest for intermediate $St$,
the smallest separations, and the highest Reynolds numbers. In all cases, the kurtosis is above
that for a Gaussian distribution ($S^p_{4\parallel} / (S^p_{2\parallel})^{2} = 3$).

\subsubsection{Inertial range relative velocity statistics}
\label{sec:wr_inertial}

We finally consider the inertial-range statistics of the relative velocities.
In figure~\ref{fig:wr_lom}, we see that the relative velocities
in the inertial range generally decrease with increasing $St$.
This implies that the filtering mechanism (which causes
the velocities to decrease with increasing $St$) dominates
the path-history mechanism (which causes the velocities to 
increase with increasing $St$), in contrast to their relative 
roles in the dissipation range.  The role reversal occurs because the 
path-history effect weakens as the separation is increased,
as explained in \cite{bragg14b}.

For $St \leq 10$, the relative velocity variances appear to scale with $r^{2/3}$, 
the same scaling predicted by K41 for $St=0$ particles. However, we observe that at $St = 30$, no clear
inertial-range scaling is present.
The lack of inertial scaling suggests that these particles
are affected by their memory of large-scale turbulence throughout the entire inertial range.

We now determine the scalings of the structure functions in the inertial range
for $St \leq 10$ by computing the scaling exponents $\zeta_{\parallel}^n$ and
$\zeta_{\perp}^n$.  Following convention \citep[e.g., see][]{ishihara09},
we consider the scaling exponents of the relative velocity magnitudes of $w_\parallel^p(t)$
and $w_\perp^p(t)$ here,
\begin{equation}
S^{p}_{|n|\parallel}(r) = \Big\langle \left|w^p_\parallel(t)\right|^n \Big\rangle_r
\propto r^{\zeta_{\parallel}^n}
\end{equation}
and 
\begin{equation}
S^{p}_{|n|\parallel}(r) =
\Big\langle \left|w^p_\perp(t)\right|^n \Big\rangle_r
\propto r^{\zeta_{\perp}^n} \mathrm{.}
\end{equation}

According to K41, for $\eta \ll r \ll \ell$ and $St=0$, 
$\zeta_{\parallel,\perp}^n = n/3$.
It is well-known, however, that for fluid particles,
the effect of intermittency leads to a nonlinear relationship
between $\zeta_{\parallel,\perp}^n$ and $n$ \citep[e.g., see][]{pope}.
Kolmogorov's refined similarity hypothesis \citep[][hereafter `K62']{kolmogorov62}
attempts to correct for the effect of intermittency, giving (for $St=0$)
\begin{equation}
\label{eq:refined_similarity_hypothesis}
\zeta_{\parallel,\perp}^n = \frac{n}{3} \left[1-\frac{\mu}{6} (n-3) \right] \mathrm{,}
\end{equation}
where $\mu$ is typically taken to be 0.25 \citep{pope}.

\begin{figure}
 \centering
 \includegraphics[width=2.6in]{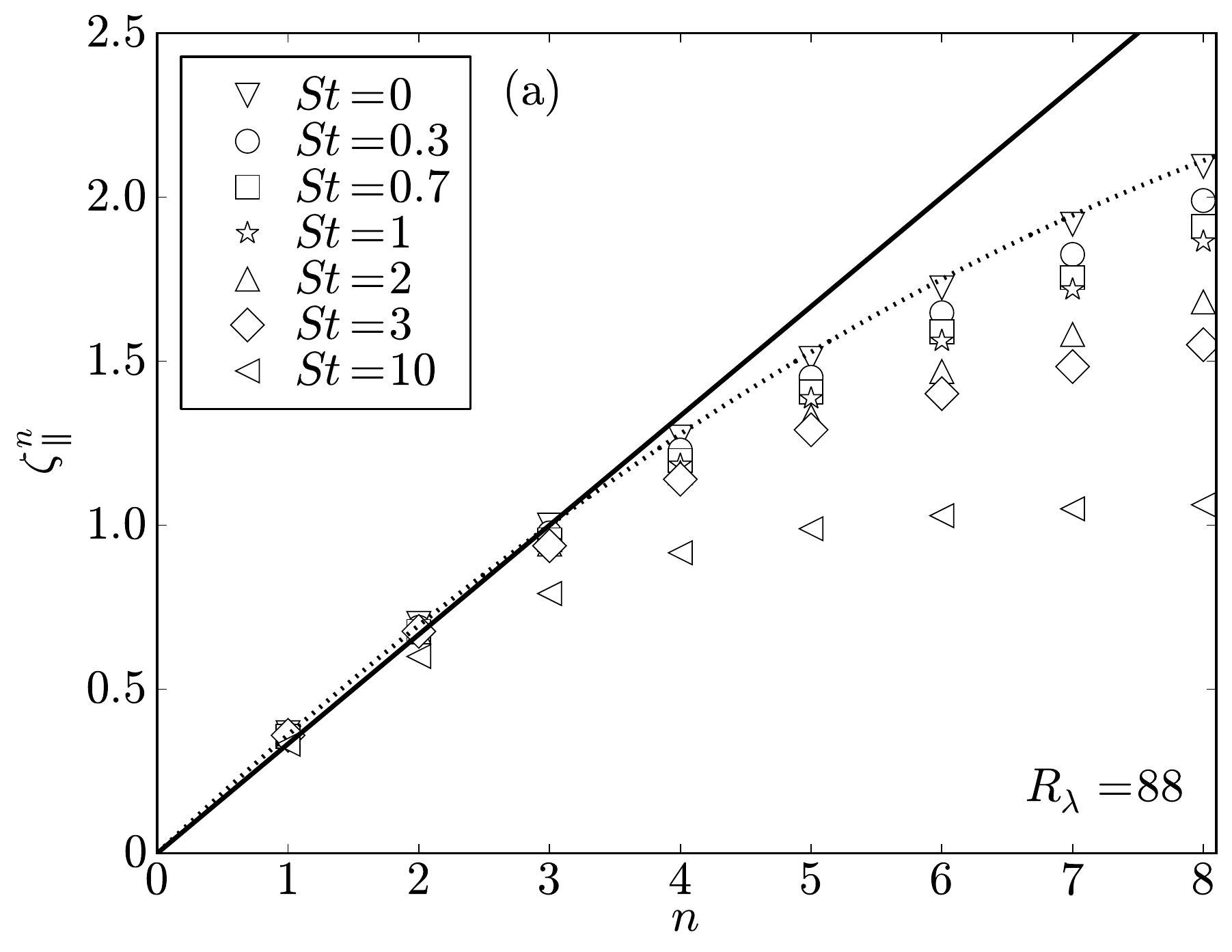}
 \includegraphics[width=2.6in]{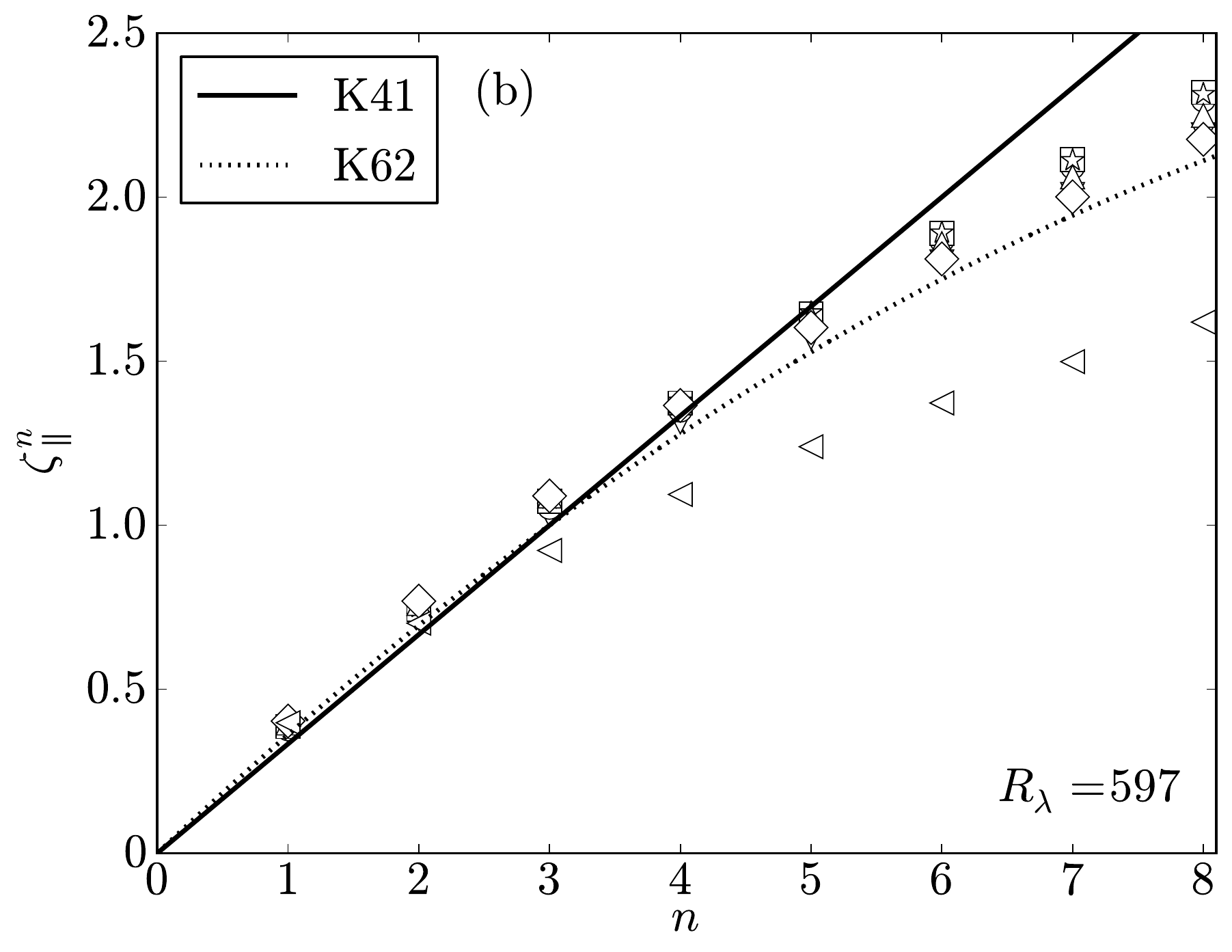}
 \includegraphics[width=2.6in]{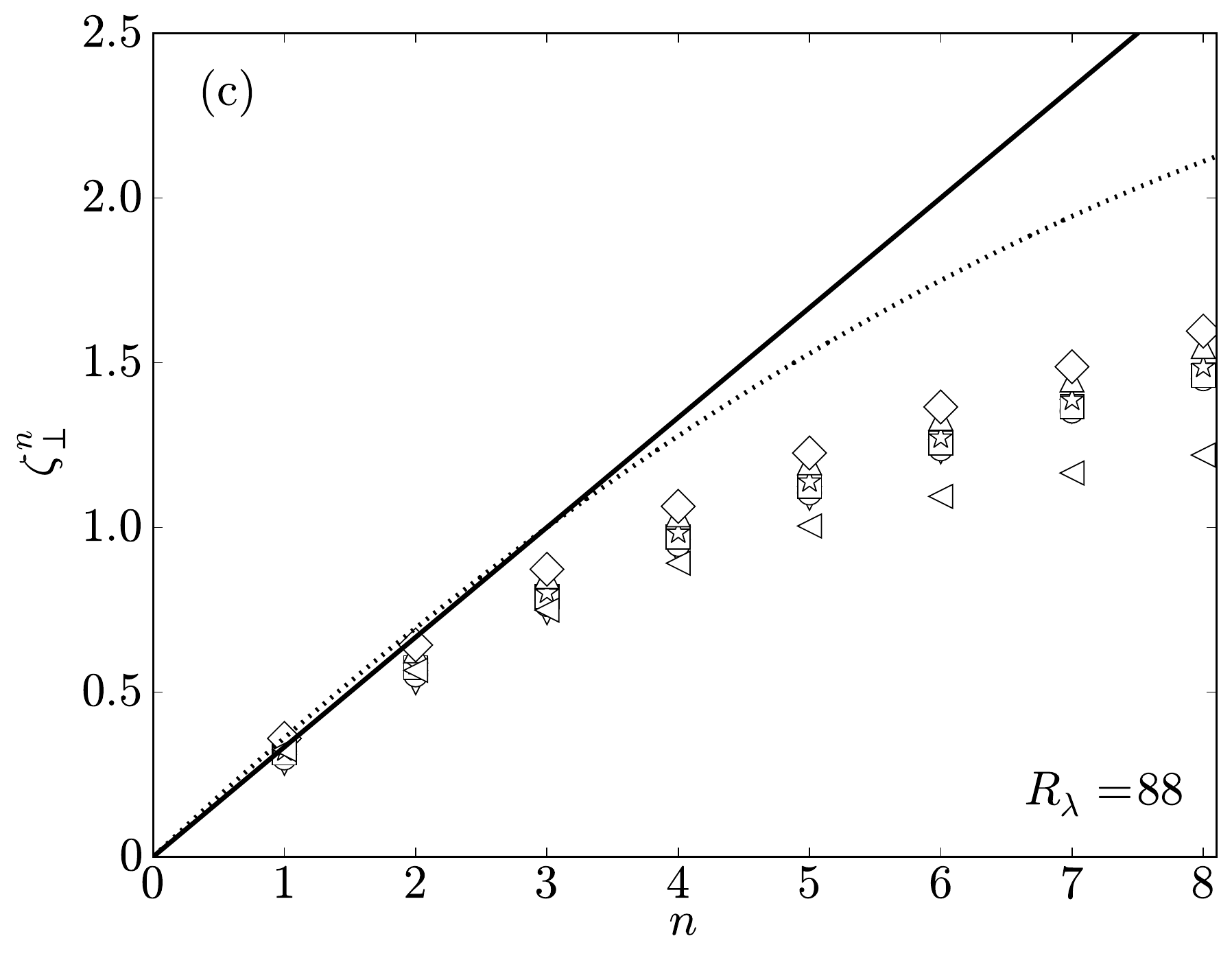}
 \includegraphics[width=2.6in]{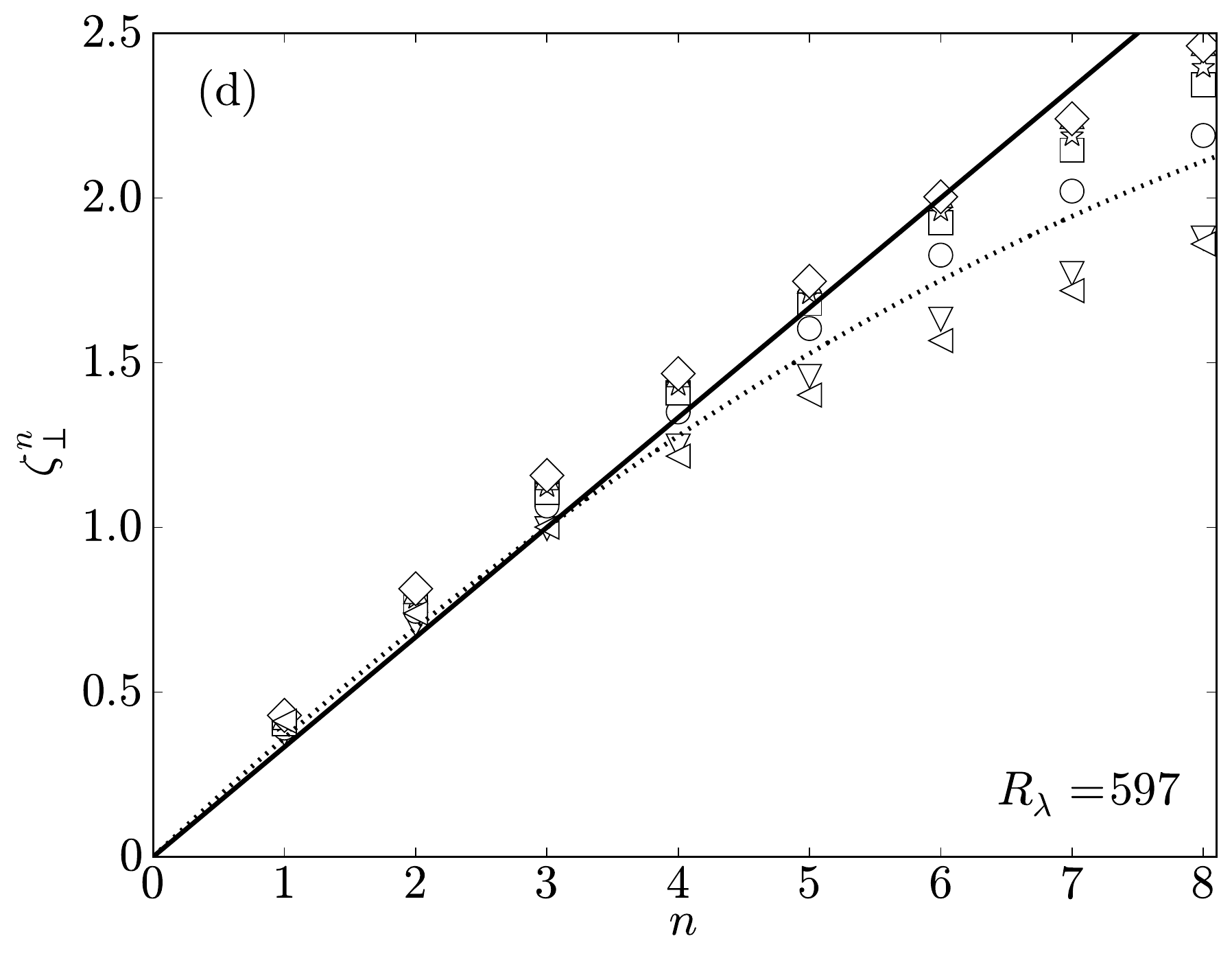}
  \caption{(a,b) Longitudinal and (c,d) transverse particle structure function 
  scaling exponents in the inertial range for various values of $St$.
  (a,c) are for $R_\lambda=88$, and (b,d) are for $R_\lambda = 597$.
  The exponents are computed from linear least-squares regression using ESS in (a,c)
  and directly in (b,d).
  The predicted scalings from from K41 and K62 
  (i.e., (\ref{eq:refined_similarity_hypothesis}) with $\mu = 0.25$) are indicated
  by the solid and dotted lines, respectively.}
 \label{fig:sf_scaling_inertial}
\end{figure}

$\zeta_{\parallel,\perp}^n$ are shown in figure~\ref{fig:sf_scaling_inertial} at $R_\lambda=88$ and $R_\lambda=597$.
For $R_\lambda = 88$, we have no clear
inertial range and therefore used extended self-similarity \citep[][hereafter `ESS']{benzi93} to
increase the scaling region for $\eta \ll r \ll \ell$.
At $R_\lambda = 597$ we have nearly a decade of inertial range scaling
($50 \lesssim r/\eta \lesssim 500$), and thus we can compute the exponents directly
over this range.

(To verify that any differences between the scaling exponents at $R_\lambda = 88$
and $R_\lambda = 597$ were in fact due to Reynolds-number effects and were not merely artifacts of ESS, 
we also computed the exponents for $R_\lambda = 597$ using ESS.
Both methods of computing the exponents (directly and with ESS) gave similar results,
with differences that were less than 8\%,
indicating the trends observed below are robust.
We also note that while the inertial scaling region varies with $St$,
we used the same fitting range for all values of $St$ for consistency.)

For $St = 0$, (\ref{eq:refined_similarity_hypothesis}) approximates
the longitudinal scaling exponents excellently for $p \leq 8$ at $R_\lambda = 88$ 
(figure~\ref{fig:sf_scaling_inertial}(a)),
while it slightly under-predicts them at $R_\lambda = 597$
(figure~\ref{fig:sf_scaling_inertial}(b)).
By comparing figure~\ref{fig:sf_scaling_inertial}(a) and figure~\ref{fig:sf_scaling_inertial}(b),
it is evident that $\zeta_\parallel^n$ increases with increasing $R_\lambda$.
For $R_\lambda = 88$, the longitudinal scaling exponents decrease monotonically with increasing $St$,
as was observed in \cite{salazar12a}.
However, for $R_\lambda = 597$, the exponents increase with $St$ up to $St \approx 1$ before
decreasing for higher values of $St$. The reason for these trends is unclear.

For most values of $St$, the transverse structure functions (figure~\ref{fig:sf_scaling_inertial}(c,d))
are more intermittent than their longitudinal counterparts
(figure~\ref{fig:sf_scaling_inertial}(a,b)), in agreement with earlier observations \citep[e.g., see][]{ishihara09}. The difference between the longitudinal and transverse
structure functions seems to decrease as $R_\lambda$ increases, however,
suggesting that it may be a low-Reynolds-number artifact \citep[see][]{kerr01,gotoh02,shen02}.

\subsection{Particle clustering}
\label{sec:particle_clustering}
As discussed in \textsection \ref{sec:introduction},
inertial particles form clusters when placed in a turbulent flow.
We first consider a theoretical framework for understanding this clustering (\textsection \ref{sec:particle_clustering_theory}),
and then analyze the clustering using DNS (\textsection \ref{sec:particle_clustering_dns}).

\subsubsection{Theoretical framework for particle clustering}
\label{sec:particle_clustering_theory}

A variety of measures have been proposed to study particle clustering,
including Vorono\"{i} diagrams \citep{monchaux10},
Lyapunov exponents \citep{bec06d}, Minkowski functionals \citep{calzavarini08},
and radial distribution functions (RDFs) \citep{mcquarrie}.
The RDF has distinct advantages over these other methods.
The RDF, unlike both Minkowski functionals \citep{calzavarini08}
and Vorono\"{i} diagrams \citep{tagawa12}, is not biased by the number of particles
simulated.  Also, as \cite{bec06d} noted,
the accurate computation of Lyapunov exponents
is numerically unfeasible for high-Reynolds-number simulations,
while computation of the RDF is relatively straightforward.
Finally, the RDF, unlike the other measures, 
has a direct relevance to particle collisions,
since it precisely corrects the collision kernel for 
particle clustering \citep{sundaram4}.

The RDF $g(r)$ is defined as the ratio of the number of particle pairs at a given separation $r$
to the expected number of particle pairs in a uniformly distributed particle field,
\begin{equation}
\label{eq:RDF}
 g(r) \equiv \frac{N_i/V_i}{N/V} \mathrm{.}
\end{equation}
Here, $N_i$ is the number of particle pairs that lie within a shell with an average radius $r$
and a radial width $\Delta r$, $V_i$ is the volume of the shell, and $N$ is the
total number of particle pairs located in the total volume $V$.
An RDF of unity corresponds to uniformly distributed particles,
while an RDF in excess of one indicates a clustered particle field.

Based on the findings of \cite{bragg14} we use the model of \cite{zaichik09} as a framework for understanding
the physical mechanisms governing particle clustering. We will validate
this model against DNS data in \textsection \ref{sec:particle_clustering_dns}.
In the following discussion, we non-dimensionalize all variables by Kolmogorov units
and use $\hat{Y}$ to denote the non-dimensionalized form of a variable $Y$.

From \cite{zaichik09}, the equation describing $g(\hat{r})$ at steady-state for an isotropic system is
\begin{equation}
\label{eq:zaichik_rdf}
 0 = -St \left( \hat{S}^p_{2\parallel} + \hat{\lambda}_\parallel \right) \nabla_{\hat{r}} g
     -St g \left(
     \nabla_{\hat{r}} \hat{S}^p_{2\parallel} + 2 \hat{r}^{-1} 
     \left[ \hat{S}^p_{2\parallel} - \hat{S}^p_{2\perp} \right] \right) \mathrm{,}
\end{equation}
where $\hat{\lambda}_\parallel$ is a diffusion coefficient describing the effect of the turbulence 
on the dispersion of the particle pairs \citep[e.g., see][]{bragg14}.  
We now consider (\ref{eq:zaichik_rdf}) in different $St$-regimes 
to consider the effect of changes in $R_\lambda$ within these regimes.

In the limit $St \ll 1$, (\ref{eq:zaichik_rdf}) can be reduced to \citep[see][]{bragg14},
\begin{equation}
\label{eq:zaichik_rdf_low_St}
 0 = -\hat{r}^2 B_{nl}
 \nabla_{\hat{r}} g 
 - \frac{St}{3} \hat{r} g \left( \langle \hat{\mathcal{S}}^2 \rangle^p - \langle \hat{\mathcal{R}}^2 \rangle^p \right) \mathrm{,}
\end{equation}
where $B_{nl}$ is a $St$-independent, non-local diffusion coefficient \citep[see][]{chun05,bragg14}.
The first term on the right-hand-side is associated with an outward
particle diffusion which reduces clustering, while the second term on the right-hand-side
is responsible for an inward particle drift which increases clustering.

We therefore see that if $B_{nl}$ is independent of $R_\lambda$, the diffusion will be independent
of $R_\lambda$. 
The drift is dependent on $\tau_\eta^2 \langle \mathcal{S}^2 \rangle^p - \tau_\eta^2 \langle \mathcal{R}^2 \rangle^p$ and 
we see from \textsection \ref{sec:topology} that
$\tau_\eta^2 \langle \mathcal{S}^2 \rangle^p - \tau_\eta^2 \langle \mathcal{R}^2 \rangle^p$ increases weakly
with $R_\lambda$ for $St \ll 1$.
We therefore expect the degree of clustering at low $St$ to increase weakly as $R_\lambda$ increases.
We will test this expectation against DNS data in \textsection \ref{sec:particle_clustering_dns}.

For particles with intermediate values of $St$, we are generally unable to simplify
(\ref{eq:zaichik_rdf}), since all terms are of comparable magnitude, and
the clustering in this range is due to both preferential sampling and path-history effects.
\cite{bragg14} showed that path-history effects induce an asymmetry in the particle inward and outward motions,
causing particles to come together more rapidly than they separate, 
generating a net inward drift and increased clustering.
The precise range of $St$ over which path-history effects increase clustering will likely vary with $R_\lambda$,
but a rough guideline \citep[based on][]{bragg14} is $0.2 \lesssim St \lesssim 0.7$.
Below this range, path-history effects have a negligible impact on particle clustering,
and above this range, the path-history mechanism acts to diminish clustering.
For the upper end of this $St$-range, path-history effects
are the dominant particle-clustering mechanism \citep{bragg14}.

We next simplify (\ref{eq:zaichik_rdf}) when $St \gtrsim 1$.
As noted in \textsection \ref{sec:wr_dissipation}, at sufficiently large $St$ and small $r/\eta$,
the relative particle velocities are dominated by path-history effects,
and $S^p_{2\parallel}\approx S^p_{2\perp}$.
Furthermore, $\lambda_\parallel\ll S^p_{2\parallel}$ in this regime \citep[see][]{bragg14b}.  
Using these results we can simplify (\ref{eq:zaichik_rdf}) in the dissipation range to the form,
\begin{equation}
\label{eq:zaichik_rdf_high_St}
 0 \approx -St \hat{S}^p_{2\parallel}\nabla_{\hat{r}} g
     -St g \nabla_{\hat{r}} \hat{S}^p_{2\parallel}
     \mathrm{.}
\end{equation}
The overall changes in the particle clustering at high $St$ will therefore be determined by 
the extent to which the drift coefficient
($\nabla_{\hat{r}} \hat{S}^p_{2\parallel}$)
and the diffusion coefficient ($\hat{S}^p_{2\parallel}$) 
are influenced by changes in $R_\lambda$.
That is, if the ratio between the drift and diffusion coefficients increases (decreases)
with increasing $R_\lambda$, the RDFs are expected to increase (decrease).

We therefore take the ratio between the drift and diffusion coefficients and obtain
\begin{equation}
\label{eq:zt_ratio}
 \frac{\nabla_{\hat{r}} \hat{S}^p_{2\parallel}}{\hat{S}^p_{2\parallel}} = \frac{\zeta^2_\parallel}{\hat{r}} \mathrm{,}
\end{equation}
where $\zeta^2_\parallel$ is the scaling exponent of the longitudinal relative velocity variance.
(\ref{eq:zt_ratio}) implies that increases (decreases) in $\zeta^2_\parallel$
are fundamentally linked to increases (decreases) in the RDFs at high $St$.
From \textsection \ref{sec:wr_dissipation}, we see that $\zeta_\parallel^2$
increases with increasing $R_\lambda$ for $1 \lesssim St \lesssim 3$, which suggests 
that $g(r/\eta)$ will increase with increasing $R_\lambda$ here. 

We also note that (\ref{eq:zaichik_rdf_high_St})
is only applicable for high-$St$ particles in the dissipation range,
and is thus unable to predict the clustering for $St > 3$ particles,
which is primarily dependent on inertial-range scales.
We will examine the RDFs for $St > 3$ from DNS data in \textsection \ref{sec:particle_clustering_dns}.

In summary, at small $St$, clustering may increase with increasing $R_\lambda$ depending 
upon whether $B_{nl}$ varies with $R_\lambda$. 
Clustering at intermediate values of $St$ will be due to both preferential sampling
and path-history effects, though it is unclear the degree to which $g(r/\eta)$ will
change with $R_\lambda$.
At high $St$, the degree of clustering is determined by the influence of path-history effects
on the scaling of the relative velocity variances,
which in turn affects the relative strengths of the drift and diffusion mechanisms.
Based on our relative velocity data in \textsection \ref{sec:relative_velocities},
we expect that clustering will increase with increasing $R_\lambda$ here. 
We next consider DNS data to test these predictions.

\subsubsection{Particle clustering results}
\label{sec:particle_clustering_dns}

\begin{figure}
 \centering
 \includegraphics[width=2.6in]{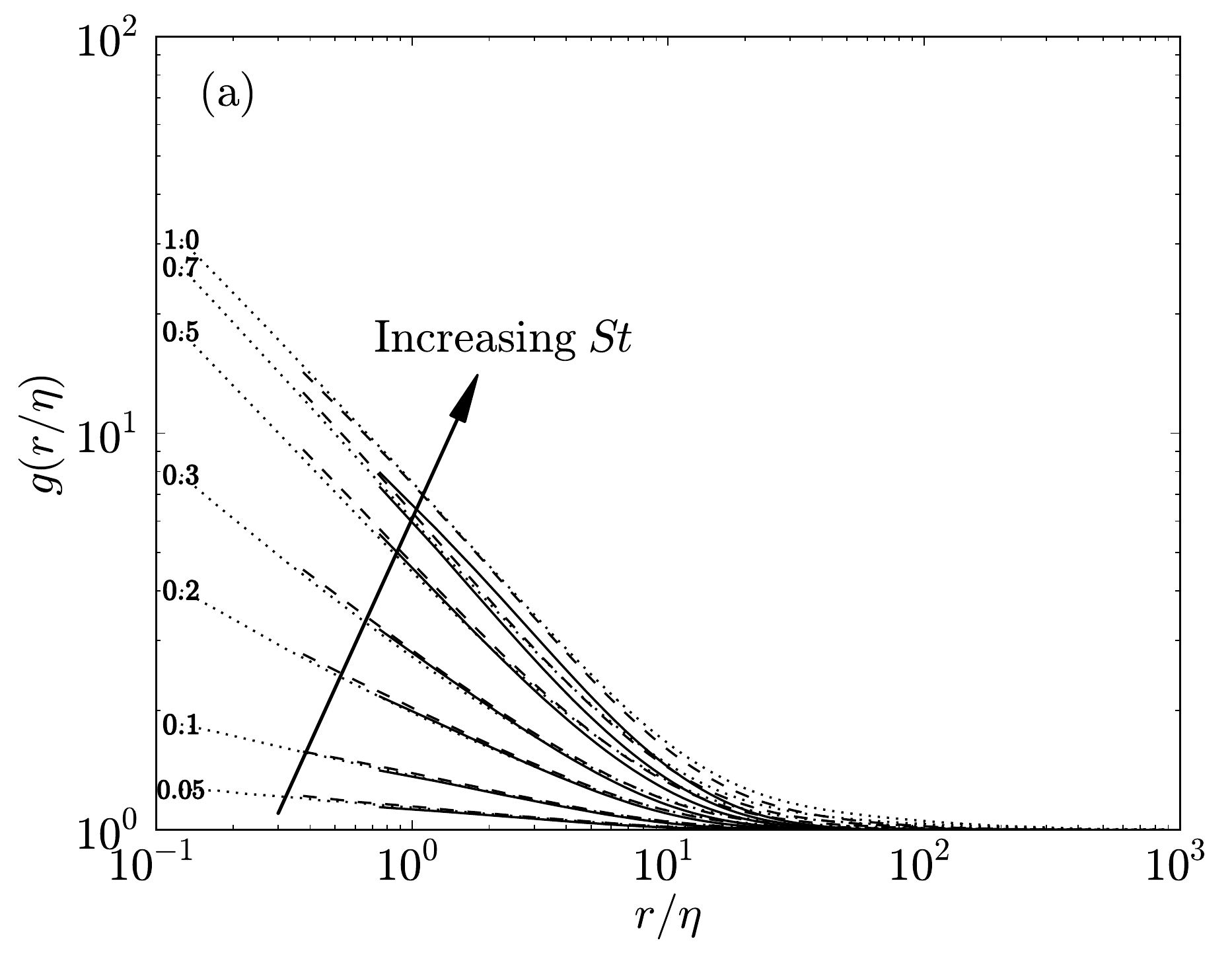}
 \includegraphics[width=2.6in]{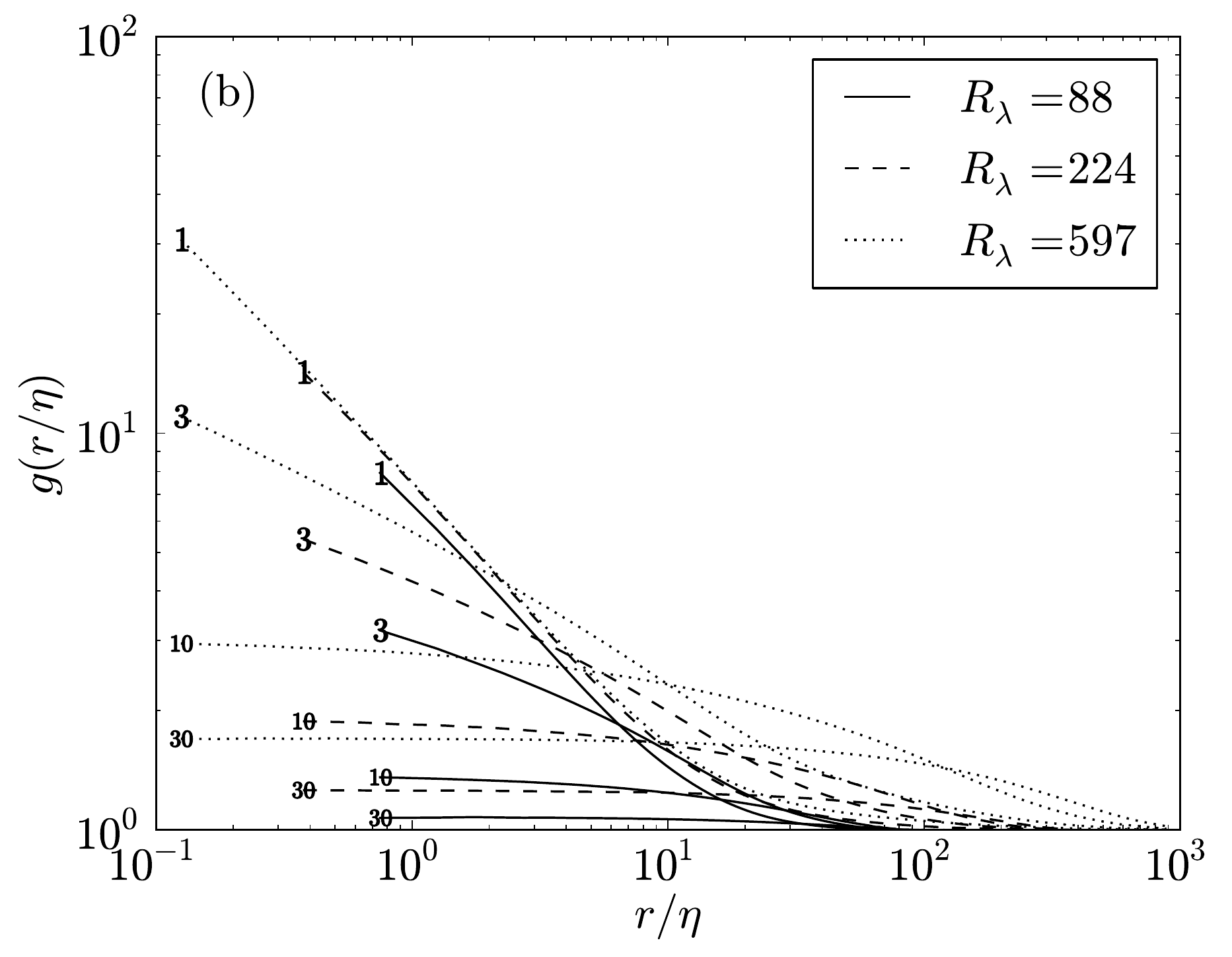}
 \caption{RDFs for (a) low-$St$ particles and (b) high-$St$ particles at three different values of $R_\lambda$,
 plotted as a function of the radial separation $r/\eta$. The Stokes numbers are indicated by the line labels.}
 \label{fig:rdfs}
\end{figure}

In figure~\ref{fig:rdfs}, we plot the RDFs for the different values of $St$ considered
at three different Reynolds numbers.  Note that as the size of the simulation (and thus $R_\lambda$)
increases, we are able to calculate $g(r/\eta)$ statistics accurately at progressively smaller values of $r/\eta$.

\begin{figure}
 \centering
  \includegraphics[height=2.5in]{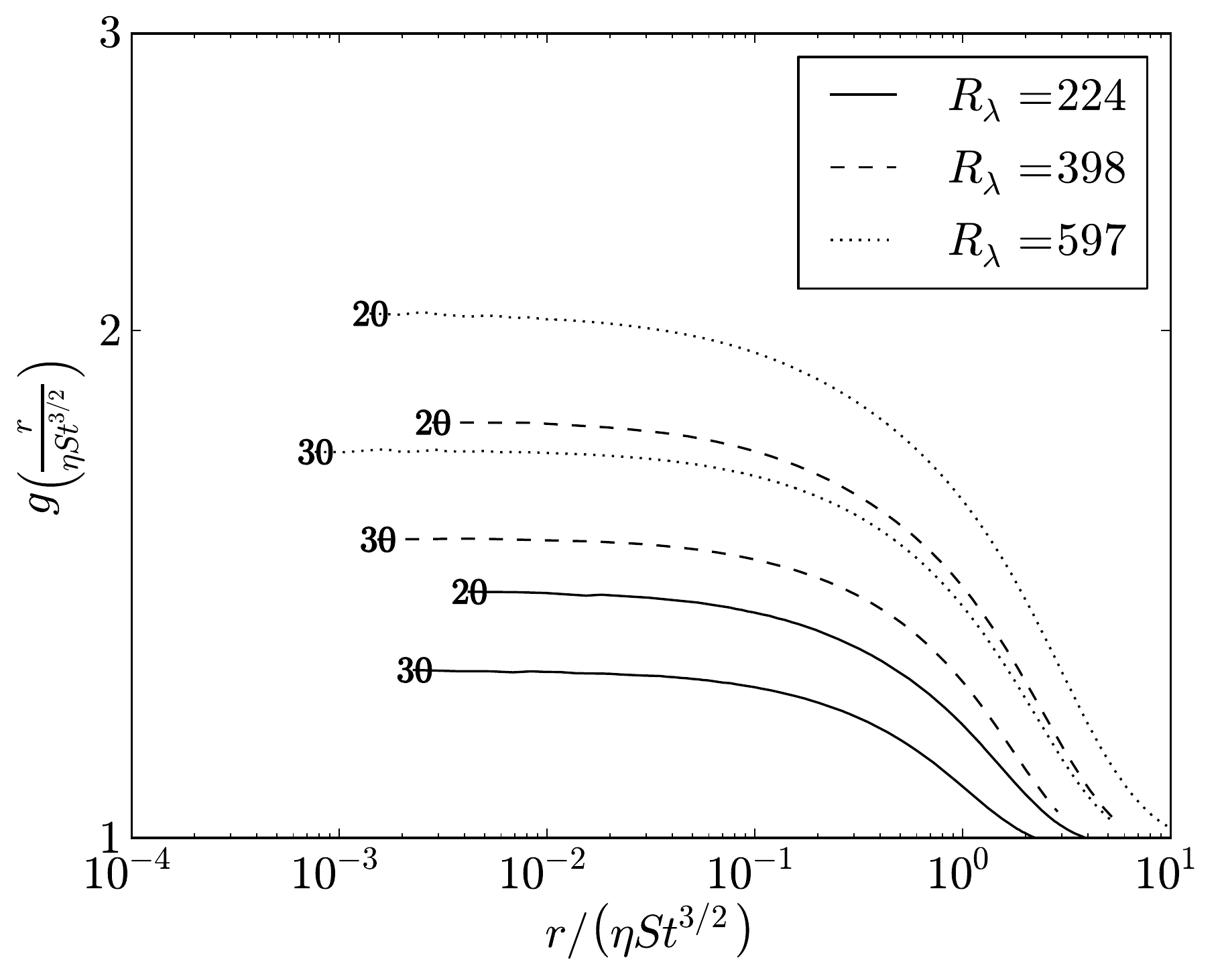}
 \caption{RDFs for $St = 20$ and $St = 30$ particles at the three highest values of $R_\lambda$.
 The separations are scaled by $\eta St^{3/2}$ to test for inertial range scaling.
 The Stokes numbers are indicated by the line labels.}
 \label{fig:rdfs_inertial_range}
\end{figure}

In agreement with past studies \citep[e.g., see][]{wang93,sundaram4,balachandar10}, 
we see that particle clustering peaks for $St \sim 1$
at all Reynolds numbers shown.  
Figure~\ref{fig:rdfs} also indicates that the largest particles ($St \geq 10$) exhibit clustering
outside of the dissipation range of turbulence, and that the degree
of clustering is independent of separation in the dissipation range.
This is because large-$St$ particles are unresponsive to the dissipative range scales 
and so move almost ballistically at these separations.
The clustering that is observed for these particles is due almost entirely to eddies
in the inertial range with timescales similar to the particle response time \citep{goto06,bec10b}.
If we make that assumption, along with the standard K41 approximations for the inertial range,
we expect the clustering will depend only on $\epsilon$ and $r$, and will occur at lengthscales
on the order of $\eta St^{3/2}$ \citep{elmaihy05,bec10b}.
We test this in figure~\ref{fig:rdfs_inertial_range} by plotting the RDFs for $St=20$ and $St=30$ particles
as a function of $r/\left( \eta St^{3/2} \right)$ at the three highest Reynolds numbers.
(The two lower Reynolds numbers do not have a well-defined inertial range,
as noted in \textsection \ref{sec:fluid_phase}, and hence the above argument would not hold.)
We see that the RDFs decrease rapidly near $r/\left( \eta St^{3/2} \right) \sim 1$,
suggesting that the particles are indeed clustering due to the influence of turbulent eddies
in the inertial range with a timescale on the order of $\tau_p$.
Refer to \cite{bragg15b} for a recent theoretical and computational analysis 
of particle clustering in the inertial range of turbulence.

We now discuss how the RDFs change with the Reynolds number.
In \textsection \ref{sec:particle_clustering_theory},
we argued that $g(r/\eta)$ might increase weakly with $R_\lambda$ for $St \ll 1$,
since $\tau_\eta^2 \langle \mathcal{S}^2 \rangle^p - \tau_\eta^2 \langle \mathcal{R}^2 \rangle^p$ 
increases with $R_\lambda$ in this limit.
In figure~\ref{fig:rdfs}(a), however, we observe that $g(r/\eta)$ is essentially 
independent of $R_\lambda$ for $St \lesssim 1$, which implies that the non-local
correction coefficient $B_{nl}$ in (\ref{eq:zaichik_rdf_low_St}) must increase weakly with $R_\lambda$
in a compensating way.
Several authors have also found the level of particle clustering to be independent of $R_\lambda$ at small $St$
(without gravity),
including \cite{keswani04} (from data at $65 \leq R_\lambda \leq 152$),
\cite{bec07} ($65 \leq R_\lambda \leq 185$),
\cite{bec10a} ($185 \leq R_\lambda \leq 400$),
\cite{ray11} ($95 \leq R_\lambda \leq 227$),
and \cite{rosa13} ($28 \leq R_\lambda \leq 304$).
Our data confirms this point up to $R_\lambda = 597$.
The fact that $g(r/\eta)$ is independent of $R_\lambda$ 
for small Stokes numbers implies that the
clustering mechanism is driven almost entirely by the small-scale turbulence,
independent of any intermittency in the turbulence that occurs at higher Reynolds numbers.
For $St \gtrsim 1$, the RDFs increase with increasing $R_\lambda$,
in agreement with our expectations in \textsection \ref{sec:particle_clustering_theory}.

We note, however, that two recent studies \citep{onishi13,onishi14}
found that $g(r/\eta)$ decreases weakly with increasing $R_\lambda$
over the range $81 \leq R_\lambda \leq 527$ at $St = 0.4$ and $St = 0.6$.
Our results do not indicate such a trend, possibly
because we are unable to analyze $g(r/\eta)$ at separations
as low as those considered in \cite{onishi13} and \cite{onishi14}.
In any case, the trends with $R_\lambda$ at low $St$ reported here, in \cite{onishi13} and \cite{onishi14}, and in the rest
of the literature are at most very weak.

It is important to note, however, that just because $g(r/\eta)$ is invariant with $R_\lambda$ for low-$St$ particles
does not necessarily imply that higher-order moments of clustering are also independent of $R_\lambda$.
For example, $g(r/\eta)$ is related to the variance of the particle density field \citep{shaw02b}.
Higher-order moments or PDFs of the particle density field \citep[e.g., see][]{pan11} could also be compared
at different values of $R_\lambda$.  However, we found that the number of particles in
our simulations was insufficient to compute such statistics accurately at small separations.
We would likely need about an order of magnitude more particles 
to test the Reynolds-number dependence of these higher-order clustering moments.
Refer to \cite{yoshimoto07} for a more complete discussion on the number of particles
necessary for accurate higher-order clustering statistics.

\begin{figure}
 \centering
 \includegraphics[width=2.6in]{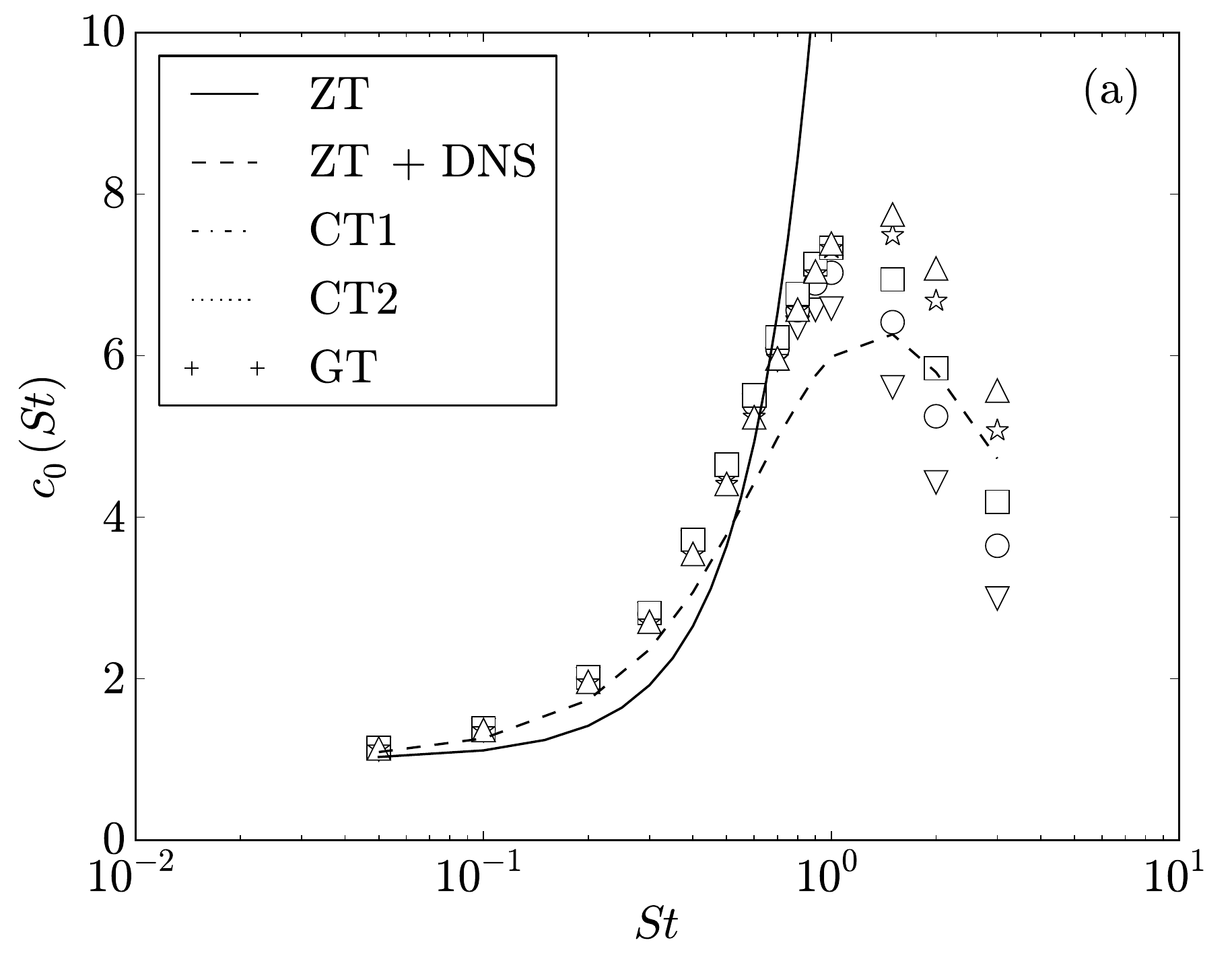}
 \includegraphics[width=2.6in]{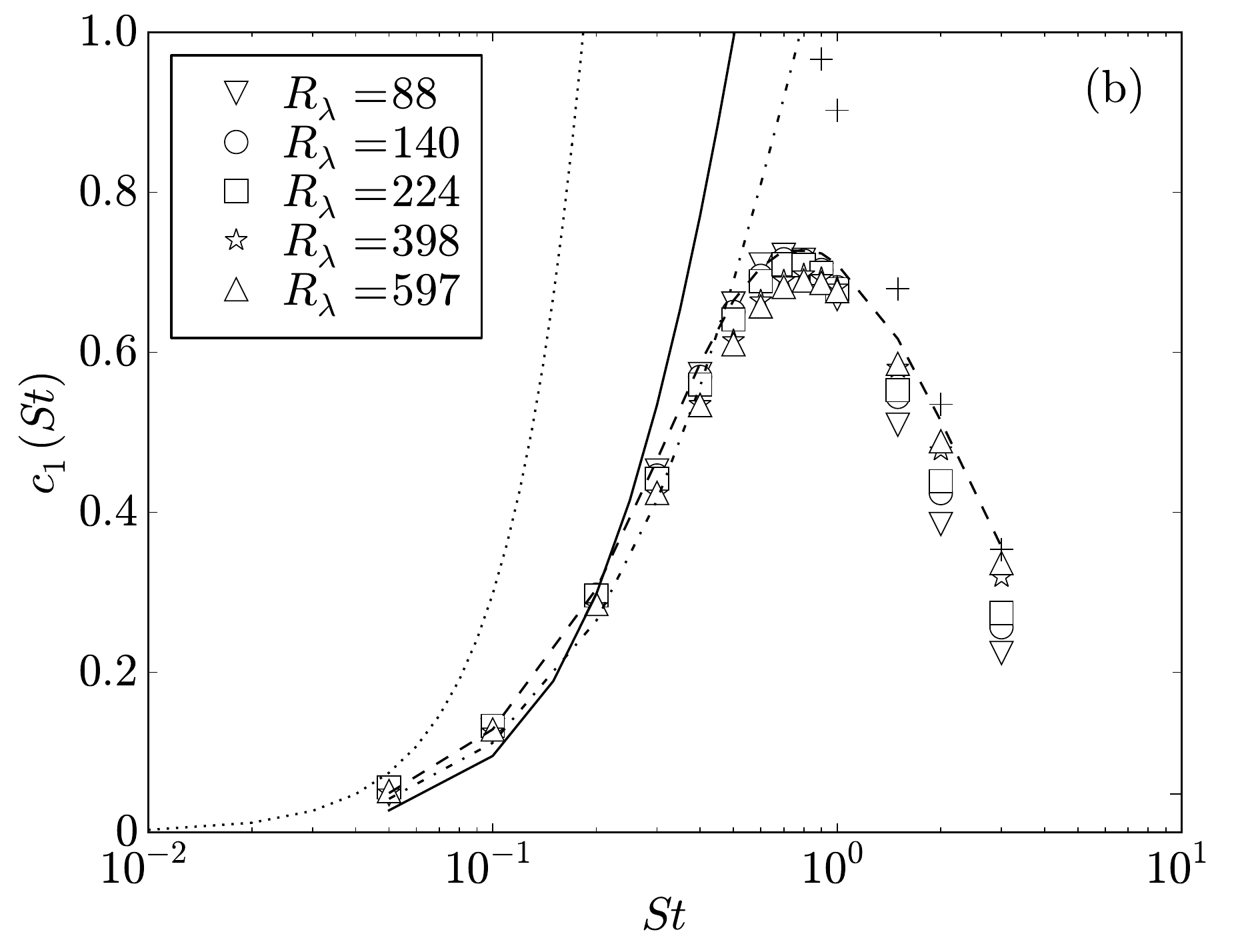}
 \caption{Power-law fits for $g(r/\eta)$ from (\ref{eq:gr_power_law}).
 (a) shows the coefficient $c_0$, and (b) shows the exponent $c_1$.
 DNS data are shown with symbols,
 and the theoretical predictions from \cite{zaichik09} (`ZT' and `ZT + DNS'),
 \cite{chun05} (`CT1' and `CT2'),
 and \cite{gustavsson11} (`GT') at $R_\lambda = 597$ are shown with lines and plus signs. The details of each 
 of the theoretical models are discussed in the text.}
 \label{fig:c0_c1}
\end{figure}

Following \cite{reade00}, we fit the RDFs by a power law of the form
\begin{equation}
\label{eq:gr_power_law}
 g(r/\eta) \approx c_0 \left( \frac{\eta}{r} \right)^{c_1} \mathrm{.}
\end{equation}
(Note that $c_1$ is related to the correlation dimension $\mathcal{D}_2$ \citep{bec07}
by the relation $c_1 = 3-\mathcal{D}_2$.)
This allows us to compare the DNS data to several theoretical predictions
in figure~\ref{fig:c0_c1}.
For each value of $R_\lambda$,
we computed $c_0$ and $c_1$ by fitting $g(r/\eta)$ in the range
$0.75 \leq r/\eta \leq 2.75$ using linear least-squares regression.
For $St \geq 10$, we do not observe power-law scaling for the RDF, 
and thus no values of $c_0$ and $c_1$ are plotted here.

To verify the arguments presented in \textsection \ref{sec:particle_clustering_theory},
we compare the DNS values of $c_0$ and $c_1$ to the predicted values from \cite{zaichik09} 
at $R_\lambda = 597$.
The comparisons are performed in two ways. In the first way (which we denote as `ZT'),
we use the model of \cite{zaichik09} to compute the relative velocities,
and then use these predicted relative velocities in (\ref{eq:zaichik_rdf})
to solve for the RDFs. In this manner, we can test the quantitative predictions
of the model when no additional inputs are used. In the second approach (which 
we denote as `ZT + DNS'), we solve (\ref{eq:zaichik_rdf}) 
with the particle velocities and the strain rate timescales along particle trajectories specified
using DNS data.
(The strain rate timescales are used in computing the dispersion tensor $\boldsymbol{\lambda}$.
To maintain consistency in the model, we also adjusted the inertial range timescales
through (18) in \cite{zaichik03}.) In both cases, we used the non-local diffusion correction
discussed in \cite{bragg14}, with $B_{nl} = 0.056$.

As expected, `ZT' is only able to provide a reasonable prediction for $c_0$ and $c_1$ 
for $St \lesssim 0.3$. Above this point, inaccuracies in the predicted
relative velocities lead to inaccurate clustering predictions, as discussed in \cite{bragg14}.
However, `ZT + DNS' predicts $c_1$ almost perfectly, with only slight discrepancies
at $St \sim 1$, in agreement with the findings of \cite{bragg14} at a lower Reynolds number.
We expect that these discrepancies are due to an additional
drift term that was omitted in \cite{zaichik09}, as discussed in \cite{bragg14}.
`ZT + DNS' also provides reasonable predictions for $c_0$, though the agreement
is not as good as that for $c_1$, possibly because $c_0$ is influenced by the inertial-range scales,
which are generally more difficult to model. From these comparisons,
we see that the model presented in \textsection \ref{sec:particle_clustering_theory} is accurate,
validating its use in interpreting the physical mechanisms responsible for particle clustering.

We next compare our results for $c_1$ against two relations derived in \cite{chun05}
in the limit of small $St$.
The first (which we denote as `CT1') uses DNS data for the strain and rotation
rates sampled along inertial-particle trajectories to compute $c_1$, giving
\begin{equation}
\label{eq:chun_S2-R2}
 c_1 = \frac{St \tau_\eta^2}{3 B_{nl}} \left(\langle \mathcal{S}^2 \rangle^p - \langle \mathcal{R}^2 \rangle^p \right) \mathrm{.}
\end{equation}
The second (which we denote as `CT2') requires only DNS data for quantities
sampled along fluid-particle trajectories and predicts,
\begin{equation}
\label{eq:chun_fluid_only}
\begin{split}
 c_1 = \frac{St^2}{12 B_{nl}} \Big[ & \frac{(\sigma^p_{\mathcal{S}^2})^2}{(\langle \mathcal{S}^2 \rangle^p)^2}
 \frac{T^p_{\mathcal{S}^2 \mathcal{S}^2}}{\tau_\eta} - \rho^p_{\mathcal{S}^2 \mathcal{R}^2}
 \frac{\sigma^p_{\mathcal{S}^2}}{\langle \mathcal{S}^2 \rangle^p}
 \frac{\sigma^p_{\mathcal{R}^2}}{\langle \mathcal{R}^2 \rangle^p}
 \left( \frac{T^p_{\mathcal{S}^2 \mathcal{R}^2}}{\tau_\eta} + \frac{T^p_{\mathcal{R}^2 \mathcal{S}^2}}{\tau_\eta} \right) \\
 &+ \frac{(\sigma^p_{\mathcal{R}^2})^2}{(\langle \mathcal{R}^2 \rangle^p)^2}
 \frac{T^p_{\mathcal{R}^2 \mathcal{R}^2}}{\tau_\eta}\Big] \mathrm{.}
\end{split}
\end{equation}

`CT1' agrees well with the DNS up to $St \approx 0.5$, 
while `CT2' only agrees well for $St = 0.05$, in agreement with \cite{chun05,bragg14}.
At higher values of $St$, both models from \cite{chun05} over-predict $c_1$.
As explained in \cite{bragg14},
this over-prediction is because the theory of \cite{chun05} fails to
account for the contribution of the path-history effects on the
drift and diffusion mechanisms that govern the clustering.

Finally, we compare our DNS values for $c_1$ against the theory from \cite{gustavsson11},
here denoted as `GT.'
The theory in \cite{gustavsson11} predicts that in the limit of small $r/\eta$,
\begin{equation}
\label{eq:mehlig_Sp2}
 S^p_{n\parallel} \propto r^{c_1} \mathrm{,}
\end{equation}
for $n > c_1$. 
(Note that the predictions of \cite{zaichik09} and \cite{gustavsson11} are equivalent
when $St$ is large, as explained in \cite{bragg14}.)
It therefore follows that for sufficiently small $r/\eta$, 
$c_1 = \zeta_\parallel^2$, where $\zeta_\parallel^2$
is the scaling exponent of the relative velocity variance in the dissipation range,
as computed in \textsection \ref{sec:wr_dissipation}. 

We include the prediction $c_1 = \zeta_2^\parallel$ in figure~\ref{fig:c0_c1},
and see that while `GT' is in excellent agreement with the DNS for $St=2,3$,
significant discrepancies exist at low $St$, as explained in \cite{bragg14}.

\subsection{Collision kernel}
\label{sec:collision_kernel}

We now consider the kinematic collision kernel $K$ for inertial particles,
which has been shown to depend on both the radial distribution function
and the radial relative velocities,
\begin{equation}
 K(d) = 4 \pi d^2 S^p_{-\parallel}(r=d) g(r=d) \mathrm{,}
\end{equation}
where $d$ is the particle diameter \citep[see][]{sundaram4,wwz98a}.
While we simulate only point-particles (refer to \textsection \ref{sec:particle_phase}), 
we compute $d$ from $St$ by assuming a given $\rho_p/\rho_f$.
To study the dependence of $K(d)$ on $\rho_p/\rho_f$, we consider
three different values for this parameter: $250$, $1000$, and $4000$. 
(Note that for droplets in atmospheric clouds, $\rho_p/\rho_f \approx 1000$.)

In general, we do not have adequate statistics to calculate $g(r)$ or 
$S^p_{-\parallel}(r)$ at $r=d$ at low values of $St$
($St \leq 3$ for $\rho_p/\rho_f = 250$ and $1000$, and $St \leq 10$ for $\rho_p/\rho_f = 4000$)
and so we extrapolate from the power-law fits 
in \textsection \ref{sec:wr_dissipation} and \textsection \ref{sec:particle_clustering_dns}
down to these separations, as was also done in \cite{rosa13}.
For larger $St$ ($St \geq 10$ for $\rho_p/\rho_f = 250$ and $1000$, and $St \geq 20$ for $\rho_p/\rho_f = 4000$), 
the particle diameters are sufficiently large such that
we can compute $g(d)$ and $S^p_{-\parallel}(d)$ 
by interpolating between data at smaller and larger separations.

\begin{figure}
 \centering
 \includegraphics[height=1.9in]{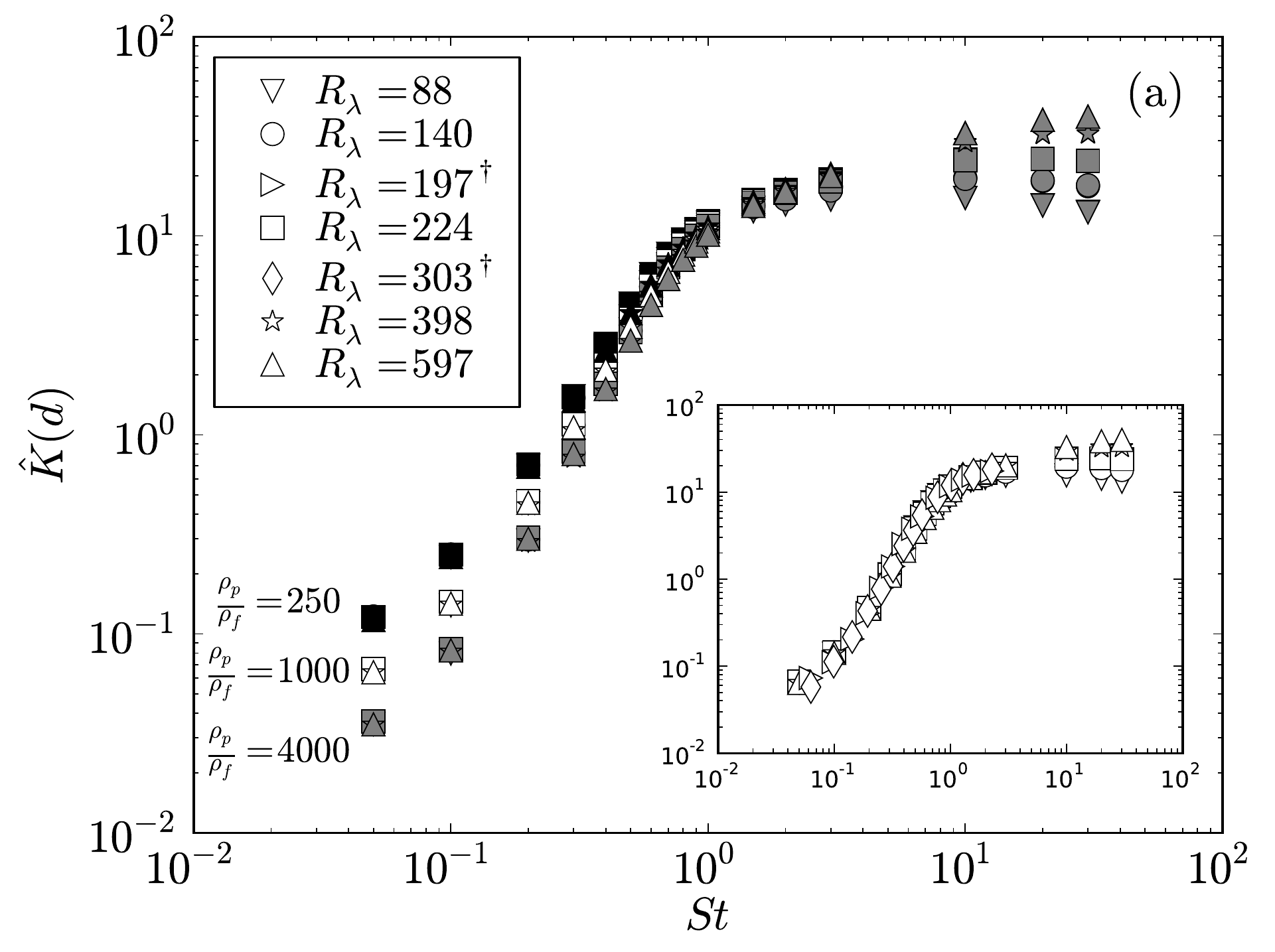}
 \includegraphics[height=1.9in]{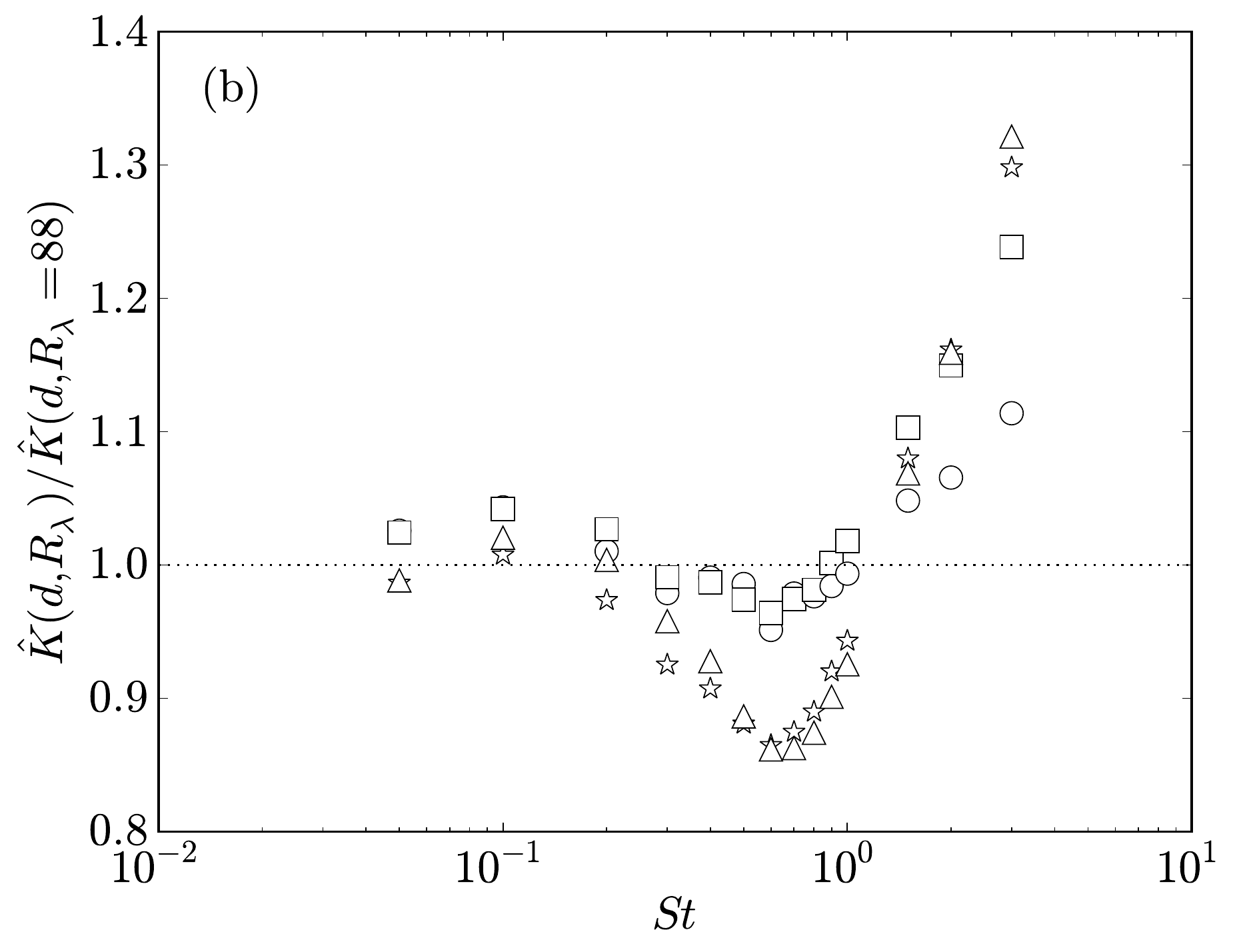}
  \caption{(a) The non-dimensional collision kernel $\hat{K}(d)$ as a function of $St$
  for different values of $R_\lambda$. Data are shown for $\rho_p/\rho_f = 250$ (filled black symbols),
  $\rho_p/\rho_f = 1000$ (open symbols), and $\rho_p/\rho_f = 4000$ (filled gray symbols).
  Legend entries marked with $\dagger$ indicate data taken from \cite{rosa13} (deterministic forcing scheme, no gravity)
  at $\rho_p/\rho_f = 1000$. These data are only included in the inset, where they are compared with 
  our results at $\rho_p/\rho_f = 1000$.
  (b) The ratio between $\hat{K}(d)$ at a given value of $R_\lambda$ to that
  at $R_\lambda = 88$, to highlight any Reynolds-number effects for $St \leq 3$.
  All data correspond to $\rho_p/\rho_f = 1000$.}
 \label{fig:collision_kernel}
\end{figure}

Following \cite{vosskuhle14}, we compute the non-dimensional collision kernel
$\hat{K}(d) \equiv K(d)/(d^2 u_\eta) = 4 \pi g(d) S^p_{-\parallel}(d)/u_\eta$.
Figure~\ref{fig:collision_kernel}(a) shows $\hat{K}(d)$
for different values of $\rho_p/\rho_f$.
Results from \cite{rosa13} (deterministic forcing scheme, no gravity, $\rho_p/\rho_f = 1000$) 
are included in the inset to Figure~\ref{fig:collision_kernel}(a).

For $St \geq 10$, the collision kernels increase strongly with increasing $R_\lambda$,
since both the relative velocities and the RDFs increase with $R_\lambda$ here
(see \textsection \ref{sec:wr_dissipation} and \textsection \ref{sec:particle_clustering_dns}).
$\hat{K}(d)$ is also independent of $\rho_p/\rho_f$ here. The physical explanation
is that while changes in $\rho_p/\rho_f$ lead to changes $d$, $S^p_{-\parallel}(d)/u_\eta$ 
and $g(d)$ are largely independent of $d$ here
(see \textsection \ref{sec:wr_dissipation} and \textsection \ref{sec:particle_clustering_dns}).

Such particles, however, are generally above the size range of droplets in atmospheric clouds 
\citep[e.g., see][]{ayala08a},
and thus our primary focus is on the collision rates of smaller ($St \lesssim 3$) particles.
$\hat{K}(d)$ is independent of $\rho_p/\rho_f$ for $1 \lesssim St \leq 3$,
in agreement with the findings of \cite{vosskuhle14}. In this case, while both $g(d)$ and $S^p_{-\parallel}/u_\eta$
are dependent on $d$, these two quantities have opposite scalings
(see \textsection \ref{sec:particle_clustering_dns}), causing
their product to be independent of $d$ (and thus of $\rho_p/\rho_f$).

For $St \lesssim 3$, our data show very little effect of $R_\lambda$ on the collision rates,
and are in good agreement with the collision statistics from \cite{rosa13} at $\rho_p/\rho_f = 1000$
(shown in the inset to figure~\ref{fig:collision_kernel}(a)).
However, since the Reynolds numbers in clouds ($R_\lambda \sim 10,000$) are at least an order of
magnitude larger than those in the DNS, it is important to discern even weak trends
in the collision kernel with the Reynolds number.
We therefore plot the ratio of $\hat{K}(d)$ at a given Reynolds number to that at $R_\lambda = 88$ for $St \leq 3$
in figure~\ref{fig:collision_kernel}(b).

At $St \lesssim 0.2$, the collision statistics are almost completely independent of $R_\lambda$, since 
both $S^p_{-\parallel}/u_\eta$ and $g$ are independent of $R_\lambda$ here
(refer to \textsection \ref{sec:wr_dissipation} and \textsection \ref{sec:particle_clustering_dns}).
For larger $St$, the collision kernel very weakly decreases with increasing $R_\lambda$,
since the mean inward relative velocities decrease with increasing $R_\lambda$
here (see \textsection \ref{sec:wr_dissipation}). Finally, for $1 < St \leq 3$,
the collision kernel increases weakly as $R_\lambda$ increases.
In this case, the increase in the RDFs with increasing $R_\lambda$ (\textsection \ref{sec:particle_clustering_dns})
overwhelms the decrease in the relative velocities (\textsection \ref{sec:wr_dissipation}),
causing the collision kernel to increase weakly.

These findings suggest that lower-Reynolds-number studies may in fact
capture the essential physics responsible for droplet collisions in highly
turbulent clouds. However, the results must be interpreted with caution for two reasons.
First, the collision rates for $St \leq 3$ were computed by extrapolating power-law fits
to very small separations, and it is not known if the functional form of the relative
velocities and the RDFs remains the same at these separations. 
Second, even the highest Reynolds numbers in this study
are still at least an order of magnitude smaller than those in atmospheric clouds.
It is thus possible that the turbulence could exhibit different characteristics at much
higher Reynolds numbers, or that the above trends in the Reynolds number,
though weak, could lead to substantially different collision rates when $R_\lambda$
is increased by another order of magnitude.
\section{Conclusions}
\label{sec:conclusions}
We have studied the effect of particle inertia and the flow Reynolds number on particle dynamics
at the highest Reynolds number ($R_\lambda \approx 600$)
and largest number of particles ($\sim 2.5$ billion) to date.
These simulations have provided new insights into both single- 
and two-particle statistics in homogeneous isotropic turbulence.

We first analyzed the statistics of individual inertial particles.
At large $St$, the particle motions were seen to be influenced primarily by inertial filtering.
The theoretical models of \cite{abrahamson75} and \cite{zaichik08}
were able to quantify the effect of filtering on kinetic energies and
particle accelerations, respectively, in this limit,
and provided us with a clear physical understanding of the effect of Reynolds number
on these quantities.

In the opposite limit ($St \ll 1$), the particle
motions were influenced primarily by preferential sampling,
and we used the theoretical model of \cite{chun05} to understand
and predict the statistics here.
For $St \ll 1$, the mean rotation rate sampled by the particles
decreased with increasing $St$ and $R_\lambda$, since intense rotation regions
became more prevalent and more efficient
at ejecting particles \citep[see][]{keswani04}. As $R_\lambda$ increased,
intense rotation regions tended to occur together
with intense strain regions in `vortex sheets,' in agreement with \cite{yeung12},
and particles were also ejected from these regions,
decreasing the mean strain rate sampled by the particles.
In agreement with \cite{salazar12a}, the particle kinetic energy increased
with $St$ for $St \ll 1$ due to preferential sampling of the flow field.
However, since ejections from vortex sheets
tend to reduce the particle kinetic energy, 
this trend was reduced as the Reynolds number was increased.
Fluid particle accelerations were seen to be extremely intermittent
at high $R_\lambda$, and the trends in the acceleration variance
were well-captured by the model of \cite{sawford03}.
The particle acceleration variances decreased rapidly with increasing $St$,
as inertial particles tended to be ejected from vortex tubes
and vortex sheets, which were both characterized by very high fluid accelerations.

We then studied the relative velocity, clustering, and collision statistics of inertial particles.
For $St \ll 1$, preferential sampling led to an increase in the longitudinal relative velocities
and to a decrease in the transverse relative velocities, and the relative velocities
were generally independent of $R_\lambda$ for $St \lesssim 0.1$.
At higher values of $St$, the particle motions
were influenced more by path-history interactions, leading to a sharp increase
in the relative velocities with increasing $St$. While the mean inward relative velocities
were generally independent of $R_\lambda$ for $0.2 \lesssim St \lesssim 1$, the relative velocity variances
increased weakly with increasing $R_\lambda$ here, a trend we attributed to 
either the increased scale separation at higher Reynolds numbers,
the increased intermittency of the turbulence at higher Reynolds numbers, or some combination of the two.
For intermediate $St$ ($1 \lesssim St \lesssim 3$), the relative velocities
decreased with increasing $R_\lambda$, which we argued was related to the decrease
in the Lagrangian rotation timescales with increasing $R_\lambda$.
We observed that the relative velocities of particles with $St \gtrsim 10$
increased with increasing $R_\lambda$, since inertial filtering effects
diminish and $u'/u_\eta$ increases as the Reynolds number increases.

We also analyzed the dissipation-range scaling exponents of the relative velocities,
and found that particles with higher relative velocities generally had lower scaling exponents,
since the particles were more influenced by path-history effects.
Relative velocities in the dissipation range were seen to be strongly non-Gaussian, with the
degree of non-Gaussianity being largest for $St \sim 1$, $r/\eta \rightarrow 0$,
and high $R_\lambda$, suggesting that theories which 
assume a Gaussian distribution to relate the velocity variances to the mean inward velocities
provide poor predictions for the mean inward
relative velocities at particle contact.
Higher-order inertial range structure functions were also examined
and were observed to follow similar trends to those reported in \cite{salazar12a}.

We then used these trends in the relative velocities
to predict the degree of clustering through the model of \cite{zaichik09},
and compared the results to DNS data.
The trends in the RDFs at low $St$ were tied to preferential sampling
effects, which increased the inward particle drift, as was found in \cite{chun05}. The RDFs were
independent of $R_\lambda$ here, in agreement with \cite{keswani04,ray11,rosa13},
suggesting that the non-local coefficient $B_{nl}$
\citep[see][]{chun05,bragg14} must weakly increase as $R_\lambda$ increases.
(We were unable to test higher-order measures of clustering
to determine if they were affected by changes in $R_\lambda$ due
to the limitations in the number of particles that could be simulated.)

At high $St$, the degree of clustering was tied to the influence
of path-history effects on the particle drift and diffusion, as explained in \cite{bragg14}.
By simplifying the model of \cite{zaichik09} in this limit,
we showed that changes in the scaling exponents of the relative velocity
variances directly affected the drift and diffusion mechanisms,
which in turn altered the clustering. The scaling exponents
generally increased with increasing $R_\lambda$ (suggesting
that path-history effects became less important), which in turn
led to increased levels of clustering.
For $St \geq 10$ and $R_\lambda \geq 224$, particles were seen to cluster in the inertial range
of turbulence, and the separation at which clustering decreased was predicted
accurately by inertial-range scaling arguments.

For $St \lesssim 3$, the RDFs exhibited power-law scaling,
consistent with \cite{reade00}. The full model of \cite{zaichik09}
(without any inputs from the DNS) was able to predict the power-law coefficient
$c_0$ and power-law exponent $c_1$ accurately only for $St \lesssim 0.4$
due to errors in the predicted relative velocities. However,
when these relative velocities (and the associated Lagrangian timescales) 
were specified from the DNS, the model in \cite{zaichik09} provided
excellent predictions for $c_1$ and reasonable predictions for $c_0$,
as was also found in \cite{bragg14} at a lower Reynolds number.
We also tested the DNS against two model predictions from \cite{chun05},
one which required only fluid particle statistics from the DNS,
and one which required strain and rotation statistics along particle trajectories.
The former prediction was in acceptable agreement with the DNS only for $St = 0.05$,
while the latter prediction was in good agreement up to $St \approx 0.5$, 
in agreement with \cite{chun05,bragg14}.
Finally, we found that the theory of \cite{gustavsson11} was able to predict $c_1$ well for $St = 2$ and $St = 3$.

We used the relative velocity and RDF data to compute 
the kinematic collision kernel for inertial particles \citep{sundaram4},
and found that this quantity varied only slightly with Reynolds number 
(under 50\% when $R_\lambda$ changed by a factor of 7) for $0 \leq St \leq 3$.
Our collision kernels were in good agreement with those computed by \cite{rosa13}.

As mentioned in \textsection \ref{sec:introduction},
one of the primary motivations for this study was to determine
the extent to which turbulence-induced collisions are responsible to the rapid
growth rate of droplets observed in warm, cumulus clouds.
Our observations indicate that the collision rates of like particles
are generally unaffected by changes in the Reynolds number,
which suggests that relatively low-Reynolds-number simulations
may allow us to study the essential physics of droplet collisions in 
highly turbulent atmospheric clouds. One promising avenue of future
work would be to determine the droplet growth rates predicted
by these collision kernels, either by solving an associated kinetic equation \citep{xue08,wang09}
or by simulating the particle collision and coalescence process directly \citep{reade00a}.

Finally, we note that it is unclear to what extent these conclusions
would be altered if gravity were incorporated in the particle dynamics,
since the introduction of gravity will likely cause particles
to preferentially sample certain regions of the flow,
and will alter the residence time of particles around certain flow features
\citep[e.g., see][]{wang93,davila01,good14}.
We will analyze the effect of gravity on inertial particle motion in turbulence in Part II \citep{ireland15b}.
\section*{Acknowledgements}
\label{sec:acknowledgements}
The authors gratefully acknowledge Parvez Sukheswalla for helpful discussions 
regarding this work.
This work was supported by the National Science Foundation through CBET grants 0756510 and 0967349,
and through a graduate research fellowship awarded to PJI. Additional funding was provided
by Cornell University. We would also like to acknowledge high-performance computing support 
from Yellowstone (ark:/85065/d7wd3xhc) provided by NCAR's Computational 
and Information Systems Laboratory through grants ACOR0001 and P35091057, 
sponsored by the National Science Foundation.

\bibliographystyle{jfm2}
\bibliography{refs}

\begin{thebibliography}{100}
\expandafter\ifx\csname natexlab\endcsname\relax\def\natexlab#1{#1}\fi

\bibitem[Abrahamson(1975)]{abrahamson75}
{\sc Abrahamson, J.} 1975 Collision rates of small particles in a vigorously
  turbulent fluid. {\em Chem. Eng. Sci.\/} {\bf 30}, 1371--1379.

\bibitem[Ashurst {\em et~al.\/}(1987)Ashurst, Kerstein, Kerr \&
  Gibson]{ashurst87}
{\sc Ashurst, W.~T., Kerstein, A.~R., Kerr, R.~M. \& Gibson, C.~H.} 1987
  Alignment of vorticity and scalar gradient with strain rate in simulated
  {Navier-Stokes} turbulence. {\em Phys. Fluids\/} {\bf 30}~(8), 2343--2353.

\bibitem[Ayala {\em et~al.\/}(2008)Ayala, Rosa, Wang \& Grabowski]{ayala08a}
{\sc Ayala, O., Rosa, B., Wang, L.-P. \& Grabowski, W.~W.} 2008 Effects of
  turbulence on the geometric collision rate of sedimenting droplets. part 1.
  results from direct numerical simulation. {\em New J. Phys.\/} {\bf 10},
  075015.

\bibitem[Ayyalasomayajula {\em et~al.\/}(2008)Ayyalasomayajula, Warhaft \&
  Collins]{sathya08a}
{\sc Ayyalasomayajula, S., Warhaft, Z. \& Collins, L.~R.} 2008 Modeling
  inertial particle acceleration statistics in isotropic turbulence. {\em Phys.
  Fluids\/} {\bf 20}, 094104.

\bibitem[Balachandar \& Eaton(2010)]{balachandar10}
{\sc Balachandar, S. \& Eaton, J.~K.} 2010 Turbulent dispersed multiphase flow.
  {\em Annu. Rev. Fluid Mech.\/} {\bf 42}, 111--133.

\bibitem[Bec {\em et~al.\/}(2006{\natexlab{{\em a\/}}})Bec, Biferale, Boffetta,
  Celani, Cencini, Lanotte, Musacchio \& Toschi]{bec06a}
{\sc Bec, J., Biferale, L., Boffetta, G., Celani, A., Cencini, M., Lanotte,
  A.~S., Musacchio, S. \& Toschi, F.} 2006{\natexlab{{\em a\/}}} Acceleration
  statistics of heavy particles in turbulence. {\em J. Fluid Mech.\/} {\bf
  550}, 349--358.

\bibitem[Bec {\em et~al.\/}(2006{\natexlab{{\em b\/}}})Bec, Biferale, Boffetta,
  Cencini, Musacchio \& Toschi]{bec06d}
{\sc Bec, J., Biferale, L., Boffetta, G., Cencini, M., Musacchio, S. \& Toschi,
  F.} 2006{\natexlab{{\em b\/}}} Lyapunov exponents of heavy particles in
  turbulence. {\em Phys. Fluids\/} {\bf 18}, 091702.

\bibitem[Bec {\em et~al.\/}(2007)Bec, Biferale, Cencini, Lanotte, Musacchio \&
  Toschi]{bec07}
{\sc Bec, J., Biferale, L., Cencini, M., Lanotte, A.~S., Musacchio, S. \&
  Toschi, F.} 2007 Heavy particle concentration in turbulence at dissipative
  and inertial scales. {\em Phys. Rev. Lett.\/} {\bf 98}, 084502.

\bibitem[Bec {\em et~al.\/}(2010{\natexlab{{\em a\/}}})Bec, Biferale, Cencini,
  Lanotte \& Toschi]{bec10a}
{\sc Bec, J., Biferale, L., Cencini, M., Lanotte, A.~S. \& Toschi, F.}
  2010{\natexlab{{\em a\/}}} Intermittency in the velocity distribution of
  heavy particles in turbulence. {\em J. Fluid Mech.\/} {\bf 646}, 527--536.

\bibitem[Bec {\em et~al.\/}(2010{\natexlab{{\em b\/}}})Bec, Biferale, Lanotte,
  Scagliarini \& Toschi]{bec10b}
{\sc Bec, J., Biferale, L., Lanotte, A.~S., Scagliarini, A. \& Toschi, F.}
  2010{\natexlab{{\em b\/}}} Turbulent pair dispersion of inertial particles.
  {\em J. Fluid Mech.\/} {\bf 645}, 497--528.

\bibitem[Benzi {\em et~al.\/}(1993)Benzi, Ciliberto, Tripiccione, Baudet,
  Massaioli \& Succi]{benzi93}
{\sc Benzi, R., Ciliberto, S., Tripiccione, R., Baudet, C., Massaioli, F. \&
  Succi, S.} 1993 Extended self-similarity in turbulent flows. {\em Phys. Rev.
  E\/} {\bf 48}, R29--R32.

\bibitem[Biferale {\em et~al.\/}(2005)Biferale, Boffetta, Celani, Lanotte \&
  Toschi]{biferale05b}
{\sc Biferale, L., Boffetta, G., Celani, A., Lanotte, A. \& Toschi, F.} 2005
  Particle trapping in three-dimensional fully developed turbulence. {\em Phys.
  Fluids\/} {\bf 17}, 021701.

\bibitem[Bragg \& Collins(2014{\natexlab{{\em a\/}}})]{bragg14}
{\sc Bragg, A.~D. \& Collins, L.~R.} 2014{\natexlab{{\em a\/}}} New insights
  from comparing statistical theories for inertial particles in turbulence:
  {I}. {S}patial distribution of particles. {\em New J. Phys.\/} {\bf 16},
  055013.

\bibitem[Bragg \& Collins(2014{\natexlab{{\em b\/}}})]{bragg14b}
{\sc Bragg, A.~D. \& Collins, L.~R.} 2014{\natexlab{{\em b\/}}} New insights
  from comparing statistical theories for inertial particles in turbulence:
  {II}: {R}elative velocities. {\em New J. Phys.\/} {\bf 16}, 055014.

\bibitem[Bragg {\em et~al.\/}(2015{\natexlab{{\em a\/}}})Bragg, Ireland \&
  Collins]{bragg15}
{\sc Bragg, A.~D., Ireland, P.~J. \& Collins, L.~R.} 2015{\natexlab{{\em a\/}}}
  Forward and backward in time dispersion of fluid and inertial particles in
  isotropic turbulence. {\em Phys. Fluids\/} {S}ubmitted. eprint
  arXiv:1403.5502.

\bibitem[Bragg {\em et~al.\/}(2015{\natexlab{{\em b\/}}})Bragg, Ireland \&
  Collins]{bragg15b}
{\sc Bragg, A.~D., Ireland, P.~J. \& Collins, L.~R.} 2015{\natexlab{{\em b\/}}}
  Mechanisms for the clustering of inertial particles in the inertial range of
  isotropic turbulence. {\em Phys. Rev. E\/} {I}n review. eprint
  arXiv:1411.7422.

\bibitem[Calzavarini {\em et~al.\/}(2008)Calzavarini, Kerscher, Lohse \&
  Toschi]{calzavarini08}
{\sc Calzavarini, E., Kerscher, M., Lohse, D. \& Toschi, F.} 2008
  Dimensionality and morphology of particle and bubble clusters in turbulent
  flow. {\em J. Fluid Mech.\/} {\bf 607}, 13--24.

\bibitem[Chun {\em et~al.\/}(2005)Chun, Koch, Rani, Ahluwalia \&
  Collins]{chun05}
{\sc Chun, J., Koch, D.~L., Rani, S., Ahluwalia, A. \& Collins, L.~R.} 2005
  Clustering of aerosol particles in isotropic turbulence. {\em J. Fluid
  Mech.\/} {\bf 536}, 219--251.

\bibitem[Collins \& Keswani(2004)]{keswani04}
{\sc Collins, L.~R. \& Keswani, A.} 2004 Reynolds number scaling of particle
  clustering in turbulent aerosols. {\em New J. Phys.\/} {\bf 6}, 119.

\bibitem[{Computational and Information Systems Laboratory}(2012)]{yellowstone}
{\sc {Computational and Information Systems Laboratory}} 2012 Yellowstone:
  {IBM} i{D}ata{P}lex {S}ystem ({U}niversity {C}ommunity {C}omputing).
  http://n2t.net/ark:/85065/d7wd3xhc.

\bibitem[Cuzzi {\em et~al.\/}(2001)Cuzzi, Hogan, Paque \&
  Dobrovolskis]{cuzzi01}
{\sc Cuzzi, J.~N., Hogan, R.~C., Paque, J.~M. \& Dobrovolskis, A.~R.} 2001
  Size-selective concentration of chrondrules and other small particles in
  protoplanetary nebula turbulence. {\em Astrophysical J.\/} {\bf 546},
  496--508.

\bibitem[D\'avila \& Hunt(2001)]{davila01}
{\sc D\'avila, J. \& Hunt, J. C.~R.} 2001 Settling of small particles near
  vortices and in turbulence. {\em J. Fluid Mech\/} {\bf 440}, 117--145.

\bibitem[Devenish {\em et~al.\/}(2012)Devenish, Bartello, Brenguier, Collins,
  Grabowski, I{J}zermans, Malinowski, Reeks, Vassilicos, Wang \&
  Warhaft]{devenish12}
{\sc Devenish, B.~J., Bartello, P., Brenguier, J.-L., Collins, L.~R.,
  Grabowski, W.~W., I{J}zermans, R. H.~A., Malinowski, S.~P., Reeks, M.~W.,
  Vassilicos, J.~C., Wang, L.-P. \& Warhaft, Z.} 2012 Droplet growth in warm
  turbulent clouds. {\em Q. J. R. Meteorol. Soc.\/} {\bf 138}, 1401--1429.

\bibitem[Durham {\em et~al.\/}(2013)Durham, Climent, Barry, Lillo, Boffetta,
  Cencini \& Stocker]{durham13}
{\sc Durham, W.~M., Climent, E., Barry, M., Lillo, F.~D., Boffetta, G.,
  Cencini, M. \& Stocker, R.} 2013 Turbulence drives microscale patches of
  motile phytoplankton. {\em Nat. Commun.\/} {\bf 4}~(2148), 1--7.

\bibitem[Eaton \& Fessler(1994)]{eaton94}
{\sc Eaton, J.~K. \& Fessler, J.~R.} 1994 Preferential concentration of
  particles by turbulence. {\em Int. J. Multiphase Flow\/} {\bf 20}, 169--209.

\bibitem[Elghobashi \& Truesdell(1992)]{elghobashi92}
{\sc Elghobashi, S.~E. \& Truesdell, G.~C.} 1992 Direct simulation of particle
  dispersion in a decaying isotropic turbulence. {\em J. Fluid Mech.\/} {\bf
  242}, 655.

\bibitem[Elghobashi \& Truesdell(1993)]{elghobashi93}
{\sc Elghobashi, S.~E. \& Truesdell, G.~C.} 1993 On the two-way interaction
  between homogeneous turbulence and dispersed particles. i: Turbulence
  modification. {\em Phys. Fluids A\/} {\bf 5}, 1790--1801.

\bibitem[El{M}aihy \& Nicolleau(2005)]{elmaihy05}
{\sc El{M}aihy, A. \& Nicolleau, F.} 2005 Investigation of the dispersion of
  heavy-particle pairs and {R}ichardson's law using kinematic simulation. {\em
  Phys. Rev. E\/} {\bf 71}, 046307.

\bibitem[Falkovich {\em et~al.\/}(2002)Falkovich, Fouxon \&
  Stepanov]{falkovich02}
{\sc Falkovich, G., Fouxon, A. \& Stepanov, M.~G.} 2002 Acceleration of rain
  initiation by cloud turbulence. {\em Nature\/} {\bf 419}, 151--154.

\bibitem[Falkovich \& Pumir(2007)]{falkovich07}
{\sc Falkovich, G. \& Pumir, A.} 2007 Sling effect in collisions of water
  droplets in turbulent clouds. {\em J. Atm. Sci.\/} {\bf 64}, 4497.

\bibitem[Good {\em et~al.\/}(2014)Good, Ireland, Bewley, Bodenschatz, Collins
  \& Warhaft]{good14}
{\sc Good, G.~H., Ireland, P.~J., Bewley, G.~P., Bodenschatz, E., Collins,
  L.~R. \& Warhaft, Z.} 2014 Settling regimes of inertial particles in
  isotropic turbulence. {\em J. Fluid Mech.\/} {\bf 759}, {R}3.

\bibitem[Goto \& Vassilicos(2006)]{goto06}
{\sc Goto, S. \& Vassilicos, J.~C.} 2006 Self-similar clustering of inertial
  particles and zero-acceleration points in fully developed two-dimensional
  turbulence. {\em Phys. Fluids\/} {\bf 18}, 115103.

\bibitem[Gotoh {\em et~al.\/}(2002)Gotoh, Fukayama \& Nakano]{gotoh02}
{\sc Gotoh, T., Fukayama, D. \& Nakano, T.} 2002 Velocity field statistics in
  homogeneous steady turbulence obtained using a high-resolution direct
  numerical simulation. {\em Phys. Fluids\/} {\bf 14}, 1065--1081.

\bibitem[Grabowski \& Wang(2013)]{grabowski13}
{\sc Grabowski, W.~W. \& Wang, L.-P.} 2013 Growth of cloud droplets in a
  turbulent environment. {\em Annu. Rev. Fluid Mech.\/} {\bf 45}, 293--324.

\bibitem[Gustavsson \& Mehlig(2011)]{gustavsson11}
{\sc Gustavsson, K. \& Mehlig, B.} 2011 Distribution of relative velocities in
  turbulent aerosols. {\em Phys. Rev. E\/} {\bf 84}, 045304.

\bibitem[Hill(2002)]{hill02}
{\sc Hill, R.~J.} 2002 Scaling of acceleration in locally isotropic turbulence.
  {\em J. Fluid Mech.\/} {\bf 452}, 361--370.

\bibitem[{v}an Hinsberg {\em et~al.\/}(2013){v}an Hinsberg, {t}en
  Thije~Bookkkamp, Toschi \& Clercx]{vanhinsberg13}
{\sc {v}an Hinsberg, M. A.~T., {t}en Thije~Bookkkamp, J. H.~M., Toschi, F. \&
  Clercx, H. J.~H.} 2013 Optimal interpolation schemes for particle tracking in
  turbulence. {\em Phys. Rev. E\/} {\bf 87}, 043307.

\bibitem[I{J}zermans {\em et~al.\/}(2010)I{J}zermans, Meneguz \&
  Reeks]{ijzermans10}
{\sc I{J}zermans, R. H.~A., Meneguz, E. \& Reeks, M.~W.} 2010 Segregation of
  particles in incompressible random flows: singularities, intermittency and
  random uncorrelated motion. {\em J. Fluid Mech.\/} {\bf 653}, 99--136.

\bibitem[Ireland {\em et~al.\/}(2015)Ireland, Bragg \& Collins]{ireland15b}
{\sc Ireland, P.~J., Bragg, A.~D. \& Collins, L.~R.} 2015 The effect of
  {R}eynolds number on inertial particle dynamics in isotropic turbulence.
  {P}art {II}: Simulations with gravitational effects. {\em J. Fluid Mech.\/}
  {S}ubmitted.

\bibitem[Ireland {\em et~al.\/}(2013)Ireland, Vaithianathan, Sukheswalla, Ray
  \& Collins]{ireland13}
{\sc Ireland, P.~J., Vaithianathan, T., Sukheswalla, P.~S., Ray, B. \& Collins,
  L.~R.} 2013 Highly parallel particle-laden flow solver for turbulence
  research. {\em Comput. Fluids\/} {\bf 76}, 170--177.

\bibitem[Ishihara {\em et~al.\/}(2009)Ishihara, Gotoh \& Kaneda]{ishihara09}
{\sc Ishihara, T., Gotoh, T. \& Kaneda, Y.} 2009 Study of
  high-{R}eynolds-number isotropic turbulence by direct numerical simulation.
  {\em Annu. Rev. Fluid Mech.\/} {\bf 41}, 165--180.

\bibitem[Ishihara {\em et~al.\/}(2007)Ishihara, Kaneda, Yokokawa, Itakura \&
  Uno]{ishihara07}
{\sc Ishihara, T., Kaneda, Y., Yokokawa, M., Itakura, K. \& Uno, A.} 2007
  Small-scale statistics in high-resolution direct numerical simualtion of
  turbulence: {R}eynolds number dependence of one-point velocity gradient
  statistics. {\em J. Fluid Mech.\/} {\bf 592}, 335--366.

\bibitem[Kaneda {\em et~al.\/}(2003)Kaneda, Ishihara, Yokokawa, Itakura \&
  Uno]{kaneda03}
{\sc Kaneda, Y., Ishihara, T., Yokokawa, M., Itakura, K. \& Uno, A.} 2003
  Energy dissipation rate and energy spectrum in high resolution direct
  numerical simulations of turbulence in a periodic box. {\em Phys. Fluids\/}
  {\bf 15}, L21--L24.

\bibitem[Kerr {\em et~al.\/}(2001)Kerr, Meneguzzi \& Gotoh]{kerr01}
{\sc Kerr, R.~M., Meneguzzi, M. \& Gotoh, T.} 2001 An inertial range crossover
  in structure functions. {\em Phys. Fluids\/} {\bf 13}, 1985--1994.

\bibitem[Kolmogorov(1941)]{kolmogorov41a}
{\sc Kolmogorov, A.~N.} 1941 The local structure of turbulence in an
  incompressible viscous fluid for very large {R}eynolds numbers. {\em Dokl.
  Akad. Nauk. SSSR\/} {\bf 30}, 299--303.

\bibitem[Kolmogorov(1962)]{kolmogorov62}
{\sc Kolmogorov, A.~N.} 1962 A refinement of previous hypotheses concerning the
  local structure of turbulence in a viscous incompressible fluid at high
  {R}eynolds number. {\em J. Fluid Mech.\/} {\bf 13}, 82--85.

\bibitem[Maxey(1987)]{maxey87b}
{\sc Maxey, M.~R.} 1987 The motion of small spherical particles in a celluar
  flow field. {\em Phys. Fluids\/} {\bf 30}, 1915--1928.

\bibitem[Maxey \& Riley(1983)]{maxey83}
{\sc Maxey, M.~R. \& Riley, J.~J.} 1983 Equation of motion for a small rigid
  sphere in a nonuniform flow. {\em Phys. Fluids\/} {\bf 26}, 883--889.

\bibitem[{McQuarrie}(1976)]{mcquarrie}
{\sc {McQuarrie}, D.~A.} 1976 {\em Statistical Mechanics\/}. Harper \& Row, New
  York.

\bibitem[Meneveau(2011)]{meneveau11}
{\sc Meneveau, C.} 2011 Lagrangian dynamics and models of the velocity gradient
  tensor in turbulent flows. {\em Annu. Rev. Fluid Mech.\/} {\bf 43}, 219--245.

\bibitem[Monchaux {\em et~al.\/}(2010)Monchaux, Bourgoin \&
  Cartellier]{monchaux10}
{\sc Monchaux, R., Bourgoin, M. \& Cartellier, A.} 2010 Preferential
  concentration of heavy particles: {A} {V}orono\"{i} analysis. {\em Phys.
  Fluids\/} {\bf 22}, 103304.

\bibitem[Onishi {\em et~al.\/}(2013)Onishi, Takahashi \& Vassilicos]{onishi13}
{\sc Onishi, R., Takahashi, K. \& Vassilicos, J.~C.} 2013 An efficient parallel
  simulation of interacting inertial particles in homogeneous isotropic
  turbulence. {\em J. Comput. Phys.\/} {\bf 242}, 809--827.

\bibitem[Onishi \& Vassilicos(2014)]{onishi14}
{\sc Onishi, R. \& Vassilicos, J.~C.} 2014 Collision statistics of inertial
  particles in two-dimensional homogeneous isotropic turbulence with an inverse
  cascade. {\em J. Fluid Mech.\/} {\bf 745}, 279--299.

\bibitem[Orszag \& Patterson(1972{\natexlab{{\em a\/}}})]{orszag72}
{\sc Orszag, S.~A. \& Patterson, G.~S.} 1972{\natexlab{{\em a\/}}} Numerical
  simulation of three-dimensional homogeneous isotropic turbulence. {\em Phys.
  Rev. Lett.\/} {\bf 28}, 76--79.

\bibitem[Orszag \& Patterson(1972{\natexlab{{\em b\/}}})]{orszag}
{\sc Orszag, S.~A. \& Patterson, G.~S.} 1972{\natexlab{{\em b\/}}} {\em
  Numerical simulation of turbulence\/}. Springer-Verlag, New York.

\bibitem[Pan \& Padoan(2010)]{pan10}
{\sc Pan, L. \& Padoan, P.} 2010 Relative velocity of inertial particles in
  turbulent flows. {\em J. Fluid Mech.\/} {\bf 661}, 73--107.

\bibitem[Pan \& Padoan(2013)]{pan13}
{\sc Pan, L. \& Padoan, P.} 2013 Turbulence-induced relative velocity of dust
  particles i: identical particles. {\em Ap{J}\/} {\bf 776}, 12.

\bibitem[Pan {\em et~al.\/}(2011)Pan, Padoan, Scalo, Kritsuk \& Norman]{pan11}
{\sc Pan, L., Padoan, P., Scalo, J., Kritsuk, A.~G. \& Norman, M.~L.} 2011
  Turbulent clustering of protoplanetary dust and planetesimal formation. {\em
  Ap{J}\/} {\bf 740}, 6.

\bibitem[Pekurovsky(2012)]{p3dfft}
{\sc Pekurovsky, D.} 2012 {P3DFFT}: A framework for parallel computations of
  {Fourier} transforms in three dimensions. {\em SIAM J. Sci. Comput.\/} {\bf
  34}~(4), C192--C209.

\bibitem[Pope(2000)]{pope}
{\sc Pope, S.~B.} 2000 {\em Turbulent Flows\/}. Cambridge University Press, New
  York.

\bibitem[Pruppacher \& Klett(1997)]{prupp97}
{\sc Pruppacher, H.~R. \& Klett, J.~D.} 1997 {\em Microphysics of Clouds and
  Precipitation\/}. Kluwer, Dordrecht.

\bibitem[Ray \& Collins(2011)]{ray11}
{\sc Ray, B. \& Collins, L.~R.} 2011 Preferential concentration and relative
  velocity statistics of inertial particles in {N}avier-{S}tokes turbulence
  with and without filtering. {\em J. Fluid Mech.\/} {\bf 680}, 488--510.

\bibitem[Ray \& Collins(2013)]{ray13}
{\sc Ray, B. \& Collins, L.~R.} 2013 Investigation of sub-kolmogorov inertial
  particle pair dynamics in turbulence using novel satellite particle
  simulations. {\em J. Fluid Mech.\/} {\bf 720}, 192--211.

\bibitem[Reade \& Collins(2000{\natexlab{{\em a\/}}})]{reade00}
{\sc Reade, W.~C. \& Collins, L.~R.} 2000{\natexlab{{\em a\/}}} Effect of
  preferential concentration on turbulent collision rates. {\em Phys. Fluids\/}
  {\bf 12}, 2530--2540.

\bibitem[Reade \& Collins(2000{\natexlab{{\em b\/}}})]{reade00a}
{\sc Reade, W.~C. \& Collins, L.~R.} 2000{\natexlab{{\em b\/}}} A numerical
  study of the particle size distribution of an aerosol undergoing turbulent
  coagulation. {\em J. Fluid Mech.\/} {\bf 415}, 45--64.

\bibitem[Rosa {\em et~al.\/}(2013)Rosa, Parishani, Ayala, Grabowski \&
  Wang]{rosa13}
{\sc Rosa, B., Parishani, H., Ayala, O., Grabowski, W.~W. \& Wang, L.~P.} 2013
  Kinematic and dynamic collision statistics of cloud droplets from
  high-resolution simulations. {\em New J. Phys.\/} {\bf 15}, 045032.

\bibitem[Salazar \& Collins(2012{\natexlab{{\em a\/}}})]{salazar12b}
{\sc Salazar, J. P. L.~C. \& Collins, L.~R.} 2012{\natexlab{{\em a\/}}}
  Inertial particle acceleration statistics in turbulence: effects of
  filtering, biased sampling, and flow topology. {\em Phys. Fluids\/} {\bf 24},
  083302.

\bibitem[Salazar \& Collins(2012{\natexlab{{\em b\/}}})]{salazar12a}
{\sc Salazar, J. P. L.~C. \& Collins, L.~R.} 2012{\natexlab{{\em b\/}}}
  Inertial particle relative velocity statistics in homogeneous isotropic
  turbulence. {\em J. Fluid Mech.\/} {\bf 696}, 45--66.

\bibitem[Sawford {\em et~al.\/}(2003)Sawford, Yeung, Borgas, {La Porta},
  Crawford \& Bodenschatz]{sawford03}
{\sc Sawford, B.~L., Yeung, P.-K., Borgas, M.~S., {La Porta}, P. V.~A.,
  Crawford, A.~M. \& Bodenschatz, E.} 2003 Conditional and unconditional
  acceleration statistics in turbulence. {\em Phys. Fluids\/} {\bf 15},
  3478--3489.

\bibitem[Shaw(2003)]{shaw03}
{\sc Shaw, R.~A.} 2003 Particle-turbulence interactions in atmospheric clouds.
  {\em Annu. Rev. Fluid Mech.\/} {\bf 35}, 183--227.

\bibitem[Shaw {\em et~al.\/}(2002)Shaw, Kostinski \& Larsen]{shaw02b}
{\sc Shaw, R.~A., Kostinski, B. \& Larsen, M.~L.} 2002 Towards quantifying
  droplet clustering in clouds. {\em Q. J. R. Meteorol. Soc.\/} {\bf 128},
  1043--1057.

\bibitem[Shen \& Warhaft(2002)]{shen02}
{\sc Shen, X. \& Warhaft, Z.} 2002 Longitudinal and transverse structure
  functions in sheared and unsheared wind-tunnel turbulence. {\em Phys.
  Fluids\/} {\bf 14}, 370--381.

\bibitem[Siebert {\em et~al.\/}(2006)Siebert, Lehmann \& Wendisch]{siebert06}
{\sc Siebert, H., Lehmann, K. \& Wendisch, M.} 2006 Observations of small-scale
  turbulence and energy dissipation rates in the cloudy boundary layer. {\em J.
  Atmos. Sci.\/} {\bf 63}, 1451--1466.

\bibitem[Soria {\em et~al.\/}(1994)Soria, Sondergaard, Cantwell, Chong \&
  Perry]{soria94}
{\sc Soria, J., Sondergaard, R., Cantwell, B.~J., Chong, M.~S. \& Perry, A.~E.}
  1994 A study of the fine-scale motions of incompressible time-developing
  mixing layers. {\em Phys. Fluids\/} {\bf 6}~(2), 871--884.

\bibitem[Spalart(1988)]{spalart88}
{\sc Spalart, P.~R.} 1988 Direct simulation of a turbulent boundary layer up to
  ${R}_\theta = 1410$. {\em J. Fluid Mech.\/} {\bf 187}, 61--98.

\bibitem[Squires \& Eaton(1991)]{squires91a}
{\sc Squires, K.~D. \& Eaton, J.~K.} 1991 Preferential concentration of
  particles by turbulence. {\em Phys. Fluids A\/} {\bf 3}, 1169--1178.

\bibitem[Sundaram \& Collins(1997)]{sundaram4}
{\sc Sundaram, S. \& Collins, L.~R.} 1997 Collision statistics in an isotropic,
  particle-laden turbulent suspension {I}. {D}irect numerical simulations. {\em
  J. Fluid Mech.\/} {\bf 335}, 75--109.

\bibitem[Sundaram \& Collins(1999)]{sundaram6}
{\sc Sundaram, S. \& Collins, L.~R.} 1999 A numerical study of the modulation
  of isotropic turbulence by suspended particles. {\em J. Fluid Mech.\/} {\bf
  379}, 105--143.

\bibitem[Tagawa {\em et~al.\/}(2012)Tagawa, Mercado, Prakash, Calzavarini, Sun
  \& Lohse]{tagawa12}
{\sc Tagawa, Y., Mercado, J.~M., Prakash, V.~N., Calzavarini, E., Sun, C. \&
  Lohse, D.} 2012 Three-dimensional {L}agrangian {V}orono\"{i} analysis for
  clustering of particles and bubbles in turbulence. {\em J. Fluid Mech.\/}
  {\bf 693}, 201--215.

\bibitem[Tavoularis {\em et~al.\/}(1978)Tavoularis, Bennett \&
  Corrsin]{tavoularis78}
{\sc Tavoularis, S., Bennett, J.~C. \& Corrsin, S.} 1978 Velocity-derivative
  skewness in small {R}eynolds number, nearly isotropic turbulence. {\em J.
  Fluid Mech.\/} {\bf 88}, 63--69.

\bibitem[{van Hinsberg} {\em et~al.\/}(2012){van Hinsberg}, {Thije Boonkkamp},
  {Toschi} \& {Clercx}]{vanhinsberg12}
{\sc {van Hinsberg}, M.~A.~T., {Thije Boonkkamp}, J.~H.~M., {Toschi}, F. \&
  {Clercx}, H.~J.~H.} 2012 On the efficiency and accuracy of interpolation
  methods for spectral codes. {\em SIAM J. Sci. Comput.\/} {\bf 34}~(4),
  B479--B498.

\bibitem[Vo{\ss}kuhle {\em et~al.\/}(2014)Vo{\ss}kuhle, Pumir, L\'ev\^eque \&
  Wilkinson]{vosskuhle14}
{\sc Vo{\ss}kuhle, M., Pumir, A., L\'ev\^eque, E. \& Wilkinson, M.} 2014
  Prevalence of the sling effect for enhancing collision rates in turbulent
  suspensions. {\em J. Fluid Mech.\/} {\bf 749}, 841--852.

\bibitem[Voth {\em et~al.\/}(2002)Voth, {La Porta}, Crawford, Alexander \&
  Bodenschatz]{voth02}
{\sc Voth, G.~A., {La Porta}, A., Crawford, A.~M., Alexander, J. \&
  Bodenschatz, E.} 2002 Measurement of particle accelerations in fully
  developed turbulence. {\em J. Fluid Mech.\/} {\bf 469}, 121--160.

\bibitem[Wang \& Grabowski(2009)]{wang09}
{\sc Wang, L.-P. \& Grabowski, W.~W.} 2009 The role of air turbulence in warm
  rain initiation. {\em Atmos. Sci. Let.\/} {\bf 10}, 1--8.

\bibitem[Wang \& Maxey(1993)]{wang93}
{\sc Wang, L.-P. \& Maxey, M.~R.} 1993 Settling velocity and concentration
  distribution of heavy particles in homogeneous isotropic turbulence. {\em J.
  Fluid Mech.\/} {\bf 256}, 27--68.

\bibitem[Wang {\em et~al.\/}(1998)Wang, Wexler \& Zhou]{wwz98a}
{\sc Wang, L.-P., Wexler, A.~S. \& Zhou, Y.} 1998 Statistical mechanical
  descriptions of turbulent coagulation. {\em Phys. Fluids\/} {\bf 10},
  2647--2651.

\bibitem[Wang {\em et~al.\/}(2000)Wang, Wexler \& Zhou]{wwz00}
{\sc Wang, L.-P., Wexler, A.~S. \& Zhou, Y.} 2000 Statistical mechanical
  description and modeling of turbulent collision of inertial particles. {\em
  J. Fluid Mech.\/} {\bf 415}, 117--153.

\bibitem[Wilkinson \& Mehlig(2005)]{wilkinson05}
{\sc Wilkinson, M. \& Mehlig, B.} 2005 Caustics in turbulent aerosols. {\em
  Europhys. Lett.\/} {\bf 71}, 186--192.

\bibitem[Wilkinson {\em et~al.\/}(2006)Wilkinson, Mehlig \&
  Bezuglyy]{wilkinson06}
{\sc Wilkinson, M., Mehlig, B. \& Bezuglyy, V.} 2006 Caustic activation of rain
  showers. {\em Phys. Rev. Lett.\/} {\bf 97}, 048501.

\bibitem[Witkowska {\em et~al.\/}(1997)Witkowska, Brasseur \&
  Juv\'{e}]{witkowska97}
{\sc Witkowska, A., Brasseur, J.~G. \& Juv\'{e}, D.} 1997 Numerical study of
  noise from isotropic turbulence. {\em J. Comput. Acoust.\/} {\bf 5},
  317--336.

\bibitem[Xue {\em et~al.\/}(2008)Xue, Wang \& Grabowski]{xue08}
{\sc Xue, Y., Wang, L.-P. \& Grabowski, W.~W.} 2008 Growth of cloud droplets by
  turbulent collision-coalescence. {\em J. Atmos. Sci.\/} {\bf 65}, 331--356.

\bibitem[Yeung {\em et~al.\/}(2012)Yeung, Donzis \& Sreenivasan]{yeung12}
{\sc Yeung, P.~K., Donzis, D.~A. \& Sreenivasan, K.~R.} 2012 Dissipation,
  enstrophy, and pressure statistics in turbulence simulations at high reynolds
  numbers. {\em J. Fluid Mech.\/} {\bf 700}, 5--15.

\bibitem[Yeung \& Pope(1989)]{yeung89}
{\sc Yeung, P.~K. \& Pope, S.~B.} 1989 Lagrangian statistics from direct
  numerical simulations of isotropic turbulence. {\em J. Fluid Mech.\/} {\bf
  207}, 531--586.

\bibitem[Yeung {\em et~al.\/}(2006)Yeung, Pope, Lamorgese \& Donzis]{yeung06b}
{\sc Yeung, P.~K., Pope, S.~B., Lamorgese, A.~G. \& Donzis, D.~A.} 2006
  Acceleration and dissipation statistics of numerically simulated isotropic
  turbulence. {\em Phys. Fluids\/} {\bf 18}~(6), 065103.

\bibitem[Yoshimoto \& Goto(2007)]{yoshimoto07}
{\sc Yoshimoto, H. \& Goto, S.} 2007 Self-similar clustering of inertial
  particles in homogeneous turbulence. {\em J. Fluid Mech.\/} {\bf 577},
  275--286.

\bibitem[Yudine(1959)]{yudine59}
{\sc Yudine, M.~I.} 1959 Physical considerations on heavy-particle dispersion.
  {\em Adv. Geophys.\/} {\bf 6}, 185--191.

\bibitem[Zaichik \& Alipchenkov(2003)]{zaichik03}
{\sc Zaichik, L.~I. \& Alipchenkov, V.~M.} 2003 Pair dispersion and
  preferential concentration of particles in isotropic turbulence. {\em Phys.
  Fluids\/} {\bf 15}, 1776--1787.

\bibitem[Zaichik \& Alipchenkov(2008)]{zaichik08}
{\sc Zaichik, L.~I. \& Alipchenkov, V.~M.} 2008 Acceleration of heavy particles
  in isotropic turbulence. {\em Int. J. Multiphase Flow\/} {\bf 34}~(9),
  865--868.

\bibitem[Zaichik \& Alipchenkov(2009)]{zaichik09}
{\sc Zaichik, L.~I. \& Alipchenkov, V.~M.} 2009 Statistical models for
  predicting pair dispersion and particle clustering in isotropic turbulence
  and their applications. {\em New J. Phys.\/} {\bf 11}, 103018.

\bibitem[Zaichik {\em et~al.\/}(2003)Zaichik, Simonin \&
  Alipchenkov]{zaichik03a}
{\sc Zaichik, L.~I., Simonin, O. \& Alipchenkov, V.~M.} 2003 Two statistical
  models for predicting collision rates of inertial particles in homogeneous
  isotropic turbulence. {\em Phys. Fluids\/} {\bf 15}, 2995--3005.

\end{thebibliography}

\end{document}